\documentclass[aps,prd,superscriptaddress,nofootinbib,amsmath, showkeys, showpacs,amsfonts,preprintnumbers,notitlepage,10pt,english]{revtex4-1}
\usepackage{amsmath}
\usepackage{amssymb}
\usepackage{babel}

\usepackage{bm}
\usepackage{booktabs}
\usepackage{multirow}
\usepackage{array}
\usepackage{enumitem}
\usepackage{mathtools}
\makeatletter
\usepackage{array,multirow,graphicx}
\usepackage{dcolumn}
\usepackage{newlfont}
\usepackage{bm}
\usepackage[colorlinks,citecolor=blue,urlcolor=blue,linkcolor=blue]{hyperref}
\usepackage[figtopcap]{subfigure}
\usepackage{color}

\begin{document}

\date{\today}

\title{Anisotropic compact stars  in $f(R)$ gravity}
\author{G.G.L. Nashed}%
\email{nashed@bue.edu.eg}
\affiliation{Centre for Theoretical Physics, The British University in Egypt, P.O. Box 43, El Sherouk City, Cairo 11837, Egypt}
\affiliation{Egyptian Relativity Group (ERG), Cairo University, Giza 12613, Egypt}
\author{S. Capozziello}%
\email{capozziello@na.infn.it}
\affiliation{Dipartimento di Fisica ``E. Pancini``, Universit\'a di Napoli ``Federico II'',
Complesso Universitario di Monte Sant' Angelo, Edificio G, Via Cinthia, I-80126, Napoli, Italy}
\affiliation{ Istituto Nazionale di Fisica Nucleare (INFN),  Sezione di Napoli,
Complesso Universitario di Monte Sant'Angelo, Edificio G, Via Cinthia, I-80126, Napoli, Italy}
\affiliation{ Scuola Superiore Meridionale,  Largo S. Marcellino, 10, 80138 Napoli, Italy}
\affiliation{Laboratory for Theoretical Cosmology, Tomsk State University of Control Systems and Radioelectronics (TUSUR), 634050 Tomsk, Russia.}

\begin{abstract}

We derive a new interior solution for stellar compact objects in $f\mathcal{(R)}$ gravity  assuming a   differential relation to constrain the Ricci curvature scalar. To this aim, we consider  specific  forms for the radial component of the metric  and  the first derivative of  $f\mathcal{(R)}$. After,  the time component of the metric potential and the form of $f(\mathcal R)$ function are derived. From these results, it is possible to  obtain  the radial and tangential components of  pressure and the density. The resulting  interior solution  represents  a  physically motivated anisotropic neutron star model.  It is possible to  match it with a boundary exterior solution. From this matching, the components of  metric potentials can be rewritten in terms of a compactness parameter $C$ which has to be  $C=2GM/Rc^2 <<0.5$ for physical consistency.  Other physical conditions for real stellar objects are taken into account  according to the  solution. We show that  the model accurately bypasses  conditions like the finiteness of  radial and tangential pressures, and energy density  at the center of the star,  the positivity of these components through the stellar structure, and the negativity of the gradients.  These conditions are satisfied if  the energy-conditions hold. Moreover, we study the stability of the model by showing that Tolman-Oppenheimer-Volkoff equation is at hydrostatic equilibrium. The solution is matched with observational data of millisecond pulsars with a withe dwarf companion and pulsars presenting thermonuclear bursts.
\end{abstract}


\maketitle
\section{\bf Introduction}
Compact stars are  the final stage of stellar evolution when the  radial pressure, caused by the nuclear fusion in the inner core, is no longer able to oppose the  gravitational forces. Gravitational compact objects involve white dwarfs, neutron stars, and black holes. The interior structure as well as the physical properties of compact stars are studied adopting  the Einstein field equations of General Relativity (GR). Schwarzschild presented a first solution capable of   describing the interior of the compact body in  hydrostatic equilibrium \cite{1916SPAW.......189S}.

The stellar core was first assumed to be a fluid with equivalent  radial and tangential pressures \cite{Ivanov:2002xf}. However, the chances of generating anisotropies in compact stars are considerably higher due to the existence of  strong interactions between particles.  In this situation,  the inner regions become non-uniform  as discussed in \cite{Maurya_2017}. Specifically,  it is the fundamental nature of particles which can generate  anisotropies. As a result,  anisotropic compact stars can become more compact than isotropic stars. The consequences of local variations of  relative inner fields were discussed using the equation of state and some interesting results were presented  in \cite{Bowers:1974tgi}.  Ruderman suggested that a nuclear system with very high densities indicates that anisotropic pressure can emerge in compact stars \cite{doi:10.1146/annurev.aa.10.090172.002235}. In such a case, the radial and tangential pressures are different within the stellar core.
These problems can find room in the context of modified gravity.

Due to  the issue that GR is not capable to  address the whole gravitational  phenomenology from  ultraviolet to infrared  scales, modified and extended    theories of gravity have been proposed as effective approaches towards quantum gravity, from one side, and towards   astrophysics and cosmology from the other side. These  theories are considered  very useful to investigate the expansion of the universe  as well as  the dark energy and dark matter issues which could be explained under the standard of geometrical effects considering curvature and torsion invariants into the gravitational action \cite{Shamir_2017,Shamir_2017,Li_2011,Harko_2011,Capozziello_2011,Nojiri_2011,Nojiri_2006,Lovelock:1971yv,Shirafuji:1997wy,Harko_2011,Shamir:2019bcw,Nojiri_2017,Sharif_2016}.    In  particular,  the so-called $f\mathcal{(R)}$ gravity has gained special interest due to the fact that it is a straightforward  generalization of the Hilbert-Einstein action  of GR.  Various models  of $f\mathcal{(R)}$ gravity have been studied   to investigate    problems  like the late cosmic acceleration \cite{Capozziello:2006dj,Amendola:2006eh,Amendola_2007}, the ghosts of gravitational field \cite{Santos:2005pe,Carroll_2003,Alcaniz_2004}, the inflationary epoch \cite{Barrow_2006} and the whole  expansion of the  universe \cite{Visser_1997,Santos_2006,Santos_2007}.  In  $f\mathcal{(R)}$ gravity, several  basic results have been  achieved  like  the categorizations of singularities, the causal structure and the classification of energy conditions \cite{Rebou_as_2009,Santos_2010,Wang:2012mws,Banijamali_2011,Wang:2012rw}.

Using such modified theories, it is possible  to explain the collapse and stability of astrophysical  objects which escape the standard picture of GR. For example the supermassive compact object resulting from the gravitational wave event GW190814 can be framed and interpreted in the context of $f(\mathcal{R})$ gravity (see \cite{Astashenok:2020qds} and references therein).

In this framework,  the study of compact stars has recently become an interesting research  topic  because observations are reporting several anomalous objects which cannot be explained by  GR and standard equations of state. For example, anisotropic compact stars can be analyzed considering  GR with  a cosmological constant \cite{MonowarHossein:2012ec}.  The compact structure can be discussed considering  various forms of the energy density and radial pressure  \cite{Ivanov_2017}. Singularity-free solution for compact stars  can be derived in \cite{Deb_2017}. The homotopy perturbations procedure can be  applied to obtain stable structures \cite{Aziz:2016yub}, and  perturbative models of  $f\mathcal{(R)}$  gravity have been applied in \cite{Astashenok_2013} to study  neutron stars.  Neutron stars with quarks cores can also be considered, and  results can be physically consistent  \cite{Astashenok_2016}. The quadratic and cubic forms of  $f\mathcal{(R)}$  were discussed using numerical  methods in \cite{Astashenok_2014}, while  anisotropic compact stars in the context of $f\mathcal{(R)}$ gravity were studied in \cite{zubair2014study}. Neutron and quark stars are studied in the framework of   $f\mathcal{(R)} =  \mathcal{R}+ \alpha\mathcal{R}^2$  while  the Brans-Dicke theory is taken into account in  \cite{Astashenok:2017dpo}. In \cite{Astashenok_2015}, quark star models are derived with an equation of  state in the non-perturbative form of  $f\mathcal{(R)}$ gravity.

Besides these results, it is worth noticing that several   modified gravities are  receiving much attention to cure the shortcomings  of GR like the cosmic accelerated expansion,  the flat rotation curves of galaxies, possible  wormhole solutions  and other  phenomena  \cite{DeFelice:2010aj,CAPOZZIELLO2011167,Nojiri:2006ri,universe1020123,Bamba:2012cp}. For example, a  quadratic function of the Ricci scalar was studied  by Starobinsky  to address the inflationary behavior of early universe \cite{1979ZhPmR..30..719S}. It was shown that  higher-order  $f\mathcal{(R)}$ gravity  can solve  issues related to   supermassive neutron stars \cite{Astashenok_2013,PhysRevD.89.103509,Astashenok_2015,Astashenok_2017,Astashenok:2016epm}. The easiest way to prescribe   $f\mathcal{(R)}$ gravity  is assuming an analytic  function whose first order is the Ricci scalar. In general, the  equations of motion of $f\mathcal{(R)}$ are higher-order in derivatives and supply substantial classes of solutions that are different from GR. In the framework of $f\mathcal{(R)}$ gravity, the dynamical behavior of  matter and curvature  fields has been studied in several works. See, for example, \cite{6847167,Capozziello_2013,10.1093/pasj/psx099,PhysRevD.74.046004,Shah:2019mxn,Nojiri:2019dqc,Odintsov:2019ofr,PhysRevD.99.064049,Nascimento:2018sir,Miranda:2018jhu,Astashenok:2018bol,PhysRevD.99.063506,Elizalde:2018now,PhysRevD.99.064025,PhysRevD.99.104046,
Bombacigno:2018tyw,Capozziello:2018ddp,Samanta:2019tjb}.

In particular,   spherically symmetric vacuum black hole solutions in  $f\mathcal{(R)}$  have been  derived in \cite{PhysRevD.74.064022,2018EPJP..133...18N,2018IJMPD..2750074N,Nashed:2018piz}. Furthermore, from the existence of  Noether symmetries,   it is possible to find out other  spherically symmetric solutions  \cite{Capozziello_2007,2012GReGr..44.1881C}. Using the same technique, it is possible to achieve also  axially symmetric  black hole solutions  \cite{Capozziello_2010}. Non-trivial spherically symmetric black hole solutions for  specific classes of  $f\mathcal{(R)}$ are derived also  in \cite{Elizalde:2020icc,Nashed:2019yto,Nashed:2019tuk}. Due to the higher-order curvature terms in $f\mathcal{(R)}$,  one can discuss  strong gravitational fields in local objects. In this framework,  many researches  concentrate  on the study of spherically symmetric  black holes \cite{Sultana:2018fkw,Canate:2017bao,Yu:2017uyd,Canate:2015dda,Kehagias:2015ata,PhysRevD.82.104026,delaCruzDombriz:2009et}
and neutron stars  \cite{Feng:2017hje,Resco:2016upv,Capozziello:2015yza,Nashed:2006yw,
Staykov:2018hhc,Pant:2020rqq,Tamta:2019ewt,Gedela:2019ssk,Gedela:2019ckx,Gedela:2018oox,2018EPJA...54..179F,Doneva:2016xmf,Yazadjiev:2016pcb,Nashed:2004pn,
Yazadjiev:2015zia,Yazadjiev:2014cza,Ganguly:2013taa,Astashenok:2013vza,Orellana:2013gn,Arapoglu:2010rz,Cooney:2009rr}.

Finally, it is worth  noticing  that  $f\mathcal{(R)}$ gravity can be recast as scalar-tensor gravity \cite{PhysRev.124.925} with a scalar potential of gravitational origin so the further degrees of freedom with respect to GR result as  scalar field counterparts in the Einstein field equations
\cite{Chiba:2003ir,PhysRevLett.29.137,Chakraborty:2016gpg,Chakraborty:2016ydo}.

The aim of the present paper is  deriving   a spherically symmetric inner  solution   assuming  general forms of  the Ricci scalar and  the $\mathit{f(R)}$ function. After, we want  to test the  physical reliability  of this solution.
According to the procedure,  it  is possible to construct self-consistent compact stellar models capable of describing anisotropic neutron stars.

The layout of the paper is the following: In Sec. \ref{S2},   a short summary of    $f\mathcal{(R)}$  gravity is reported. In Sec. \ref{S3},  we apply the field equations of $\mathit{f(R)}$ to a spherically symmetric metric with two different potentials for time and radial components.  Assuming the  form  of  radial component as well as of  the first derivative of $f\mathcal{(R)}$,  we are  able to obtain  the  time component  of  the metric and the  $f(\mathcal R)$ function. Starting from this  information,  we derive the   density,  the radial and  the tangential pressures.   In Sec. \ref{S4},  we impose  boundary conditions to match the interior  with the exterior solution. We succeed in matching the constants  characterizing the  solution by a compactness parameter $C$ which satisfies   the  constrain $C<<0.5$.  In Sec. \ref{S5}, we discuss   the physical  conditions to show that the model is consistent with a  realistic compact star. In Sec. \ref{S6},  the stability, adopting   the Tolman-Oppenheimer-Volkoff (TOV) approach and the adiabatic index, is discussed. We  show  that the model satisfies both these conditions. The results are matched with observational data in   Sec. \ref{S7} where we  draw  the conclusions.

\section{A summary of $f(R)$ gravity}\label{S2}
In 4-dimension,  the action  of $f\mathcal{(R)}$  gravity is  \cite{Carroll:2003wy,1970MNRAS.150....1B,Nojiri:2003ft,Nashed:2009hn,Awad:2017ign,Capozziello:2003gx,Capozziello:2011et,2017JHEP...07..136A,Nashed:2001im,Nojiri:2010wj,Nojiri:2017ncd,Capozziello:2002rd}:
\begin{eqnarray} \label{a2} {\mathop{\mathcal{ E}}}:=\frac{1}{2\kappa} \int d^4x \sqrt{-g}  f\mathcal{(R)}+\frac{1}{2\kappa} \int d^4x \sqrt{-g} {\mathcal L_M}(g_{\mu \nu},\xi)\,,\end{eqnarray}
with the  gravitational coupling ${\displaystyle \kappa=\frac{8\pi G}{c^4}}$, where $G$ is the Newtonian constant, $c$ is the speed of light and  $g$ is the determinant of the metric. Here $f(\mathcal{R})$ is an arbitrary function of the Ricci scalar $\mathcal R$ and  $ {\mathcal L_M}(g_{\mu \nu},\xi)$  is the action of generic matter fields $\xi$ which are  minimally coupled to the metric $g_{\mu \nu}$. Clearly, GR is recovered for $f({\mathcal R})={\mathcal R}$.

The variations of  action (\ref{a2}) w.r.t. the metric tensor $g_{\mu \nu}$  gives  the non-vacuum field equations  \cite{2005JCAP...02..010C}
\begin{eqnarray} \label{f1}
{\mathop{\mathcal{ E}}}_{\mu \nu}:=\mathcal{ R}_{\mu \nu} f\mathcal{ }_{_\mathcal{ R}}-\frac{1}{2}g_{\mu \nu}f\mathcal{( R)}+[g_{\mu \nu}\Box -\nabla_\mu \nabla_\nu]f\mathcal{ }_{_\mathcal{ R}} -\kappa {\mathcal T}_{\mu \nu}\equiv0\,,\end{eqnarray}
 where $\Box$ is the d'Alembert operator and $\displaystyle  f\mathcal{ }_{_\mathcal{ R}}=\frac{ {df}}{d\mathcal {R}}$. Here  $ {\mathcal T}_{\mu \nu}$ is the  energy-momentum tensor, defined as
\begin{eqnarray} \label{Tmu}
{\mathcal T}_{\mu \nu}=-\frac{2}{\sqrt{-g}}\frac{\delta {\cal L_M}}{\delta g^{\mu \nu}}.
  \end{eqnarray}
Field Eqs. ~(\ref{f1}) have the following trace:
\begin{eqnarray} \label{f3}
{\mathop{\mathcal{ E}}}:=3\Box f\mathcal{ }_{_\mathcal{ R}}+\mathcal{ R}f\mathcal{ }_{_\mathcal{ R}}-2f\mathcal{(R)} -\kappa {\mathcal T}\equiv0 \,.\end{eqnarray}
From Eq. (\ref{f3}), one can obtain $f\mathcal{(R)}$  in the form:
\begin{eqnarray}\label{f3s}  f\mathcal{(R)}=\frac{1}{2}\Big[3\Box {f\mathcal{ }_{_\mathcal{ R}}}+\mathcal{ R}{f\mathcal{ }_{_\mathcal{ R}}}-\kappa {\mathcal T}\Big]\,.\end{eqnarray}
Using Eq. (\ref{f3s}) in Eq. (\ref{f1}) we get \cite{Kalita:2019xjq}
\begin{eqnarray} \label{f3ss}
{\mathop{\mathcal{ E}}}_{\mu \nu}:=\mathcal{ R}_{\mu \nu} F-\frac{1}{4}g_{\mu \nu}\mathcal{ R}F+\frac{1}{4}g_{\mu \nu}\Box F -\nabla_\mu \nabla_\nu F-\kappa\Big({\mathcal T}_{\mu \nu}-\frac{1}{4}g_{\mu \nu}{\mathcal T}\Big)  \,,\end{eqnarray}
where
\begin{eqnarray} \label{F}
F=f\mathcal{ }_{_\mathcal{ R}}=\frac{ {df}}{d\mathcal {R}}=\frac{ {df}}{dr}\frac{dr}{d\mathcal {R}}\,.\end{eqnarray}
The energy-momentum tensor ${\mathcal T}_{\mu \nu}$ can be assumed with  the following anisotropic form
\begin{eqnarray}
&&{ {\mathcal T}}_\mu{}^\nu{}=\Big(\frac{p_{_{_t}}}{c^2}+\epsilon\Big)u_\mu u^\nu +p_{_{_t}}\delta_\mu{}^\nu+(p-p_{_{_t}})\varepsilon_\mu \varepsilon^\nu,
\end{eqnarray}
where $u_\mu$ is a  time-like vector defined as $u^\mu=[c,0,0,0]$ and $\varepsilon_\mu$ is the unit space-like vector in the radial direction. In this study $\epsilon$ is the energy-density, $p$ and $p_t$ are the radial and the tangential pressures respectively. The energy-momentum tensor takes  the diagonal form ${ {\mathcal T}}_\mu{}^\nu{}=
diag(-c^2\epsilon;\,p;\,p_{_{_t}};\, p_{_{_t}})$.


\section{\bf  Compact stars in  $f\mathcal{(R)}$ gravity}
\label{S3}
In order to consider relativistic compact stars in $f({\mathcal R})$ gravity, let us assume   the following spherically symmetric space-time:
\begin{eqnarray} \label{met12}
& &  ds^2=-e^{a(r)}dt^2+e^{-b(r)}dr^2+r^2d\Omega\,, \qquad {\mbox{with}} \qquad d\Omega=(d\theta^2+\sin^2d\phi^2)\,,  \end{eqnarray}
where $a(r)$ and $b(r)$ are unknown functions of the radius.   The Ricci scalar of the metric (\ref{met12}) takes the form:
\begin{eqnarray} \label{Ricci}
  {\mathcal R(r)}=\frac{e^{-b}r[ra'b'-2ra''-ra'^2-4a'+4b'-4]+4}{2r^2}\,,
  \end{eqnarray}
 with $a'=\frac{da}{dr}$, $a''=\frac{d^2a}{dr^2}$ and $b'=\frac{db}{dr}$.  For the line--element (\ref{met12}), the non-vanishing components of field Eqs.
 (\ref{f3}) and  (\ref{f3ss})  have the form:
 \begin{eqnarray} \label{fes}
&& {\mathop{\mathcal{ E}}}_t{}^t=\frac{e^{-a}[e^{a-b}\{Fr^2( 2a''-a'b'+a'^2)+4rF[ra'+rb'-1]+r^2[3a'F'-2F''+F'b']-4rF'\}+4e^{a}F]}{r^2}-\frac{8\pi G}{c^2}\epsilon\,,\nonumber\\
&& {\mathop{\mathcal{ E}}}_r{}^r=\frac{e^{-b}[Fr^2( 2a''-a'b'+a'^2)-4F[ra'+rb'+1]-r^2[a'F'-6F''+3F'b']-4rF'+4e^{a}F]}{r^2}+\frac{8\pi G}{c^4} p\,,\nonumber\\
&& {\mathop{\mathcal{ E}}}_\theta{}^\theta={\mathop{\mathcal{ E}}}_\phi{}^\phi=\frac{F[4-e^{-b}(4-2r^2a''+r^2a'b'-r^2a'^2)]+e^{-b}[r^2a'b'+2r^2F''-r^2F'b'-4rF']}{r^2}-\frac{8\pi G}{c^4} p_{_{_t}}\,,\nonumber\\
&&{\mathop{\mathcal{ E}}}= \frac{8\pi G}{c^4}[p+2p_{_{_t}}-c^2\epsilon]+\frac{1}{r^2}\Big[e^{-b}[r^2(6F''-2Fa''-Fa'^2)+r[r\,F\, b'+3rF'-4F]a'+r\beta'[4F-3rF']\nonumber\\
&&+12rF'-4F)+4(F-f)\Big]\,.
\end{eqnarray}
Moreover, we introduce the characteristic density
\begin{equation}\label{eq:rho_star}
\epsilon_\star \equiv \frac{c^2}{8\pi G R_0^2}\,,
\end{equation}
where $R_0$ is a characteristic radius. It can be  used to re-scale the density, the radial and the tangential pressure respectively.  We get the dimensionless variables
\begin{equation}\label{eq:rho+p_dless}
\hat{\epsilon} = \frac{\epsilon}{\epsilon_\star}\,,\; \quad
\hat{p} = \frac{p}{\epsilon_\star c^2}\,,\; \quad
\hat{p}_t = \frac{p_t}{\epsilon_\star c^2}\,.\;
\end{equation}
Eqs. (\ref{fes}) are four non--linear differential equations in six unknown variables $a$, $b$, $F$ $\epsilon$, $p$ and $p_{_{_t}}$. Using ${\mathop{\mathcal{ E}}}_r{}^r$ and ${\mathop{\mathcal{ E}}}_\theta{}^\theta$, one can  define the {\it anisotropy parameter} of  stellar structure as:
\begin{eqnarray} \label{anis}
\Delta(r)=\frac{8\pi G}{c^4}\Big[p_{_{_t}}-p\Big]=\frac{e^{-b}\Big(2Fr^2a''+4r^2F''+r^2Fa'^2-r\,F\,a'[rb'+2]-2[rb'+2][rF'+F]\Big)+4F}{4r^2}\,,
\end{eqnarray}
which reduces to $\Delta(r)=0$ for isotropic systems. In order to find a physical solution of the above system of differential equations, let us assume the following forms for the radial component of metric potential $g_{rr}$ and the $f(\mathcal R)$ derivative function $F$ as:
\begin{eqnarray} \label{const}
b(y)=c_0 y^2,\qquad \qquad F(y)=1+c_1y^2\,.
\end{eqnarray}
This choice is physically motivated for the following reasons.
Here $y$ is a dimensionless variable
\begin{eqnarray} \label{dim}
y\equiv \frac{r}{l} \in[0, 1]\,,
\end{eqnarray}
and  we assume the star  can  extend up to the maximal radius $r =l$. The parameters $c_0$ and $c_1$ are dimensionless.They  will be determined later.
Eqs.(\ref{const}) show that when $c_1=0$, we recover  GR. In other words,  the constant $c_1$ is  responsible for the deviation from GR.
Using Eqs.  (\ref{const}) into (\ref{anis}), we get the metric potential  $g_{tt}$ as
\begin{eqnarray} \label{gtt}
a(y)=2ln\Big(\frac{c_0c_2+c_3e^{c_0y^2/2}}{2c_0}\Big)\,,
\end{eqnarray}
where  $c_2$ and $c_3$ are given by combinations of the parameters $c_{0}$, $c_{1}$ and  $l$.  The anisotropy parameter $\Delta(y)$  has the form \begin{eqnarray} \label{anis1}
\Delta(y)=\frac{1+y^2l^2c_1-[1+y^2l^2c_1+c_0y^2l^2+3y^2l^4c_0c_1]e^{-c_0y^2l^2}}{y^2l^2}\,.\end{eqnarray}
Using Eqs. (\ref{const}) and  (\ref{gtt}) in Eq. (\ref{fes}),  we get the exact form of energy, radial and tangential pressures that we report in Appendix A.

Inserting Eq. (\ref{const}) in  the last equation of system (\ref{fes}), the trace  of the field Eqs. (\ref{f3ss}), we get a very lengthy differential equation in the unknown $f(\mathcal R)$. We report the full equation in Appendix C.  Developing  such a differential equation up to the first order  in ${\cal O}\Bigg(\frac{1}{y}\Bigg)$, we get all the information we need for our purposes. It is 
   \begin{eqnarray}\label{dfas(r)}
\frac{1}{2c_0{}^2ym^2}(3mm_1{}^2f'''+15c_0m_1m_2f')\approx 0\,,
   \end{eqnarray}
   where $m=5c_0{}^2c_2{}^2+7c_0c_2c_3-c_3{}^2$, $m_1=c_3+c_0c_2$ and $m_2=173c_0{}^3c_2{}^3+138c_0{}^2c_2{}^2c_3{}+48c_0c_2c_3{}^2-8c_3{}^3$. The solution of the above differential equation takes the form
    \begin{eqnarray}\label{fas(r)}
    f(y)=c_4+c_5\sin\Big(\frac{\sqrt{c_0m_2}y}{\sqrt{3m\,m_0}}\Big)+c_6\cos\Big(\frac{\sqrt{c_0m_2}y}{\sqrt{3m\,m_0}}\Big)\,,
     \end{eqnarray}
     where $c_4$, $c_5$ and $c_6$ are integration constants.
     For $c_{5}$, the solution (\ref{fas(r)})  is compatible with $F(y)$, given by Eq. (\ref{const}), up to the leading order.
Equations of  density, radial and tangential pressures show  that, for $c_1=0$, we return to  GR. This is due to the fact that,  for $F=1$, we have $f(\mathcal R)=\mathcal R$. This means that   differences with GR come from $c_1\neq 0$. See   Appendix A.

   Using Eqs. (\ref{const}) and (\ref{gtt}),   we get the Ricci scalar, up to the leading order, in the form
   \begin{equation} \label{Ris}
   {\mathcal R}\approx\frac{6c_0{}^2}{m_1}-\Bigg[\frac{6c_0{}^3c_3}{m_1{}^2}+\frac{c_0{}^2[c_0c_3+5c_0{}^2-c_3{}^2]}{m_1{}^2}\Bigg]y^2+\cdots\, , \quad \mbox{which  leads  to} \quad  y\approx\pm\sqrt{\frac{6c_0{}^2m_1-{\mathcal R}\,m_1{}^2}{6c_0{}^3c_3+c_0{}^2[c_0c_3+5c_0{}^2-c_3{}^2]}}+\cdots.
   \end{equation}
   Using Eq. (\ref{Ris}) into Eq. (\ref{fas(r)}), we get the explicit form of $f({\mathcal R}) $, that is
    \begin{equation} \label{fa(R)}
    f({\mathcal R})=c_4+c_6\cos\Bigg(\frac{\sqrt{m_2(6c_0{}^2m_1-{\mathcal R}\,m_1{}^2)}}{\sqrt{3m\,m_0({6c_0{}^2c_3+c_0[c_0c_3+5c_0{}^2-c_3{}^2]})}}\Bigg)\,.
    \end{equation}
    \section{\bf  Matching conditions }\label{S4}
  The solution given in Appendix A has a non-trivial Ricci scalar as shown by Eq. (\ref{Ris}).  Let us  match it with the exterior solution.   To this aim,  we are going to match solution given in Appendix A with the  one presented in \cite{Nashed:2019tuk} that we report here:
    \begin{eqnarray}\label{Eq1} ds^2= -\Big(\frac{1}{2}-\frac{2MG}{c^2r}\Big)dt^2+\Big(\frac{1}{2}-\frac{2MG}{c^2r}\Big)^{-1}dr^2+r^2d\Omega^2,
 \end{eqnarray}
where $M$ is the total stellar  mass  and $4MG<c^2r$. We have to match the interior spacetime metric (\ref{met12}), using Eqs. (\ref{const}) and (\ref{gtt}), with
the exterior Schwarzschild spacetime metric given by Eq. (\ref{Eq1}) at the
boundary of the star, $r =l$. The continuity of the metric
functions across the boundary $r =l$ gives
\begin{eqnarray}\label{Eq2} a(r=l)=ln\Bigg(\frac{1}{2}-\frac{2GM}{c^2l}\Bigg), \qquad \qquad b(r=l)=ln\Bigg(\frac{1}{2}-\frac{2GM}{c^2l}\Bigg)^{-1}.
 \end{eqnarray}
  Using the above conditions,  we get the constraints on the constants $c_0$, $c_2$. These constants take the form
\begin{eqnarray}\label{Eq3} c_0=-ln\Bigg(1/2-C\Bigg), \qquad \qquad c_2=\frac{c_3+2\Bigg(1/2-C\Bigg)\ln\Bigg(1/2-C\Bigg)}{\sqrt{\Bigg(1/2-C\Bigg)}\ln\Bigg(1/2-C\Bigg)}\,,\end{eqnarray}
where $C=\frac{2GM}{c^2l}$.
Eq. (\ref{Eq3}) shows that the compactness parameter must be $C<0.5$. This is also a consistency condition.

\section{\bf   The physical  viability of  solution }\label{S5}
Now we are going to test if the interior solution,  reported  in Appendix A,  can describe a realistic  physical  star. To this aim,  we have to derive some  necessary  conditions which have to hold  for any realistic star. They are the following.

\subsection{\bf Energy--momentum tensor}

\begin{figure}
\centering
\subfigure[~The density   (A.I)]{\label{fig:dnesity}\includegraphics[scale=0.3]{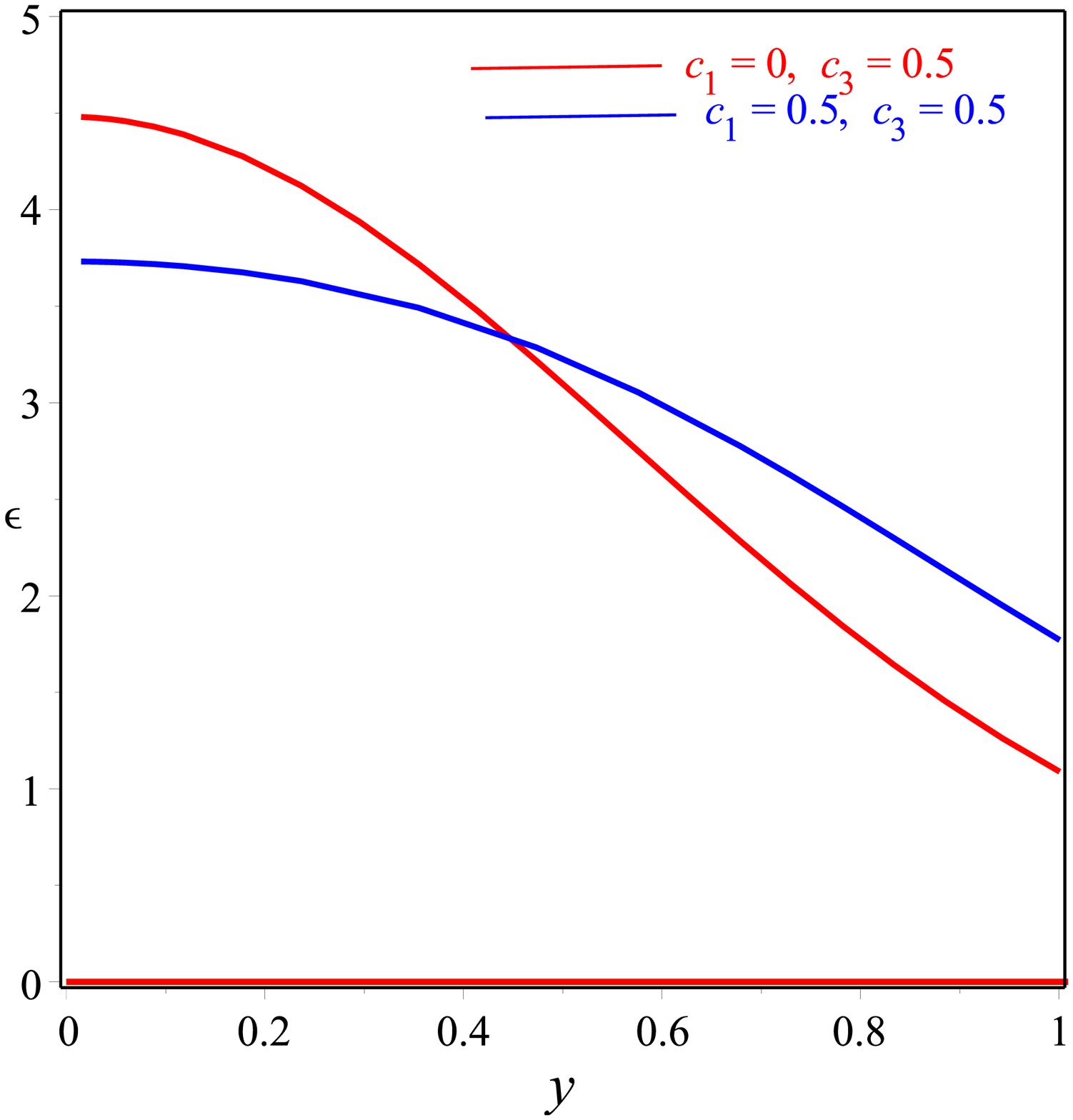}}
\subfigure[~The radial pressure (A.II) ]{\label{fig:pressure}\includegraphics[scale=.3]{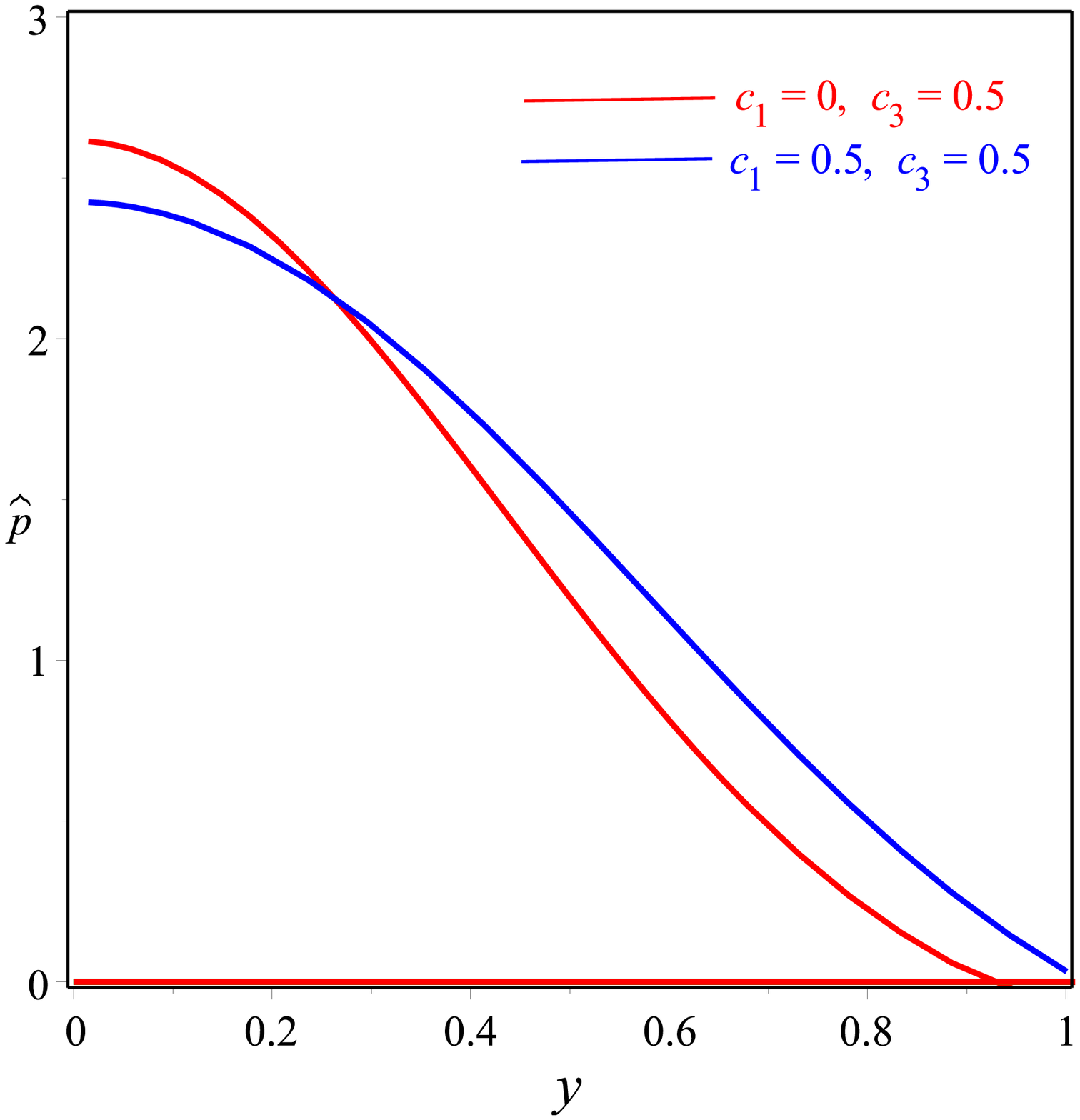}}
\subfigure[~The traverse pressure (A.III) ]{\label{fig:EoS}\includegraphics[scale=.3]{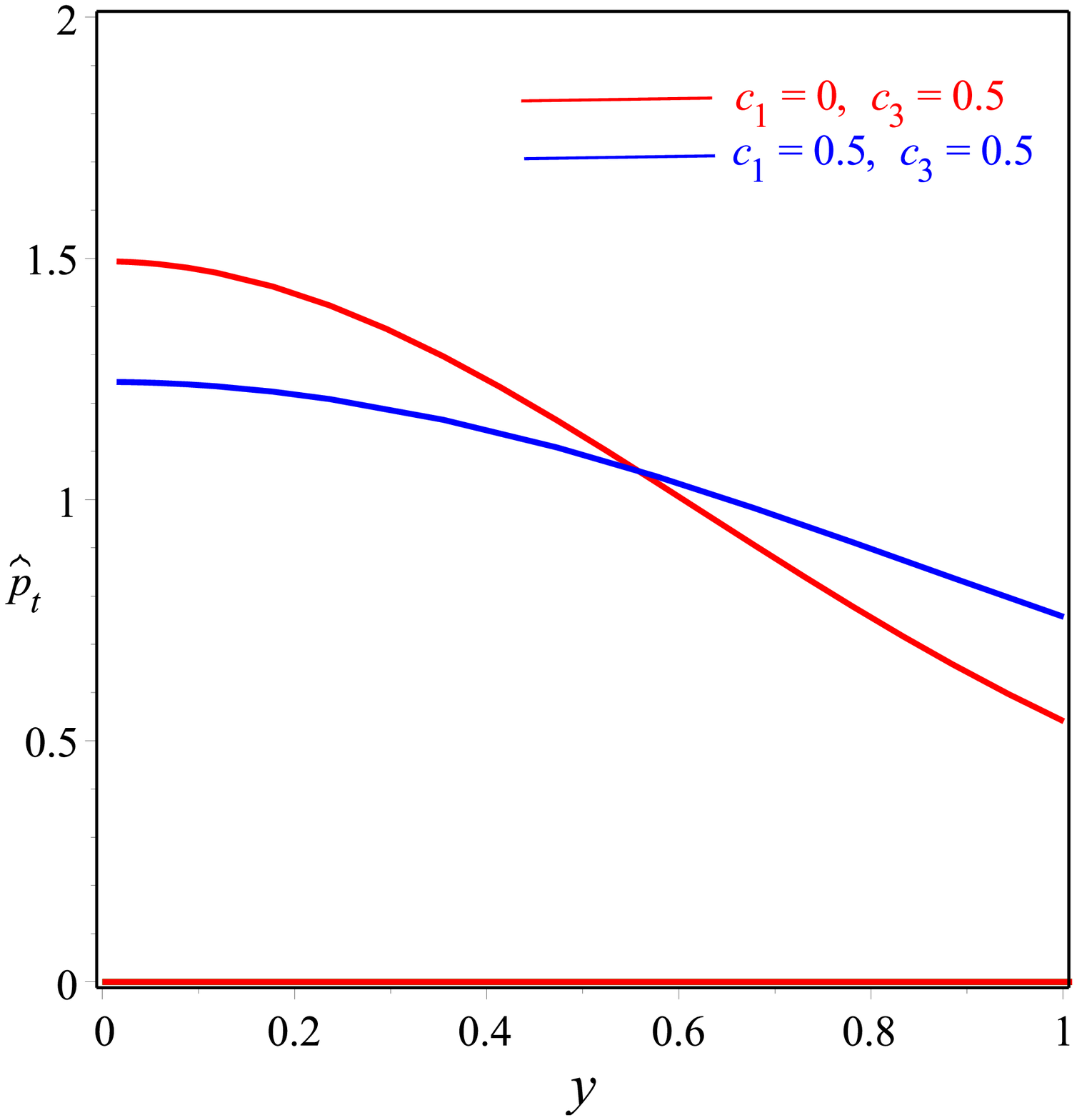}}
\caption[figtopcap]{\small{{Density, radial and transverse pressures  when  $c_1=0$ and $c_1=0.5$.}}}
\label{Fig:1}
\end{figure}

 It is well-known that, for a realistic  interior solution, the energy--density, the radial and transverse  pressures must be positively defined. Moreover, all the components of the energy-momentum  tensor should be finite  at the center of the star then become decreasing   in the direction of the surface of the stellar structure and the radial pressure should exceeds the tangential one\footnote{In this study we take the value of the constants as $C=0.3321821258$, $c_0=1.784875970$,  $c_3=0.5$ , $c_2=0.1354899376$. These values reproduce  a stellar model for the observed millisecond pulsar  PRS $J0437-4715$.}. The plots of the energy-momentum components, density, radial and transverse pressures are shown in Fig.\ref{Fig:1}. The values of parameters used in the figure are:  $\hat{\epsilon}(y=0)_{c_1=0,c_3=0.5}=4.481840660$, $\hat{\epsilon}(y=0)_{c_1=0.5,c_3=0.5}=3.731840660$, $\hat{p}(y=0)_{c_1=0,c_3=0.5}=\hat{p}_t(y=0)_{c_1=0,c_3=0.5}=1.493946887$, $\hat{p}(y=0)_{c_1=0.5,c_3=0.5}=\hat{p}_t(y=0)_{c_1=0,c_3=0.5}=1.243946887$.  Fig. \ref{Fig:1} shows that the density, radial and tangential pressures are decreasing towards the stellar surface. Moreover  Fig.\ref{Fig:1} shows that the values of  the components of the energy momentum  at the center, in case $c_1=0$, are greater than those for  $c_1\neq0$. We can also deduce, from  Fig.\ref{Fig:1},  that  density, radial and  tangential pressures converge  to the surface of the star more rapidly in case $c_1=0$  than in the case of $c_1=0.5$.  Finally,  it is easy to see that, when $c_1=0$, one gets  $\hat{p}=\hat{p}_t$ at the center however, as we reach the surface of the star, we can see that   $\hat{p}_t\geq \hat{p}$.

\begin{figure}
\centering
\subfigure[~The anisotropy $\Delta$  ]{\label{fig:dnesity}\includegraphics[scale=0.3]{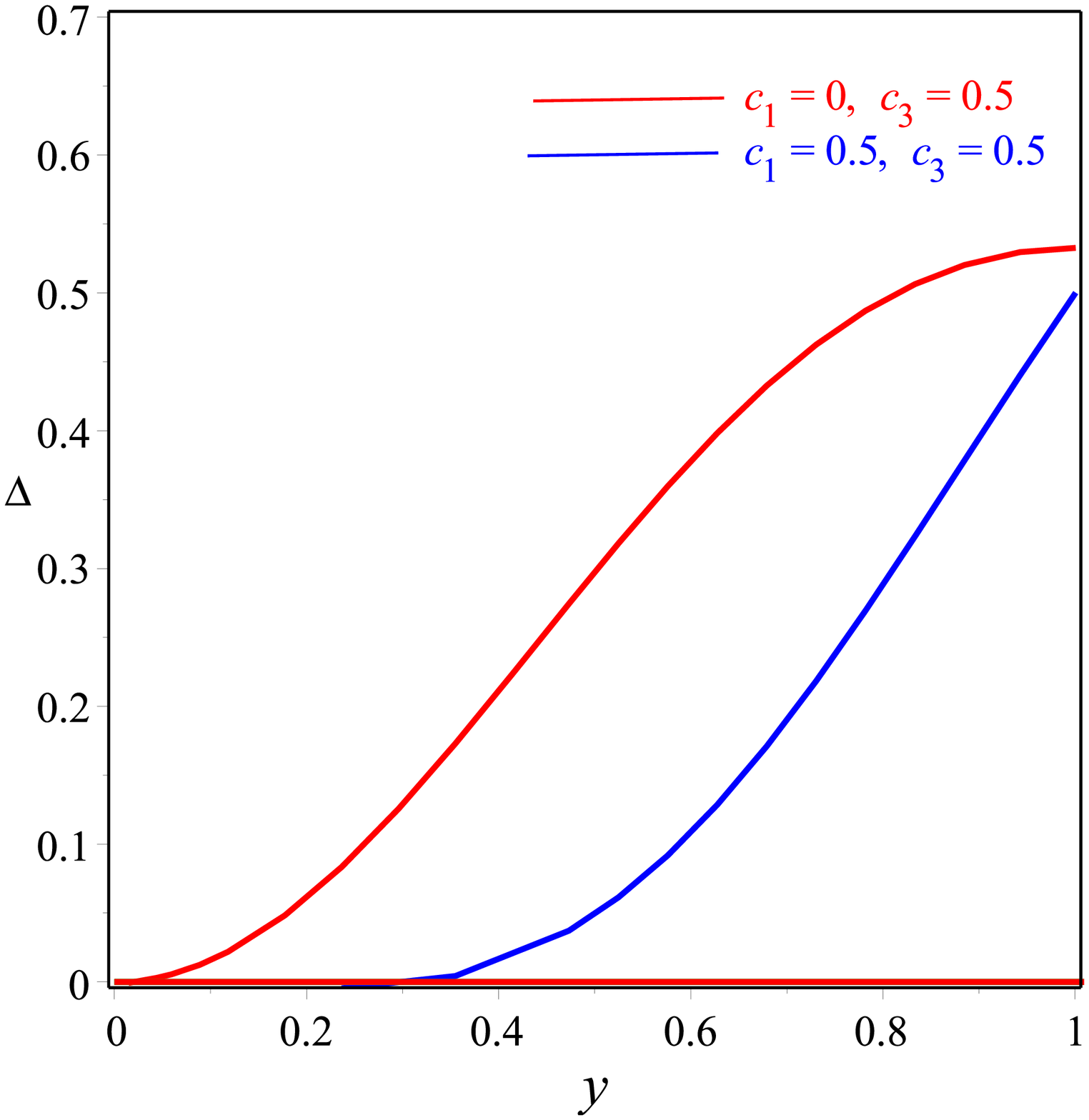}}
\subfigure[~The anisotropic force $\Delta/y$  ]{\label{fig:pressure}\includegraphics[scale=.3]{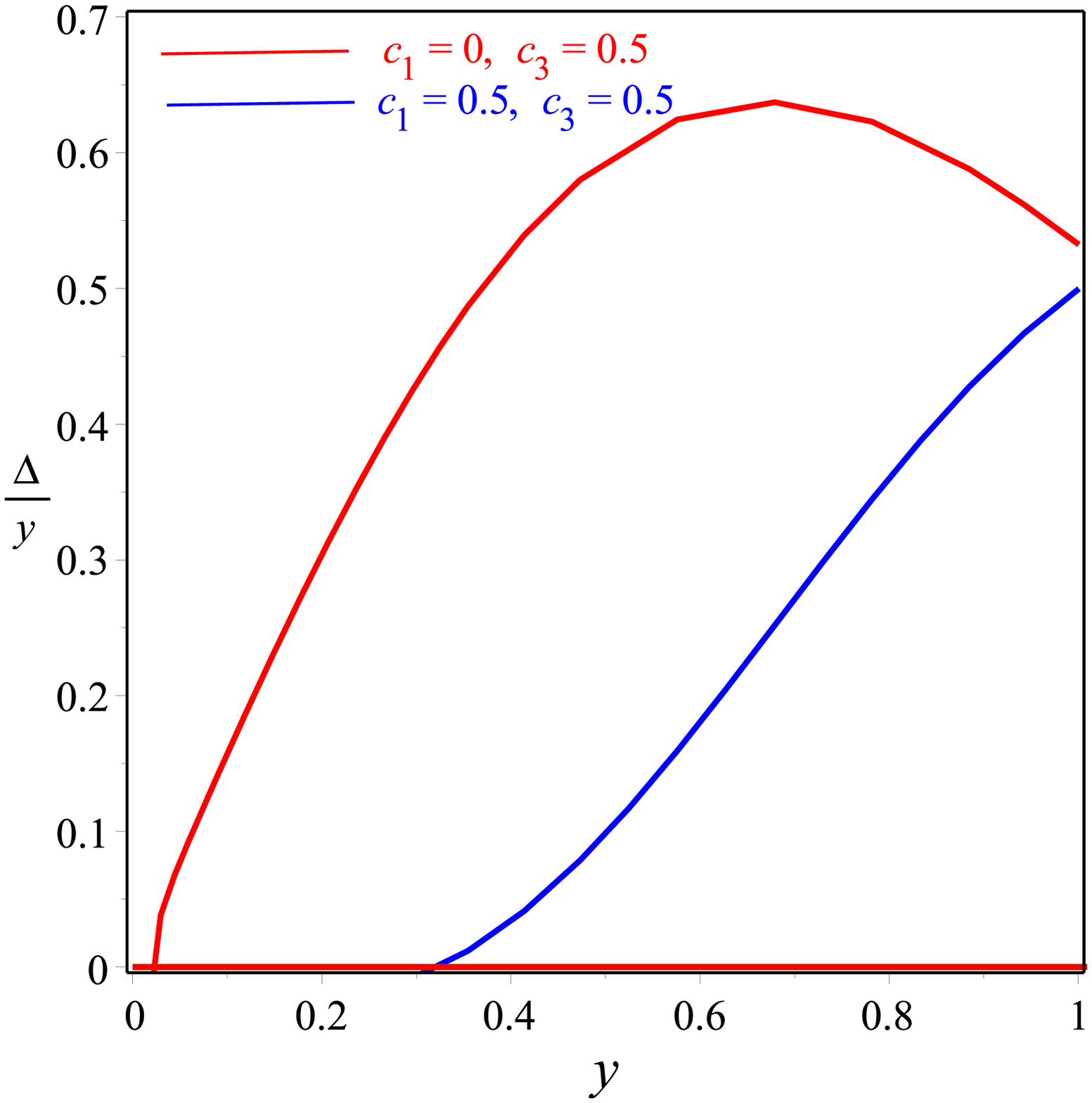}}
%
\caption[figtopcap]{\small{{Anisotropy $\Delta(y)$ and anisotropy force  when $c_1=0$ and $c_1=0.5$.}}}
\label{Fig:2}
\end{figure}

  In  Fig.\ref{Fig:2},  we  show the behavior of the anisotropy parameter  which is defined as $\Delta(y)=\hat{p}_t-\hat{p}$. Also Fig. \ref{Fig:2} shows that the anisotropic force is positive. This means that a repulsive force, due to  $p_t\geq p$, is present.


%

\begin{figure}
\centering
\subfigure[~The gradient of density ]{\label{fig:dnesity}\includegraphics[scale=0.3]{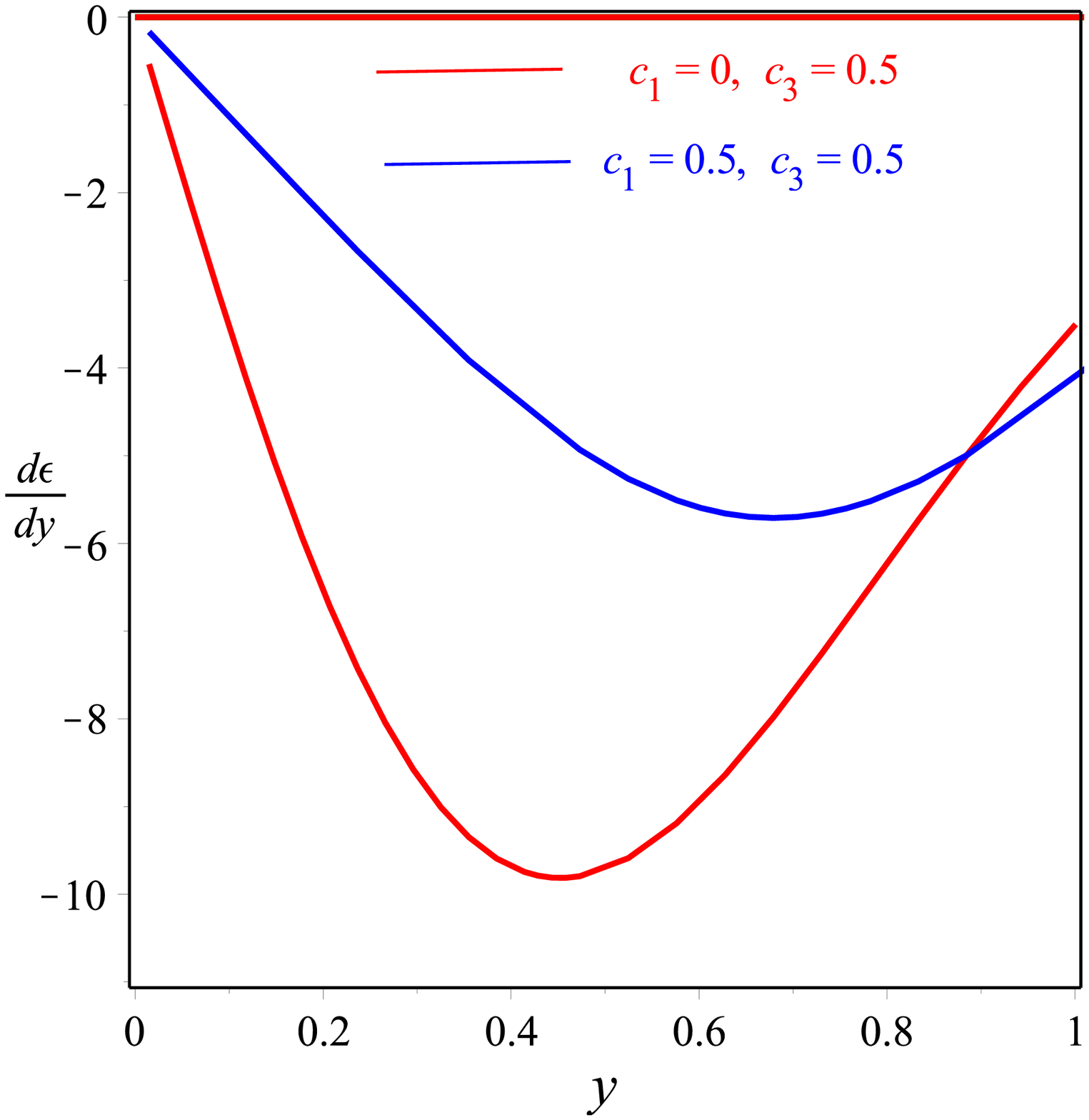}}
\subfigure[~The gradient of radial pressure ]{\label{fig:pressure}\includegraphics[scale=.3]{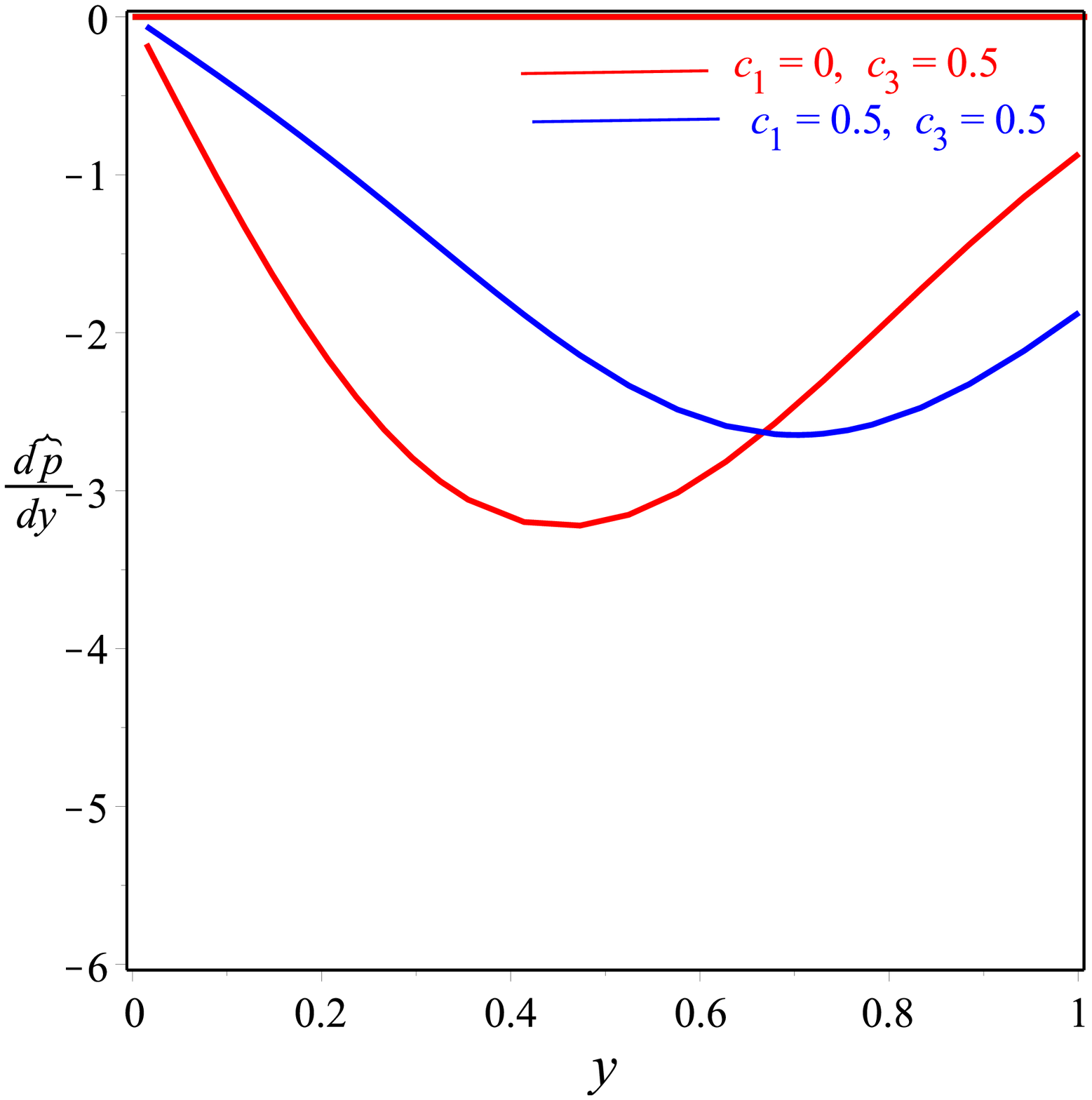}}
\subfigure[~The gradient of traverse pressure ]{\label{fig:EoS}\includegraphics[scale=.3]{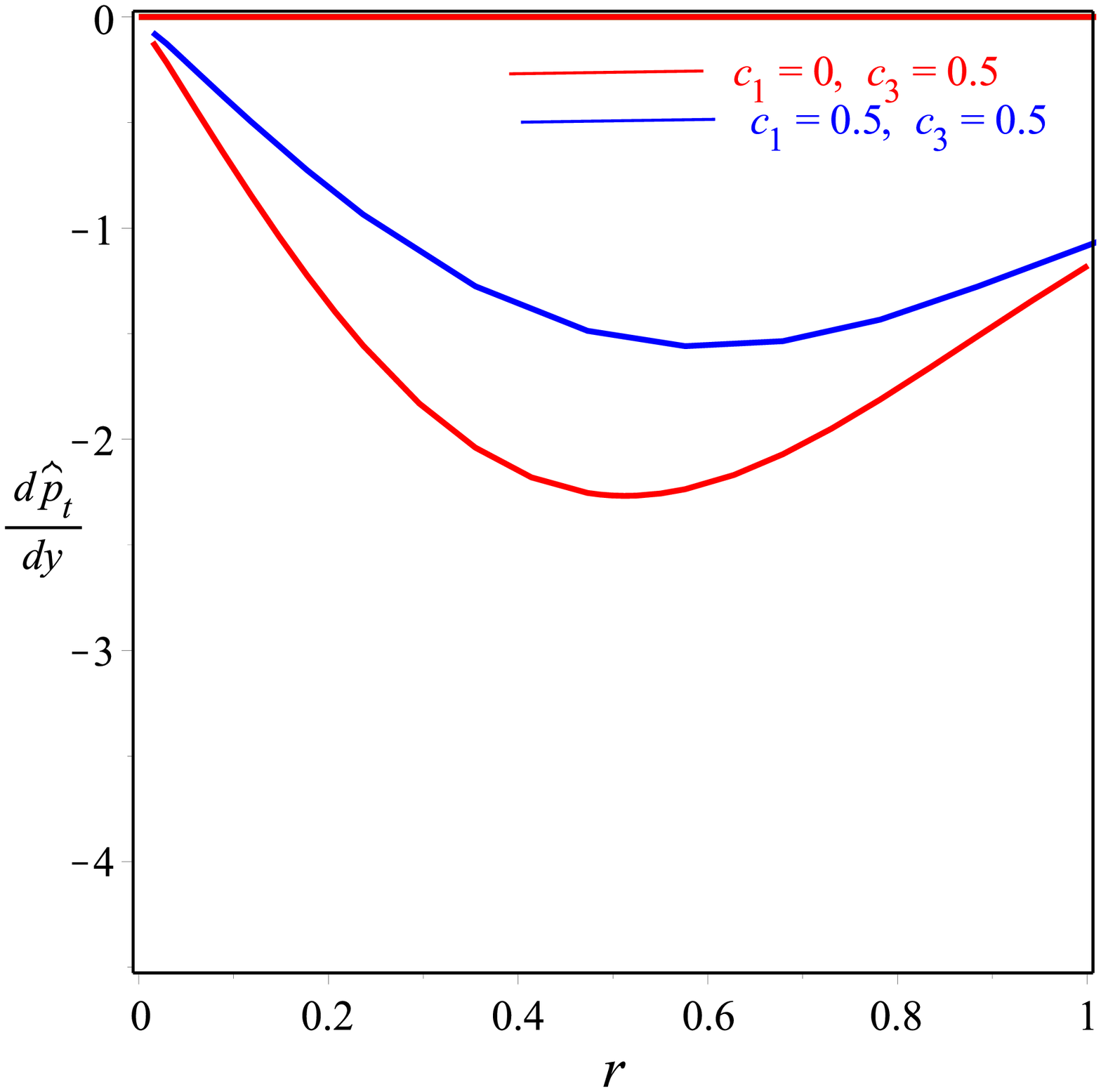}}
\caption[figtopcap]{\small{{The gradient of density, radial and transverse pressures  when $c_1=0$ and $c_1=0.5$.}}}
\label{Fig:3}
\end{figure}
\subsection{\bf Causality}
\begin{figure}
\centering
\subfigure[~The radial speed]{\label{fig:dnesity}\includegraphics[scale=0.3]{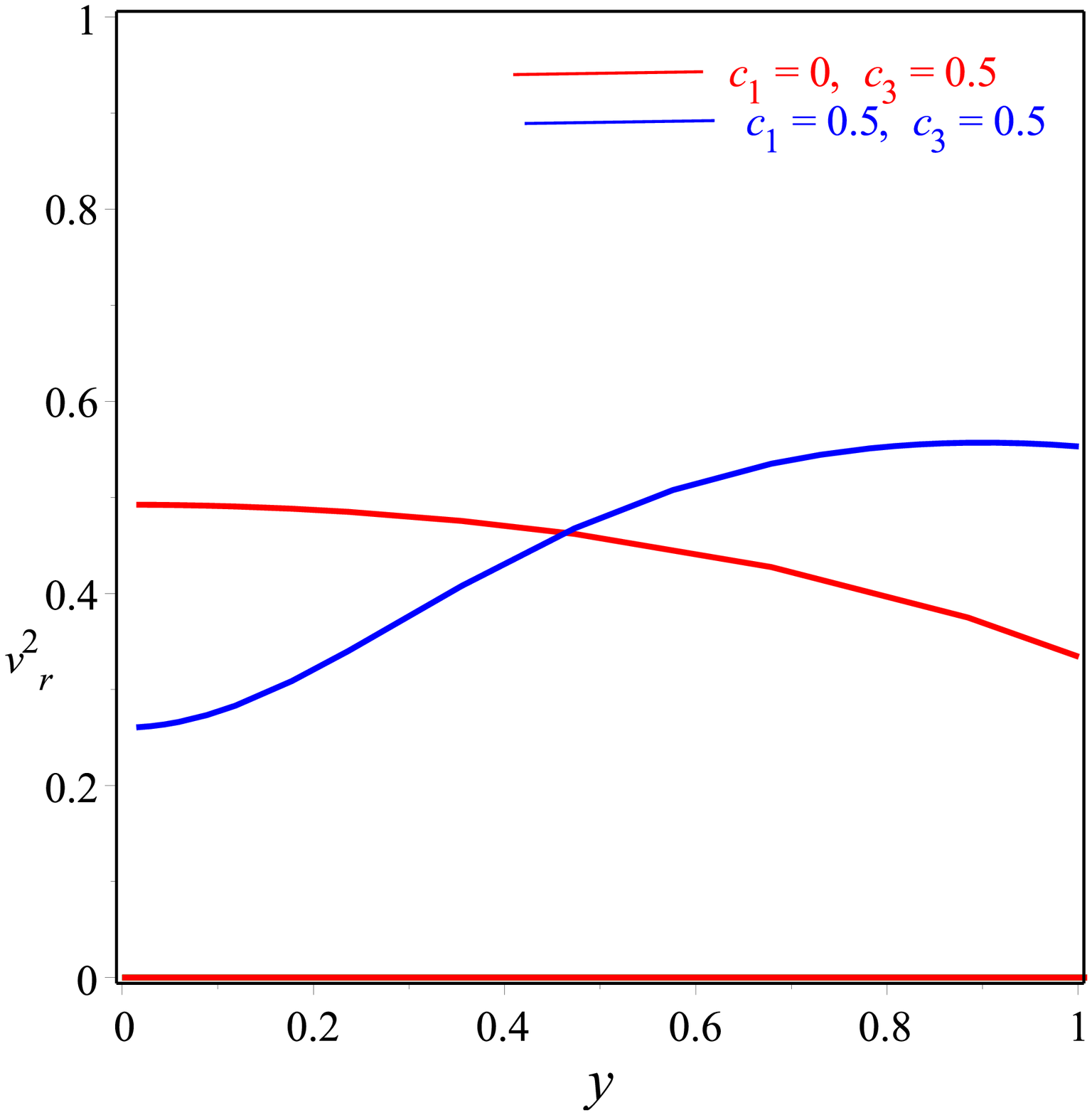}}
\subfigure[~The transverse speed  ]{\label{fig:pressure}\includegraphics[scale=.3]{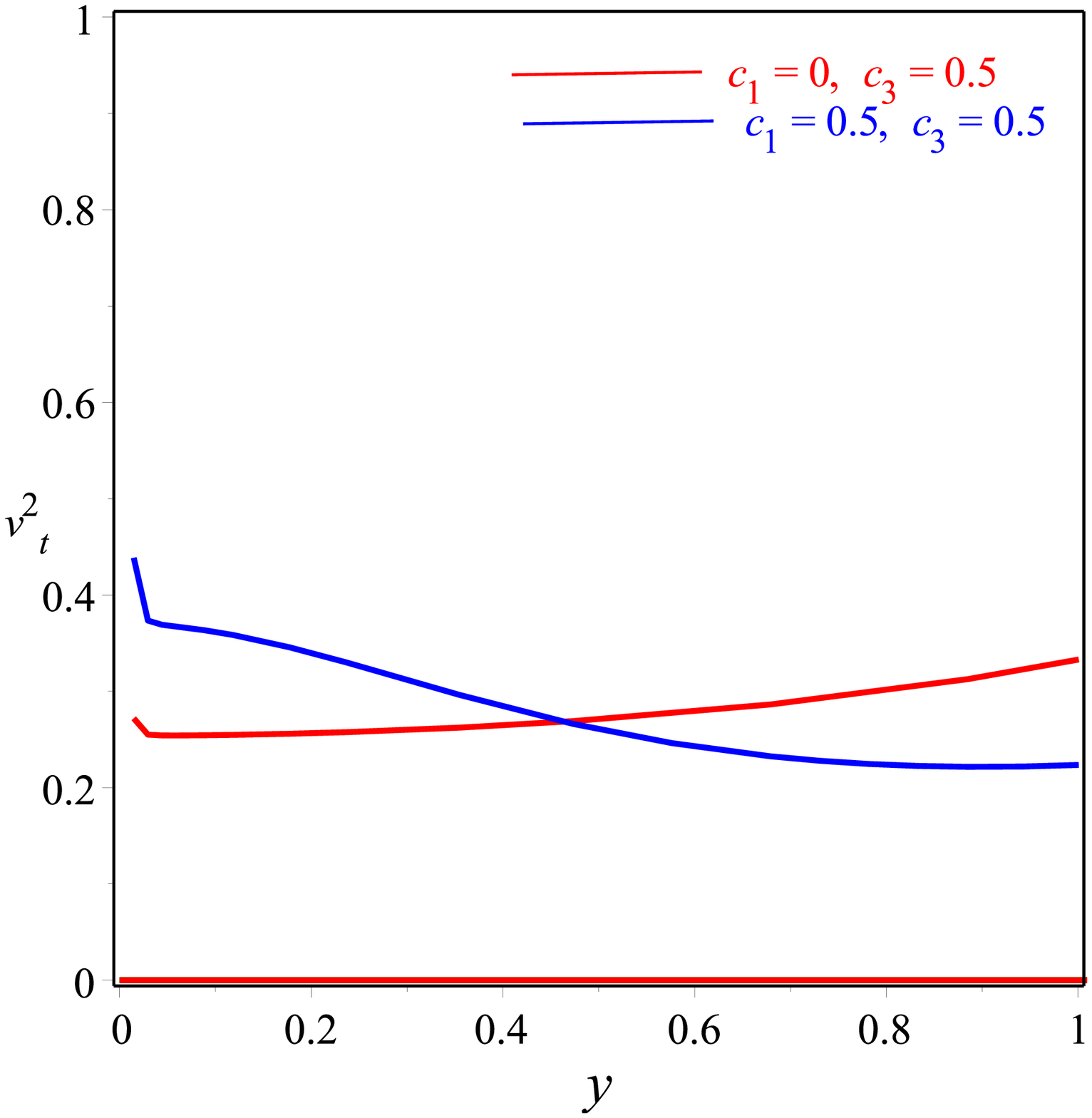}}
\subfigure[~Variation of ${\lvert}v_r{}^2-v_t{}^2{\rvert}$ ]{\label{fig:crac}\includegraphics[scale=.3]{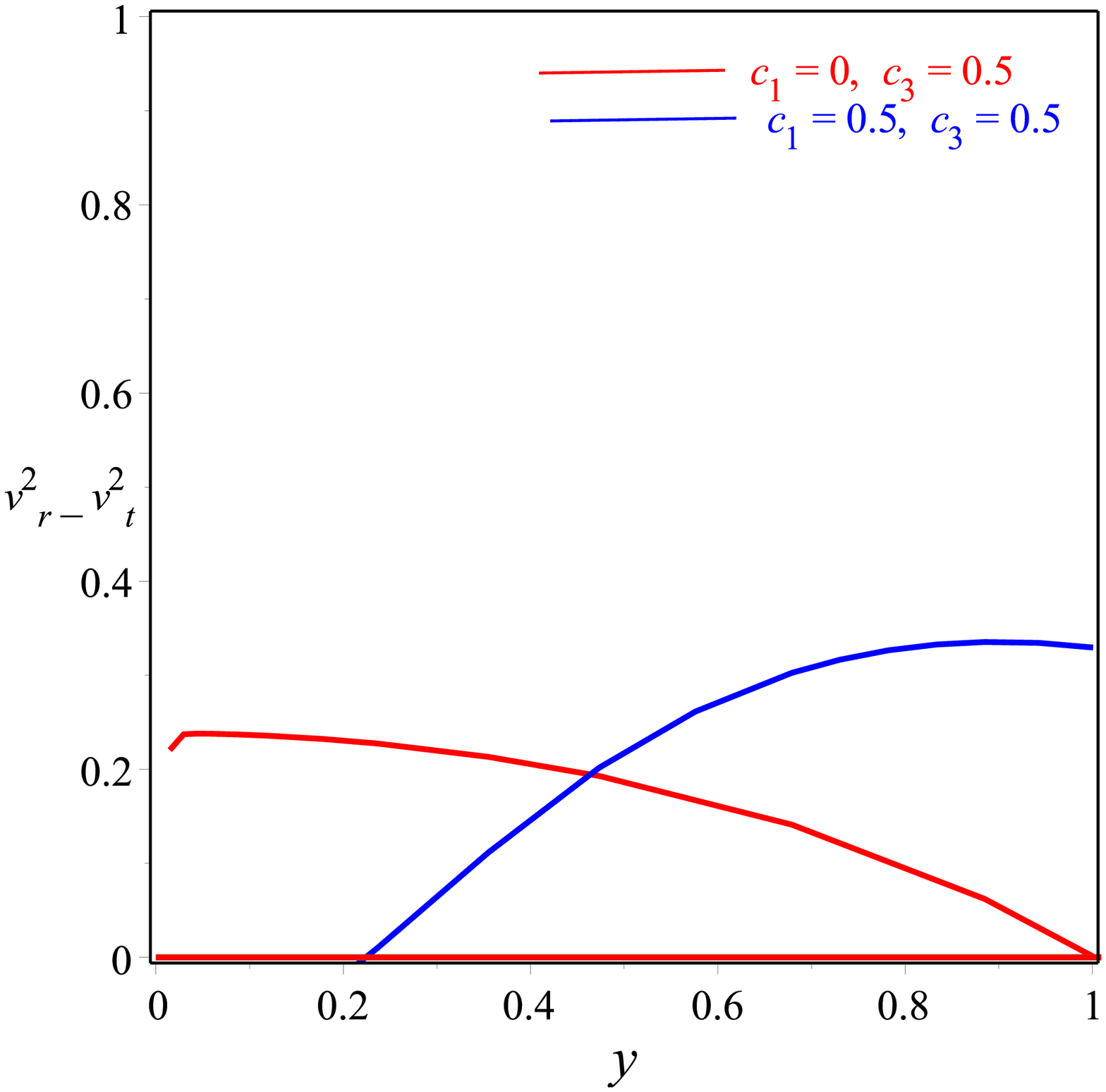}}
\caption[figtopcap]{\small{{Radial, transverse  sound speeds   and   ${\lvert}v_r{}^2-v_t{}^2{\rvert}$ for $c_1=0$ and $c_1=0.5$.}}}
\label{Fig:4}
\end{figure}

In order to show the behavior of  sound velocities, we must calculate the gradient of   energy--density, radial and transverse  pressures that take the forms  given in Appendix B. Let us analyze the gradient of  energy-momentum tensor components  reported  in Appendix B.
  The asymptotic forms   are
  \begin{eqnarray} \label{grad1}
 && \hat{\epsilon}'\approx \frac{c_0[26c_1c_3{}^2-11c_0c_3{}^2+40c_0c_1c_2c_3-13c_0{}^2c_2c_3{}^2-5c_0{}^3c_2{}^2+14c_0{}^2c_1c_2{}^2]}
 {2[c_1{}+c_0c_2]}\,,\nonumber\\
 &&\hat{p}'\approx \frac{c_0[14c_1c_2{}^2-5c_0c_3{}^2+24c_0c_1c_2c_3-7c_0{}^2c_2c_3-3c_0{}^3c_2{}^2+10c_0{}^2c_1c_2{}^2]}
 {2[c_1{}+c_0c_2]}\,,\nonumber\\
 &&\hat{p}'_t\approx \frac{c_0[6c_1c_3{}^2-3c_0c_3{}^2+8c_0c_1c_2c_3-3c_0{}^2c_2c_3-c_0{}^3c_2{}^2+2c_0{}^2c_1c_2{}^2]}
 {2[c_1{}+c_0c_2]}\,,
 \end{eqnarray}
 which give a negative gradient of the energy-momentum components provided that $c_0>c_1\geq c_2>c_3$.  This condition is satisfied according to the choice in  footnote 1. The behavior of density, radial and tangential pressure gradients are shown  in Fig. \ref{Fig:3}. To ensure  that  causality conditions,  either for the radial and
transverse sound speeds $v_r{}^2$ and $v_\bot{}^2$,  have values less than the speed of light,  we use equations reported in  given in Appendix B and get
  \begin{eqnarray} \label{grad2}
 && v_r{}^2=\frac{\hat{p}'}{\hat{\epsilon}'}\approx \frac{14c_1c_3{}^2-5c_0c_3{}^2-7c_0{}^2c_2c_3+24c_0c_1c_2c_3-3c_0{}^3c_2{}^2+10c_0{}^2c_1c_2{}^2}
 {26c_1c_3{}^2-11c_0c_3{}^2-13c_0{}^2c_2c_3+40c_0c_1c_2c_3-5c_0{}^3c_2{}^2+14c_0{}^2c_1c_2{}^2}\,, \nonumber\\
 &&
 v_t{}^2=\frac{\hat{p}'_t}{\hat{\epsilon}'}\approx \frac{6c_1c_3{}^2-3c_0c_3{}^2-3c_0{}^2c_2c_3+8c_0c_1c_2c_3-c_0{}^3c_2{}^2+2c_0{}^2c_1c_2{}^2}
 {26c_1c_3{}^2-11c_0c_3{}^2-13c_0{}^2c_2c_3+40c_0c_1c_2c_3-5c_0{}^3c_2{}^2+14c_0{}^2c_1c_2{}^2}\,.
 \end{eqnarray}
 Eqs. (\ref{grad2}) show that both the conditions  $1>v_r{}^2>0$ and $1>v_t{}^2>0$ hold using  the choice in  footnote 1. The behaviour of the radial and tangential sound speeds are shown in Fig. \ref{Fig:4} \subref{fig:dnesity} and \subref{fig:pressure}.

 The appearance of non--vanishing total radial force with
different signs in different regions of the
fluid is called
{\it gravitational cracking} when this radial force is directed
inward in the inner part of the sphere for all values of the
radial coordinate $r$ between the center and some value
beyond which the force reverses its direction \cite{1994PhLA..188..402H}. In
Ref. \cite{Abreu:2007ew}, it is stated that a simple requirement to
avoid gravitational cracking is $0< v_r{}^2-v_t{}^2<c^2$. In Fig.  \ref{Fig:4} \subref{fig:crac},  we show that the solution given in Appendix A is stable against cracking for  $c_1=0$ and $c_1=0.5$.
\subsection{\bf Energy conditions}
\begin{figure}
\centering
\subfigure[~The DEC,  $\hat{\epsilon}-\hat{p}$ ]{\label{fig:Decr}\includegraphics[scale=0.3]{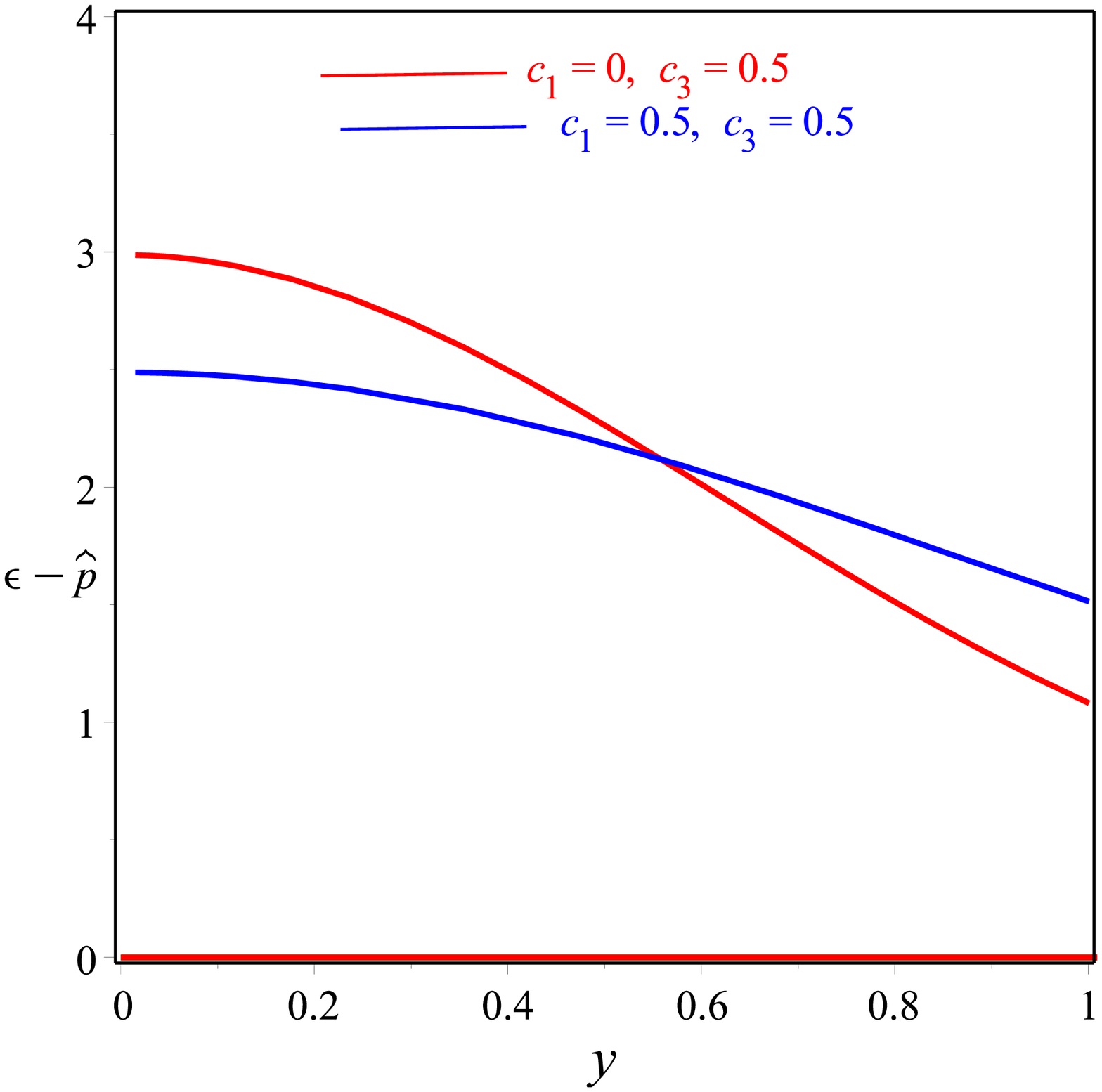}}
\subfigure[~The DEC,   $\hat{\epsilon}-\hat{p}_t$]{\label{fig:Dect}\includegraphics[scale=.3]{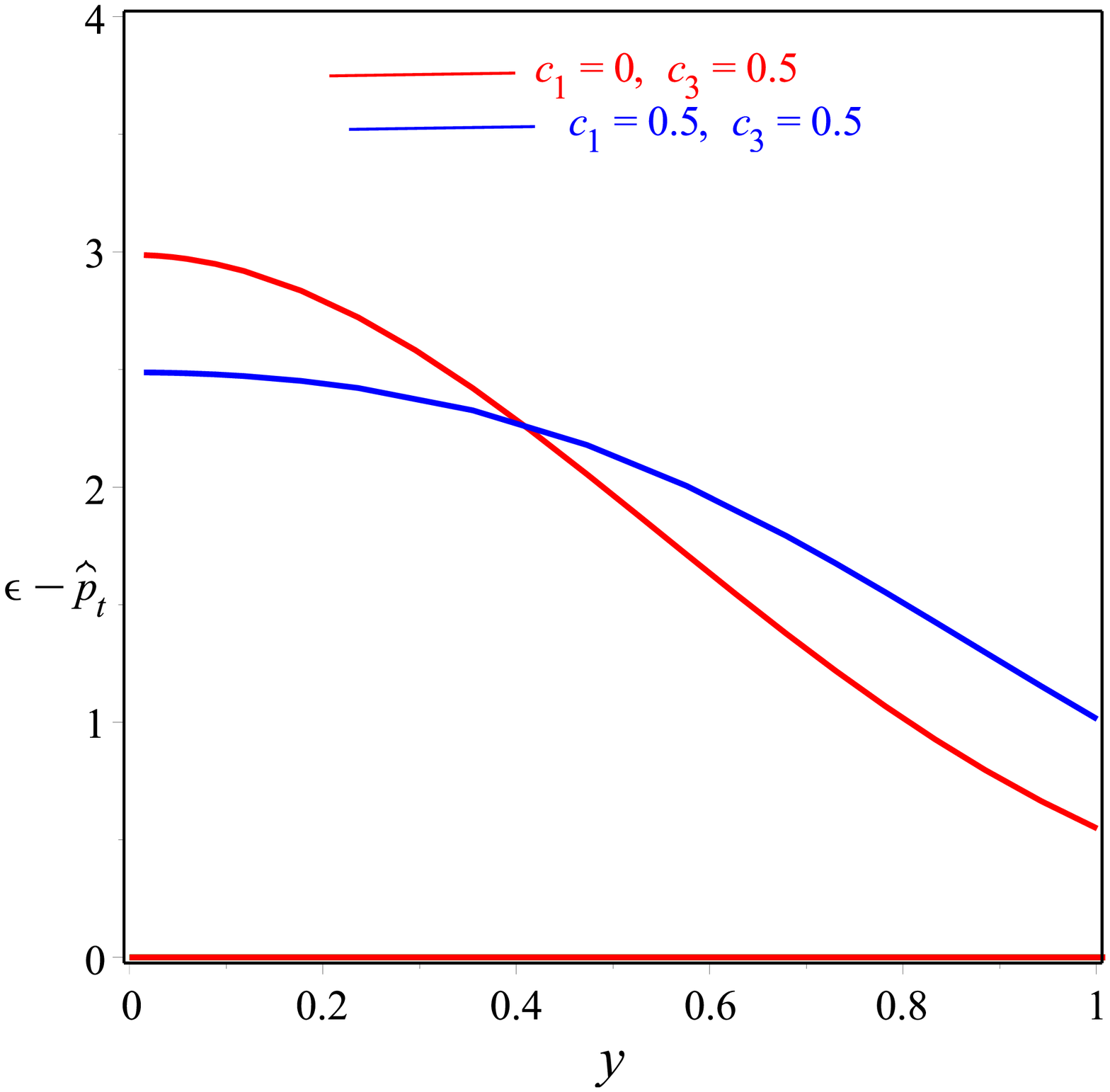}}
%
\caption[figtopcap]{\small{{DEC for $c_1=0$ and $c_1=0.5$.}}}
\label{Fig:5}
\end{figure}

\begin{figure}
\centering
\subfigure[~The WEC,  $\hat{\epsilon}+\hat{p}$]{\label{fig:Wecr}\includegraphics[scale=0.3]{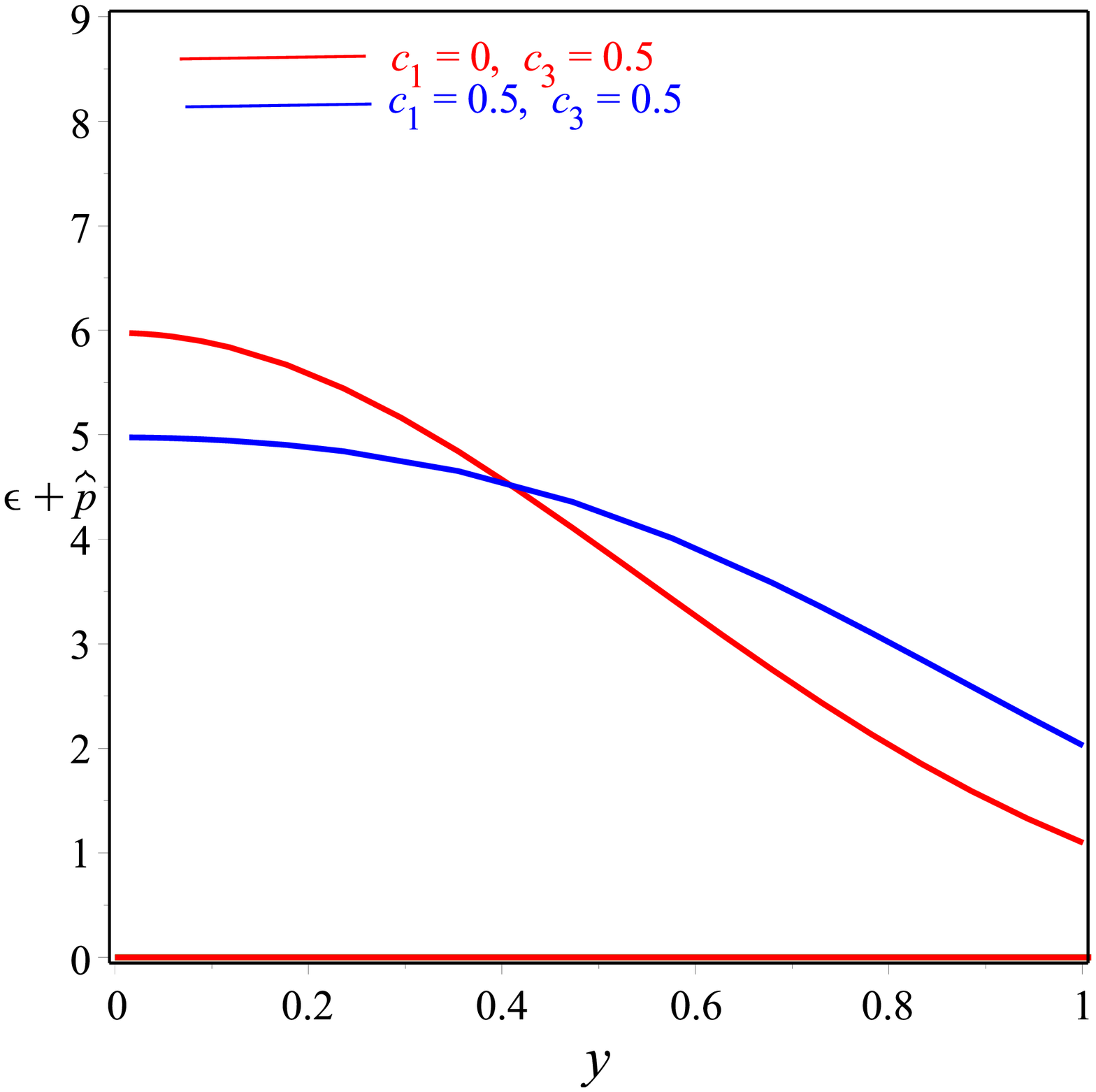}}
\subfigure[~The WEC,  $\hat{\epsilon}+\hat{p}_t$]{\label{fig:Wect}\includegraphics[scale=.3]{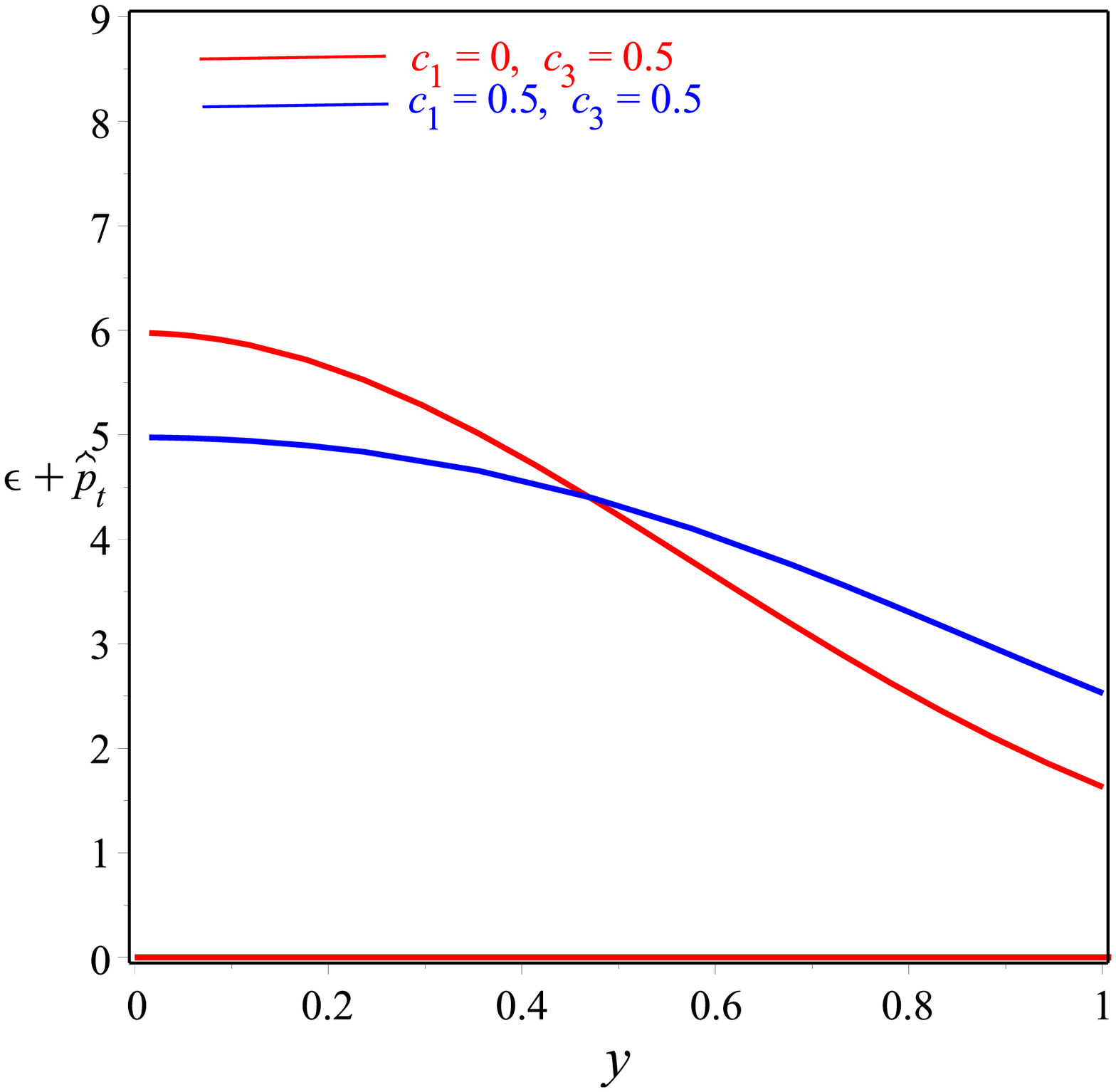}}
\subfigure[~The SEC, $\hat{\epsilon}-\hat{p}-2\hat{p}_t$]{\label{fig:Sec}\includegraphics[scale=.3]{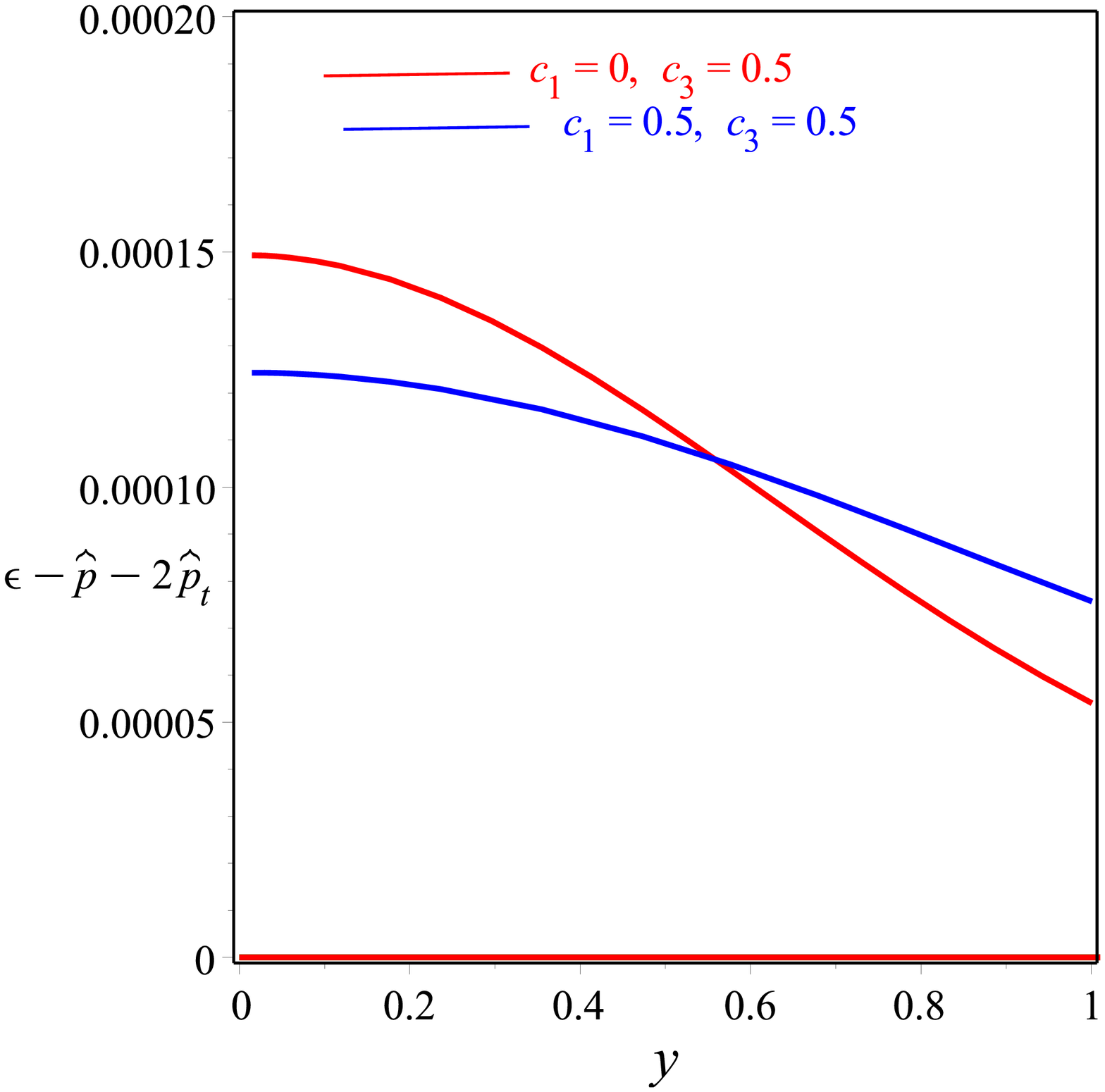}}
\caption[figtopcap]{\small{WEC and SEC of the solution  given in Appendix A for  $c_1=0$ and $c_1=0.5$.}}
\label{Fig:6}
\end{figure}

Energy conditions are considered an important test for  non--vacuum solutions.   To satisfy the dominant energy condition (DEC), we have to prove that  $\hat{\epsilon}-\hat{p} >0$ $\&$ $\hat{\epsilon}-\hat{p}_t>0$.  As shown in   Fig.  \ref{Fig:5} \subref{fig:Decr} and \subref{fig:Dect}, DEC is satisfied for suitable choices of parameter $c_1$. Furthermore, the weak energy condition (WEC), $\hat{\epsilon}+\hat{p}>0$ $\hat{\epsilon}+\hat{p}_t>0$ and the strong energy condition (SEC),  $\hat{\epsilon}-\hat{p}-2\hat{p}_t>0$   are satisfied as shown in  Figs.  \ref{Fig:6}  \subref{fig:Wecr}, \subref{fig:Wect} and \subref{fig:Sec}.

\subsection{\bf Mass-radius relation}

\begin{figure}
\centering
\subfigure[~The gravitational mass  when $c_1=0$ and $c_1=0.5$]{\label{fig:mass}\includegraphics[scale=0.3]{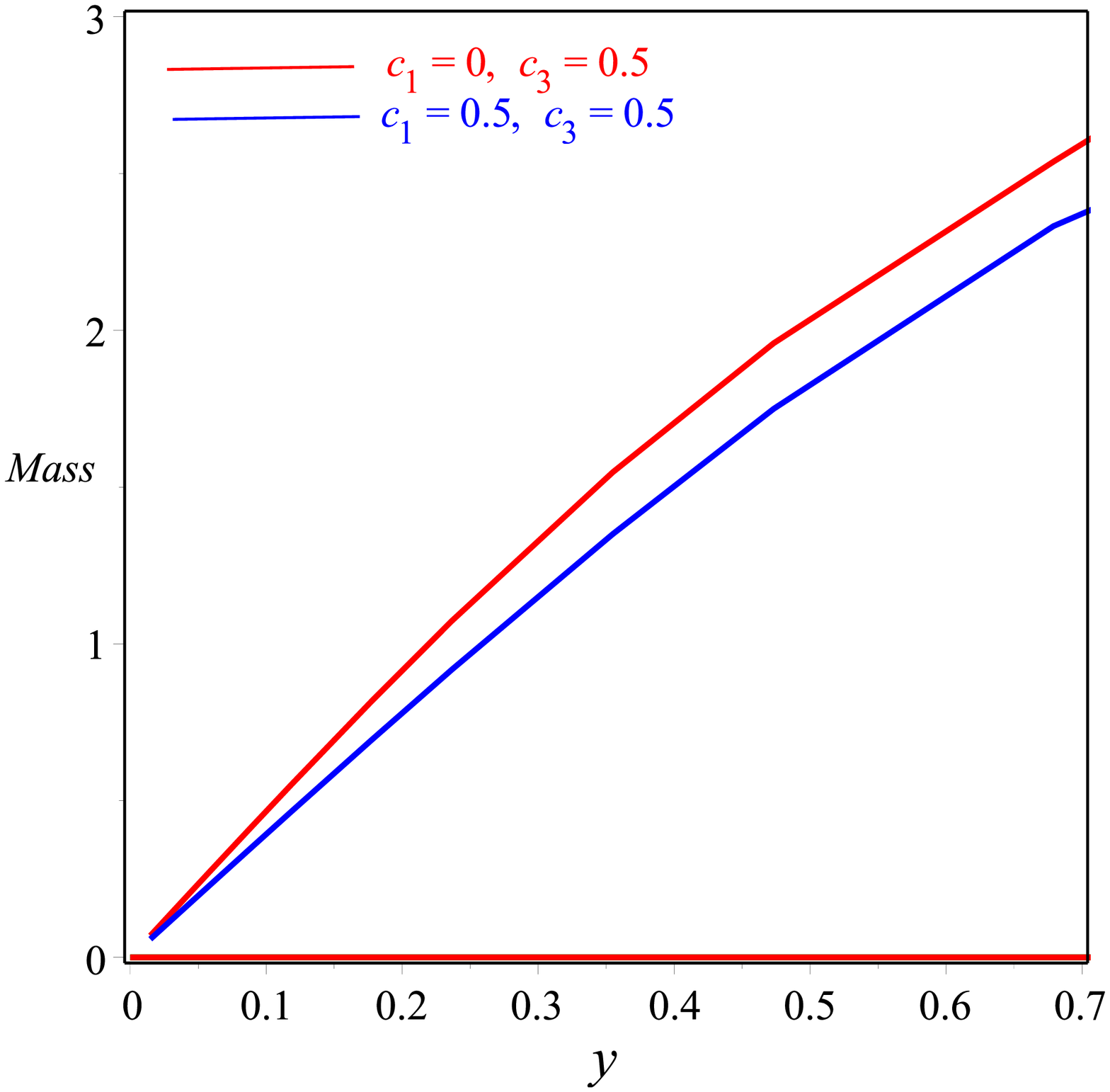}}
\subfigure[~The compactification factor when $c_1=0$ and $c_1=0.5$]{\label{fig:u}\includegraphics[scale=.3]{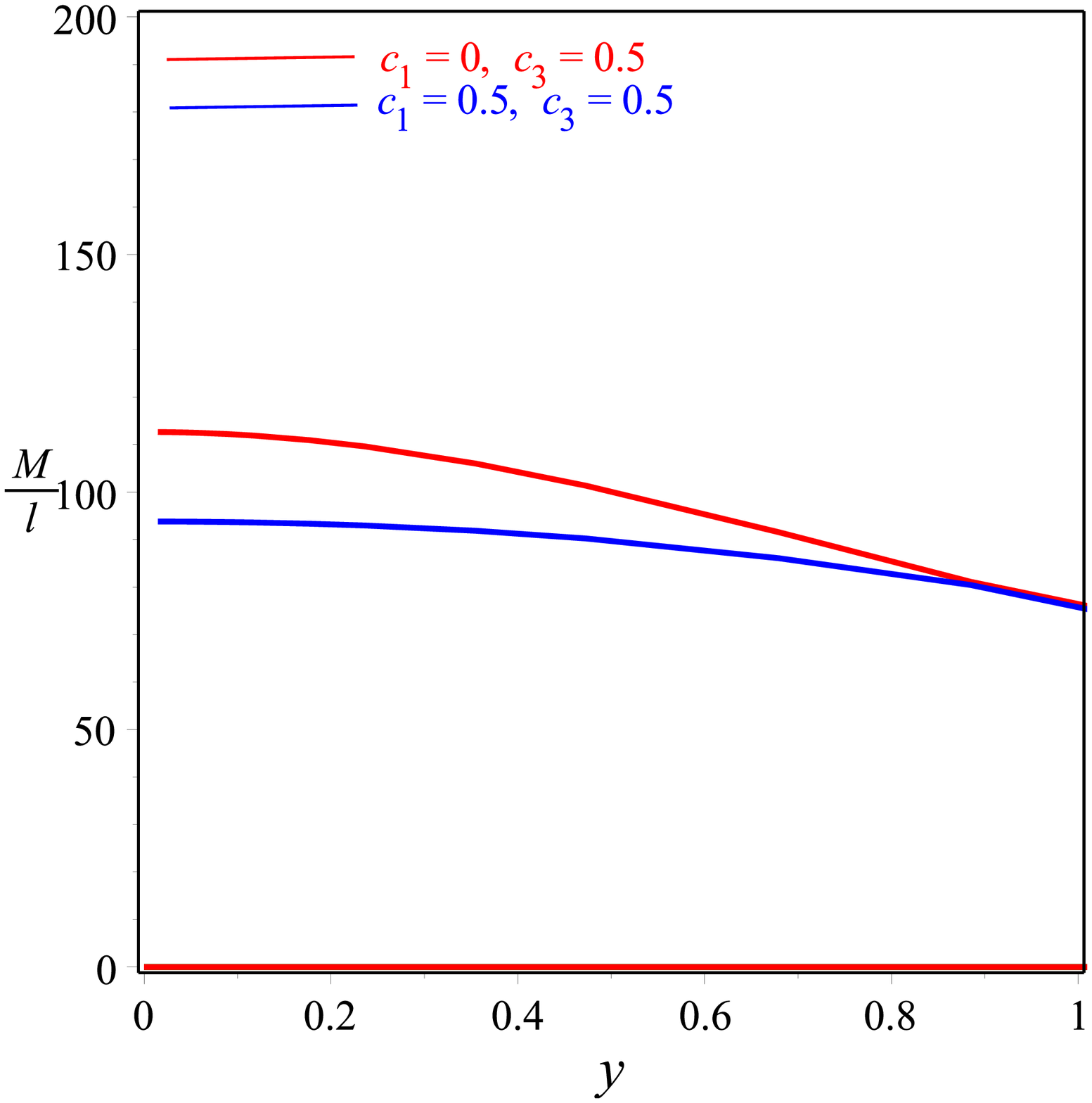}}
%
\caption[figtopcap]{\small{{The gravitational mass and compactification factor  of solution  given in Appendix A for $c_1=0$ and $c_1=0.5$.}}}
\label{Fig:7}
\end{figure}

The compactification factor $u(r)$
 is defined as the ratio between the mass and radius. It
has a key role  to understand  physical properties
of  compact objects. Starting from the solution  given in Appendix A,  we can define  the gravitational mass as
 \begin{eqnarray}\label{mass}
&& M(y)={\int_0}^y  \hat{\hat{\epsilon}} \zeta^2 d\zeta\nonumber\\
 &&=\frac{1}{8yc_0{}^{5/2}c_2{}^2}\Bigg[4yc_0{}^{5/2}c_2{}^2\int\frac{3c_3{}^3(2y^2c_1+1)}{c_0c_2{}^2(c_3+c_0c_2e^{-c_0y^2/2})}dr-24c_0{}^{3/2}c_1c_2c_3y^2
 e^{-c_0y^2/2}+6c_0c_2c_3\sqrt{2}(c_0+2c_1)\,\sqrt{\pi}\,erf(y\sqrt{c_0/2})\nonumber\\
 &&+2c_0^{5/2}c_2{}^2e^{-c_0y^2}[1-3c_1y^2]+yc_2{}^2(8c_0-5c_1)c_0{}^2\,
 \sqrt{\pi}\,erf(y\sqrt{c_0})+4c_0{}^{3/2}\{c_0c_2{}^2(y^2c_1-1)-2c_1c_2{}^2y^4-3c_3{}^2y^2\}\Bigg]\,,
\end{eqnarray}
where $erf(x)$ is the error function that is defined as
\begin{eqnarray}
erf(x)=\frac{2}{\sqrt{\pi}}\int_0^x e^{-t^2}dt\,.\end{eqnarray} The compactification factor $u(r)$ is then defined as
\begin{eqnarray}\label{comp}
&&u(y)=\frac{M(y)}{l}\,.\end{eqnarray}
Using Eq. (\ref{mass}) into (\ref{comp}), one can get the explicit form of the compactification factor.
The behavior of the gravitational mass and the  compactification factor are plotted in Fig. \ref{Fig:7}.
Fig. \ref{Fig:7} \subref{fig:mass} shows that the gravitational mass increases when the radial coordinate increases while Fig. \ref{Fig:7} \subref{fig:u} shows that the  compactification factor decreases for increasing radial coordinate.

\subsection{\bf Equation of state }
\begin{figure}
\centering
\subfigure[~Radial EoS ]{\label{fig:pr}\includegraphics[scale=0.3]{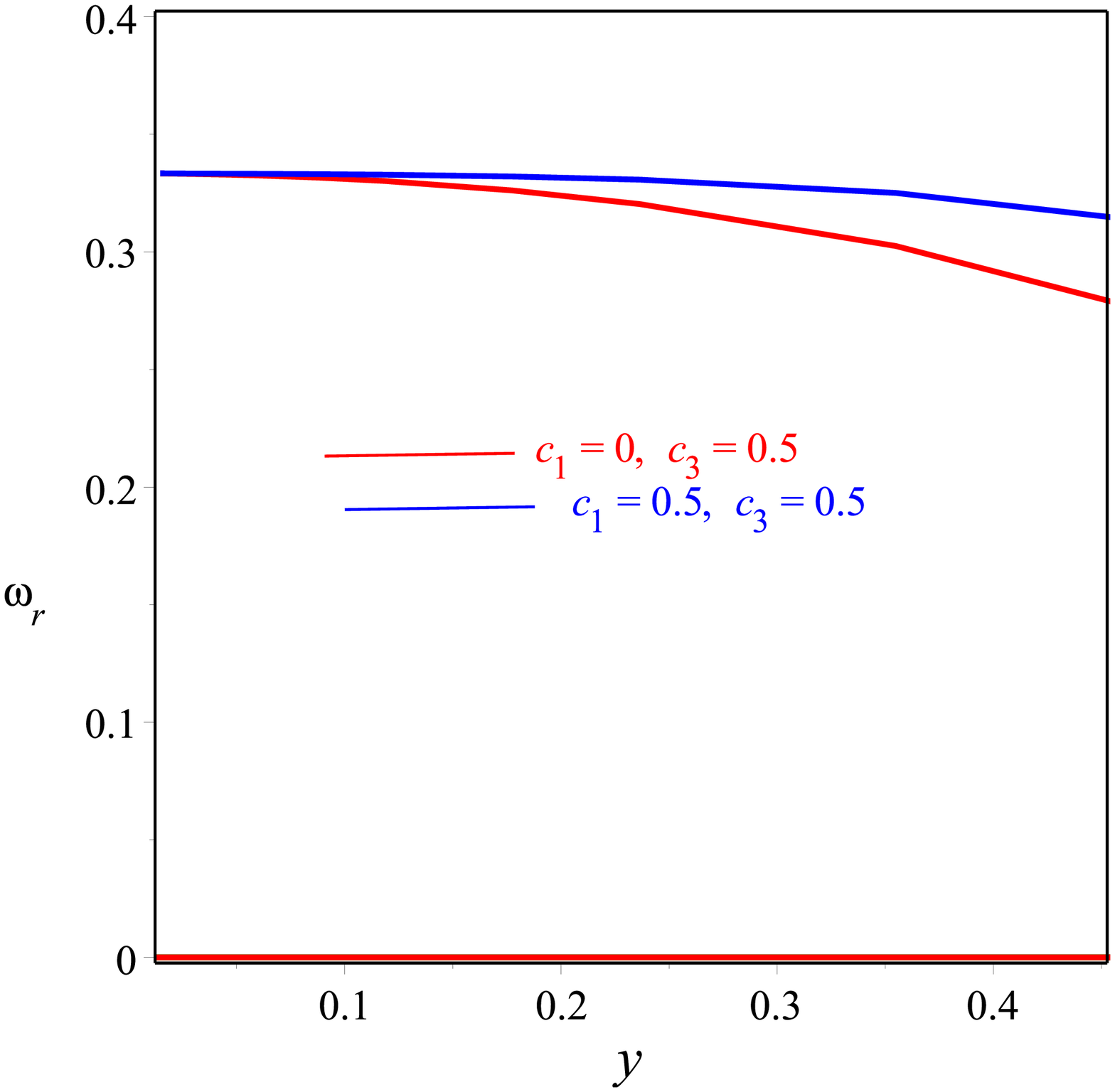}}
\subfigure[~Tangential EoS]{\label{fig:pt}\includegraphics[scale=.3]{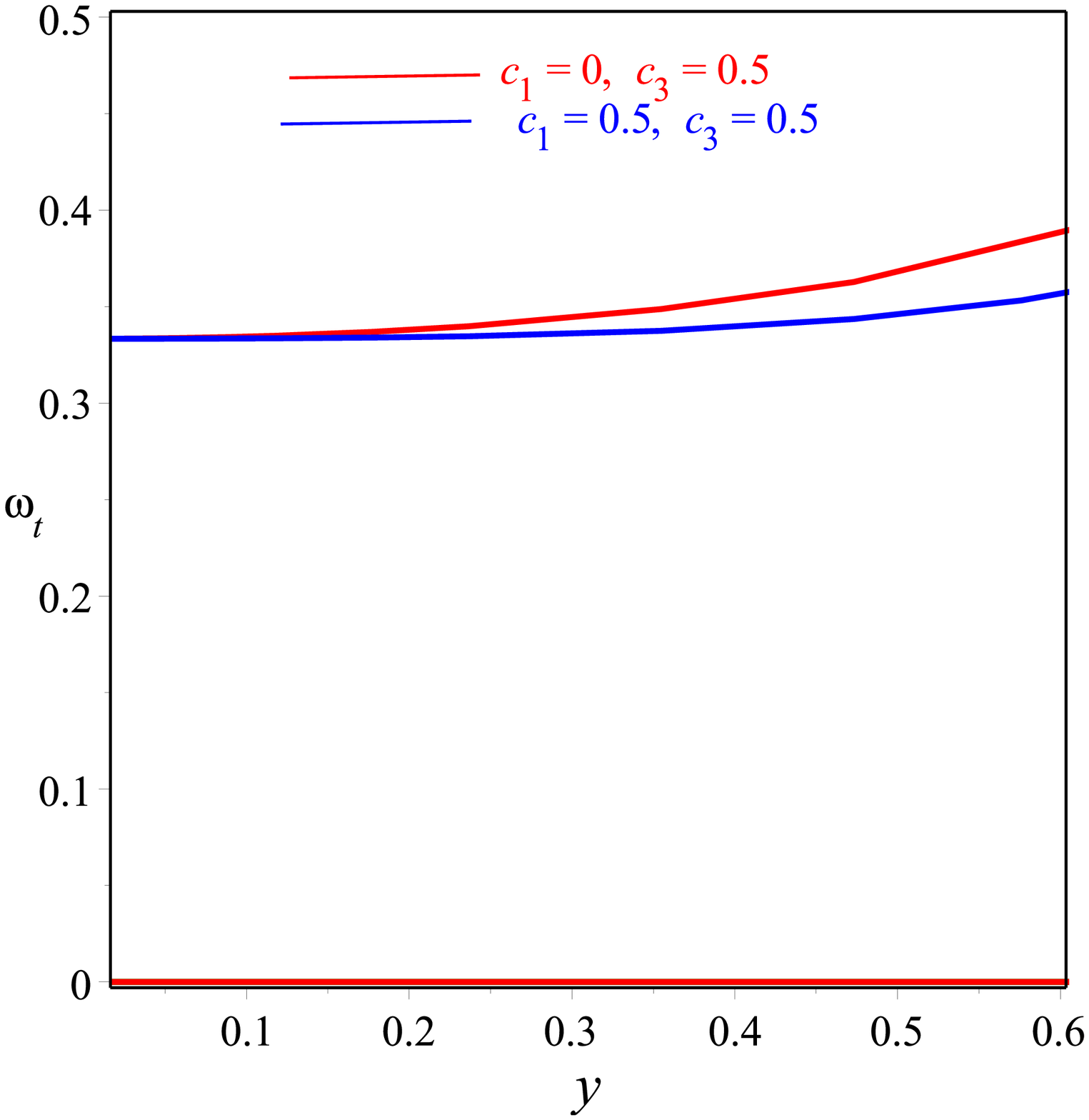}}
\caption[figtopcap]{\small{{The radial and transverse EoS   when $c_1=0$ and $c_1=-0.01$.}}}
\label{Fig:8}
\end{figure}

The study of  equation of state (EoS) for neutral compact stellar objects is reported in  \cite{Das:2019dkn} . In that case, EoS is linear.  In the present case, we will  show that  EoS cannot non-linear. To investigate this issue, we define  the radial and transverse   EoS  as:
  \begin{eqnarray} \label{sol3}
 && \omega_r=\frac{\hat{p}}{\hat{\epsilon}}\,, \qquad \qquad \omega_t=\frac{\hat{p}_t}{\hat{\epsilon}}\,,
  \end{eqnarray}
   where $\omega_r$ and $\omega_t$ are  the radial and transverse   EoS. Using Eqs.  in Appendix A,  one can  calculate the analytical form of $\omega_r$ and $\omega_t$ whose behaviors are plotted  in Fig.\ref{Fig:8}.
 As Figs. \ref{Fig:8} \subref{fig:pr} and \subref{fig:pt} show,   EoS are non-linear because of  the contribution of the higher--order curvature terms.
\section{Stability of the model} \label{S6}
In order to study the stability problem we are going to use   two different procedures; the first one is  the Tolman-Oppenheimer-Volkoff (TOV) equation and the second  is the study of the adiabatic index.
\subsection{The Tolman-Oppenheimer-Volkoff equation}
To study the stability of solution  in Appendix A, let us assume the hydrostatic equilibrium
by using the TOV equation  \cite{PhysRev.55.364,PhysRev.55.374} as  presented in \cite{PoncedeLeon1993}. It is
\begin{eqnarray}\label{TOV} \frac{2[\hat{p}_t-\hat{p}]}{y}-\frac{M_g(y)[\hat{\epsilon}(y)+\hat{p}_t(y)]e^{[a(y)-b(y)]/2}}{y}-\frac{d\hat{p}(y)}{y}=0,
 \end{eqnarray}
 where $M_g(y)$ is the gravitational mass calculated at the
radius $y$. It has the following equation
\begin{eqnarray}\label{ma} M_g(y)=4\pi{\int_0}^y\Big({{\mathcal T}_t}^t-{{\mathcal T}_r}^r-{{\mathcal T}_\theta}^\theta-{{\mathcal T}_\phi}^\phi\Big)y^2e^{[a(y)+b(y)]/2}dy=\frac{y a' e^{[b(y)-a(y)]/2}}{2}\,,
 \end{eqnarray}
Inserting Eq. (\ref{ma}) into (\ref{TOV}),  we get
\begin{eqnarray}\label{ma1} \frac{2(\hat{p}_t-\hat{p})}{y}-\frac{d\hat{p}}{dy}-\frac{a'[\hat{\epsilon}(y)+\hat{p}(y)]}{2}=F_g+F_a+F_h=0\,,
 \end{eqnarray}
 where $F_g=-\frac{a'[\hat{\epsilon}+\hat{p}]}{2}$, $F_a=\frac{2(\hat{p}_t-\hat{p})}{y}$ and $F_h=-\frac{d\hat{p}(y)}{dy}$ are the gravitational, the anisotropic and the hydrostatic forces respectively. The behavior of  TOV equation, for the model given in Appendix A, is shown in Fig.\ref{Fig:9}.
 \begin{figure}
\centering
\subfigure[~Radial EoS ]{\label{fig:pr}\includegraphics[scale=0.3]{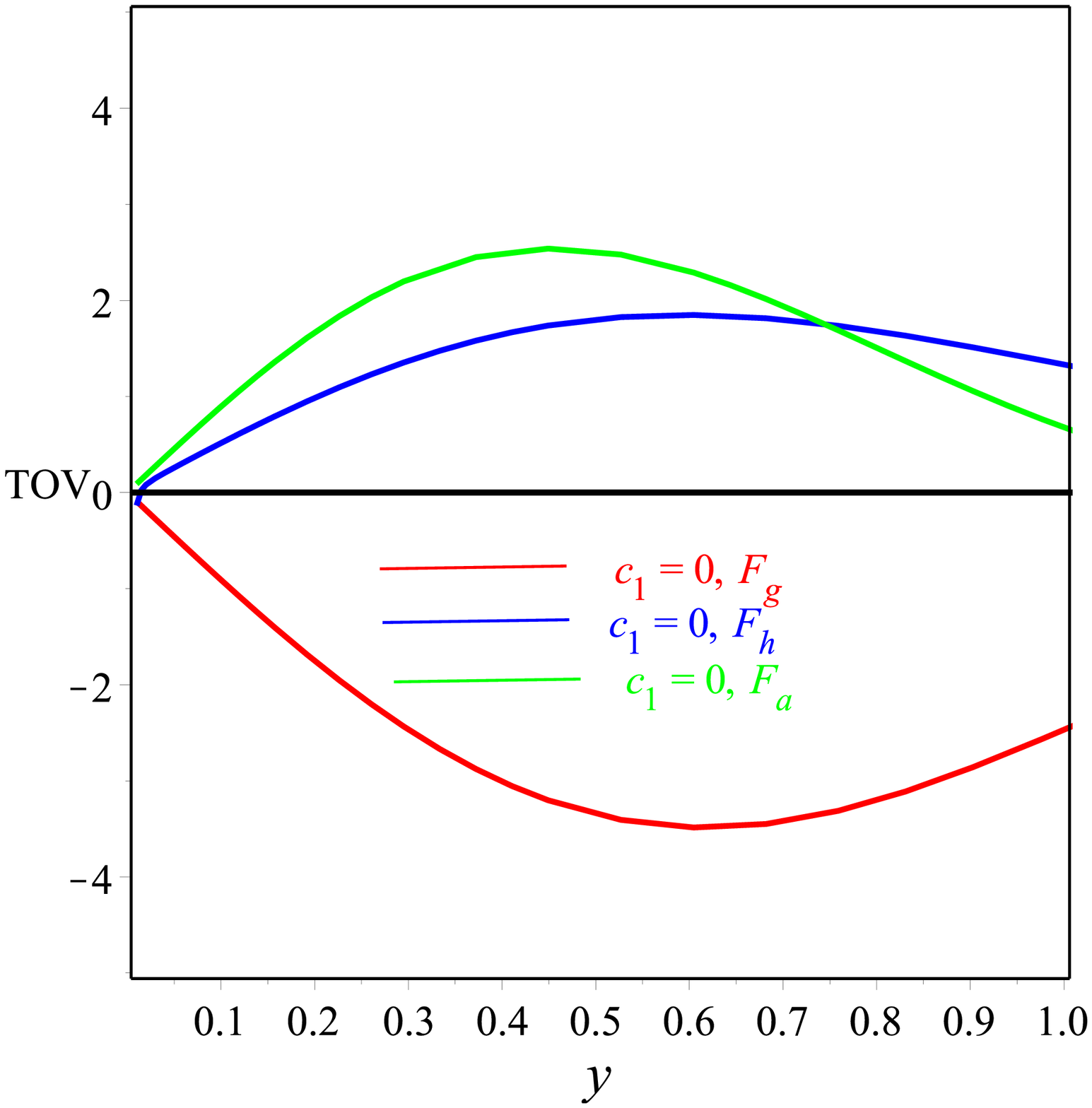}}
\subfigure[~Tangential EoS]{\label{fig:pt}\includegraphics[scale=.3]{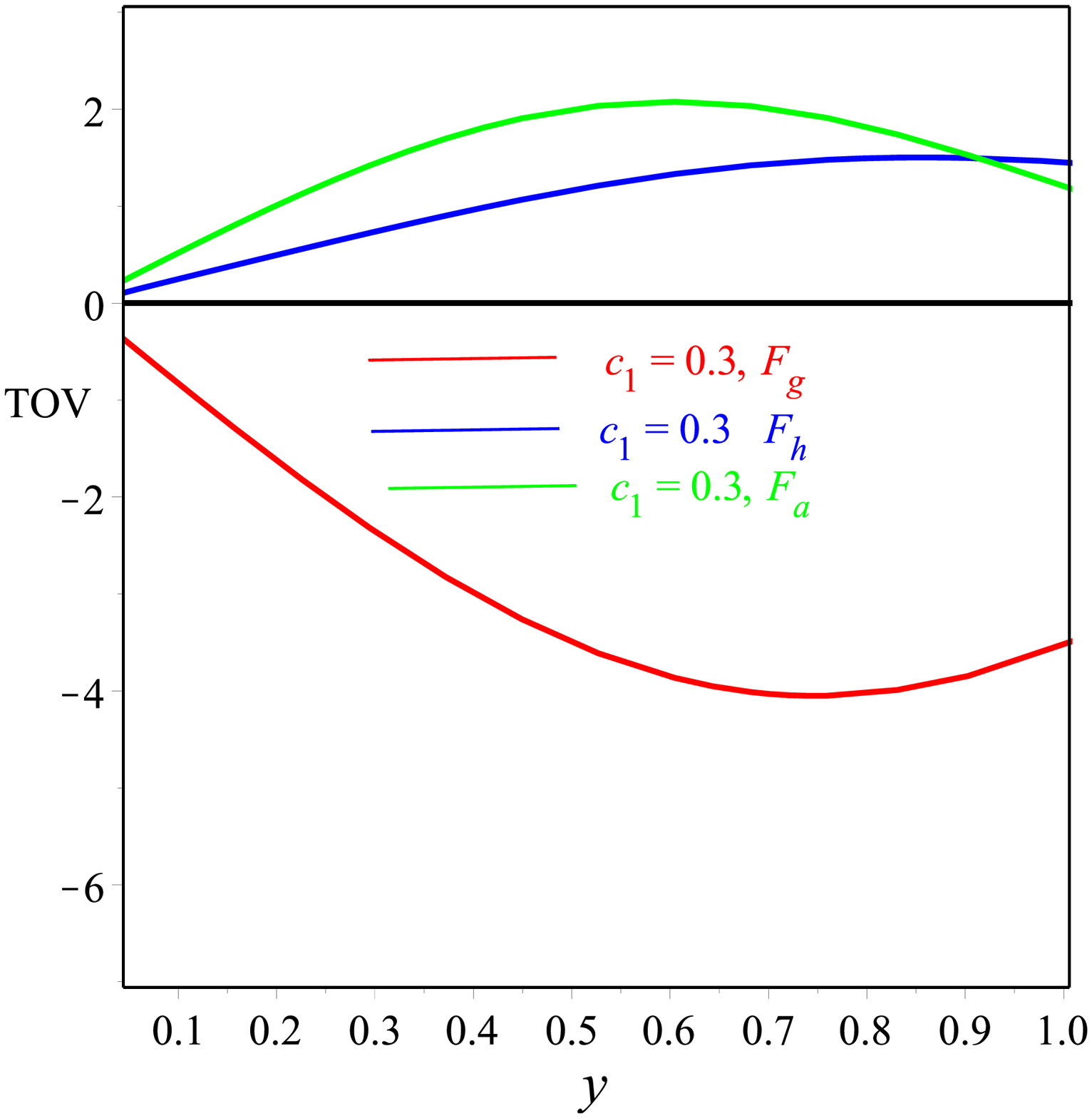}}
\caption[figtopcap]{\small{{The TOV of solution  when $c_1=0$ and $c_1=0.5$}}}
\label{Fig:9}
\end{figure}
This figure
shows that hydrostatic and anisotropic forces are positive and
 dominated on the gravitational force which is negative to
keep the system in static equilibrium either when $c_1=0$ or $c_1=0.5$.
\subsection{\bf The adiabatic index}
\begin{figure}
\centering
\subfigure[~The adiabatic index  when $c_1=0$]{\label{fig:dnesity}\includegraphics[scale=0.3]{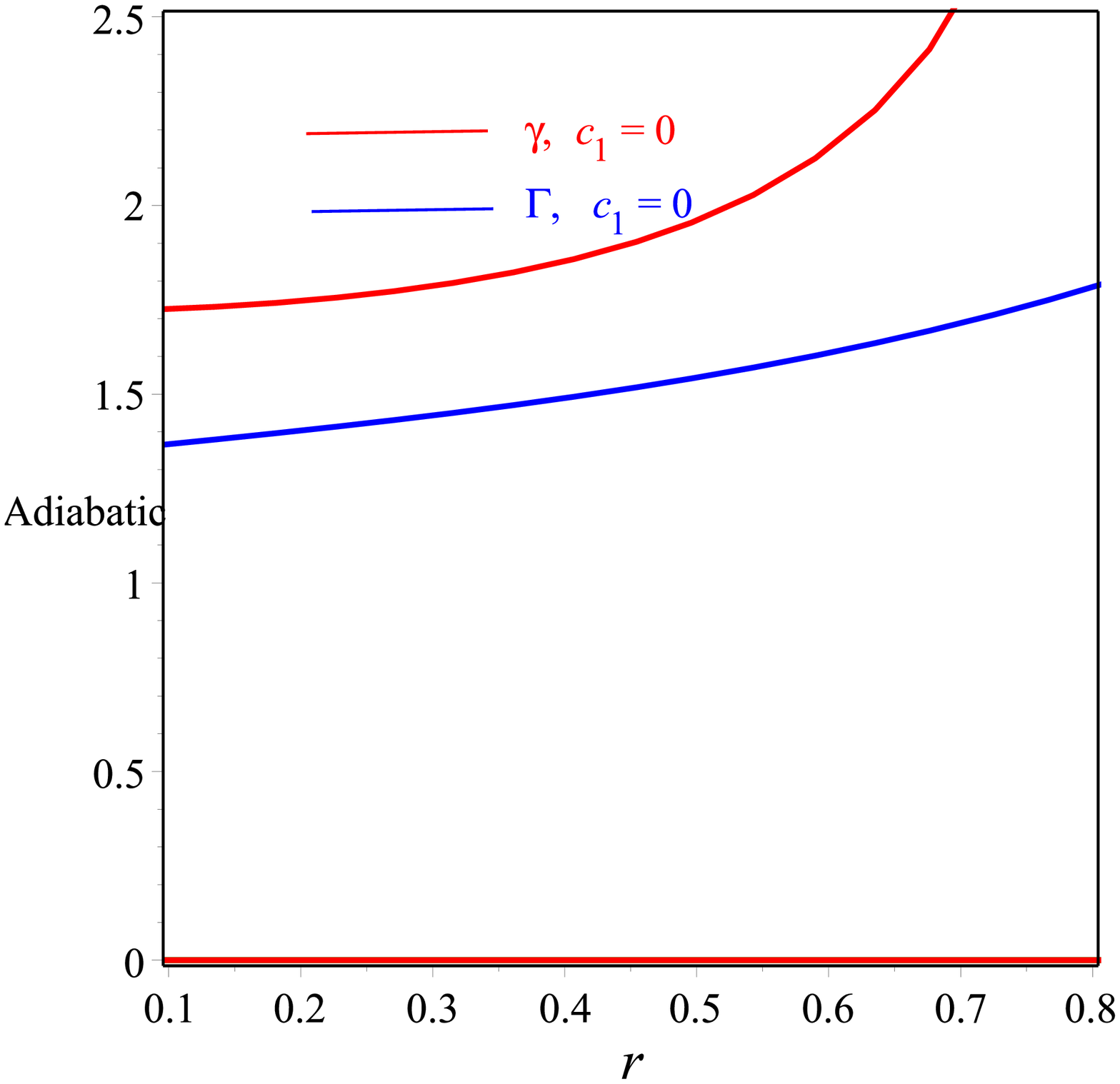}}
\subfigure[~The adiabatic index  when $c_1=0.5$]{\label{fig:pressure}\includegraphics[scale=.3]{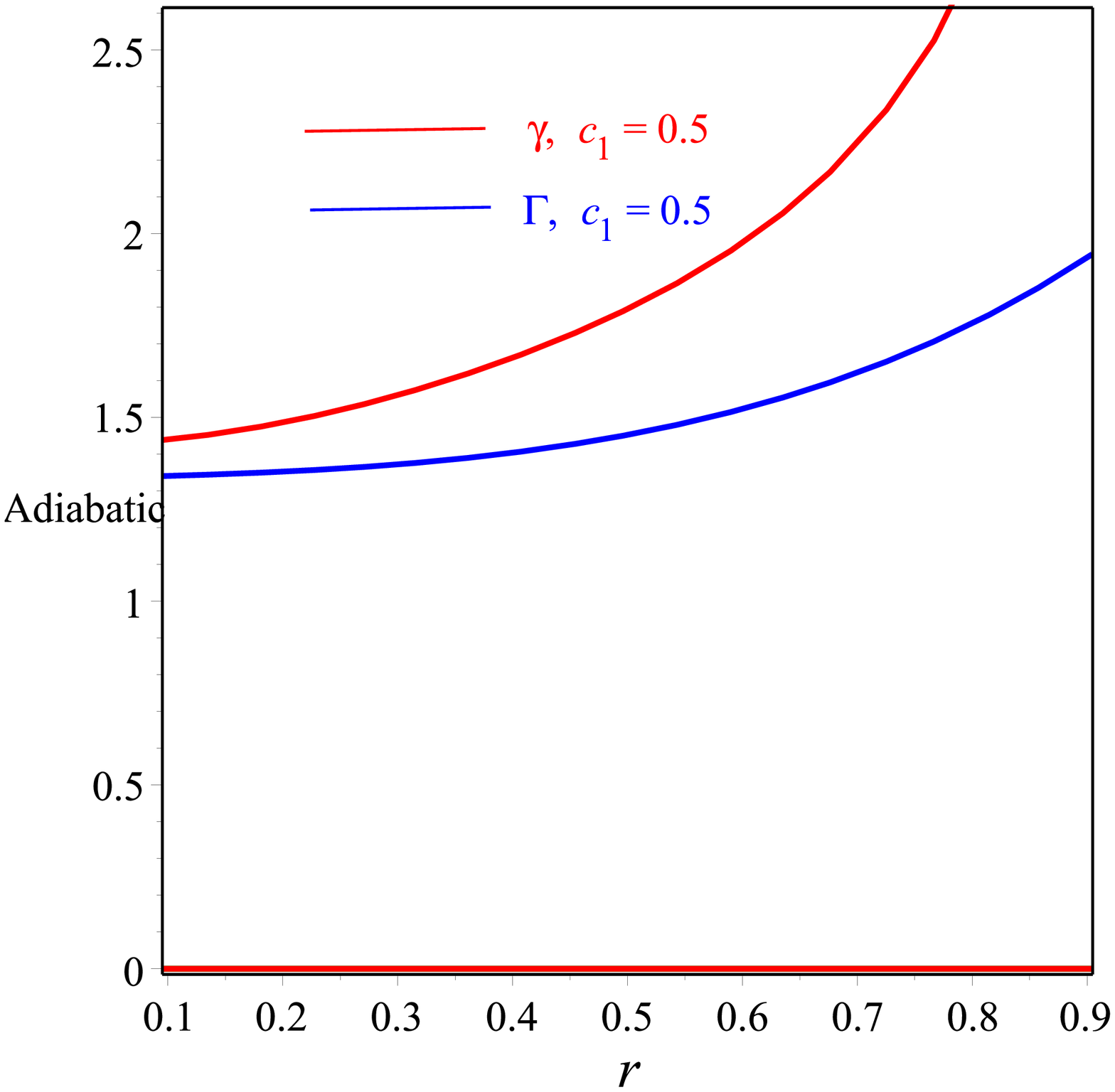}}
%
\caption[figtopcap]{\small{{The adiabatic index  of solution given in Appendix A for $c_1=0$ and $c_1=0.5$}}}
\label{Fig:10}
\end{figure}

The adiabatic index $\gamma$   is defined as \cite{1994MNRAS.267..637C,1997PhR...286...53H}
\begin{eqnarray}\label{ai} \gamma=\left(\frac{\hat{\epsilon}+\hat{p}}{\hat{p}}\right)\frac{d\hat{p}}{d\hat{\epsilon}}\,.
 \end{eqnarray}
It allows  to link the structure of a spherically symmetric
static object and the EoS of the interior solution. It can be used   to investigate the
 stability \cite{Moustakidis:2016ndw}. Any  interior solution is stable if   its adiabatic index is  greater than $4/3$ \cite{1975A&A....38...51H}.   If   $\gamma=\frac{4}{3}$,  then  the isotropic
sphere is in neutral equilibrium.  Following  Chan et al. \cite{10.1093/mnras/265.3.533},  the stability condition of a relativistic anisotropic sphere, $\gamma >\Gamma$,  must be satisfied. Here it is  \begin{eqnarray}\label{ai} \Gamma=\frac{4}{3}-\left\{\frac{4(\hat{p}-\hat{p}_t)}{3\lvert \hat{p}'\lvert}\right\}_{max}\,.
 \end{eqnarray}
 The behaviour of the adiabatic index is shown in Fig.\ref{Fig:10} which ensures the  stability condition of our solution.
\begin{small}
	\begin{table}[tbp]
		\begin{center}
			\begin{tabular}{l | c c c | c c c c} \hline
				Pulsar &
				$M (M_\odot)$ &
				$R (\textrm km)$ &
				Reference &
				${{\rho}_{_{_{_{_l}}}}}_{\lvert_{c_1=0}} (10^{15}{\textrm gr/cm^3})$ &${{\rho}_{_{_{_{_l}}}}}_{\lvert_{c_1=0.5}} (10^{15}{\textrm gr/cm^3})$ &
				$\frac{d p}{d\rho}(c^2)_{\lvert_{c_1=0}}$ &$\frac{d p}{d\rho}(c^2)_{\lvert_{c_1=0.5}}$
				 \\ [1.5ex] \hline
				\midrule
				\multicolumn{6}{c}{Millisecond Pulsars with White Dwarf Companions}
			\\ [1.5ex] \hline
				\midrule
				J0437-4715 &
				$1.44^{+0.07}_{-0.07}$ &
				$13.6^{+0.9}_{-0.8}$ &
				\cite{2016MNRAS.455.1751R,2019MNRAS.490.5848G}  &$1.27$ &$1.1$ & $0.37$ & $0.1$
				\\ [1.5ex]
				J0030+0451 &
				$1.44^{+0.15}_{-0.16}$ &
				$13.02^{+1.24}_{-1.06}$ &
				\cite{2019ApJ...887L..24M}
				&$1.7$ &$1.4$ & $0.47$ & $0.32$
				\\ [1.5ex]
				&
				$1.34^{+0.15}_{-0.16}$ &
				$12.71^{+1.14}_{-1.19}$ &
				\cite{2019ApJ...887L..21R} &
				$1.5$ & $1.3$&
				$0.3$	& $0.45$
				\\ [1.5ex] \hline
				\midrule
				\multicolumn{6}{c}{Pulsars Presenting Thermonuclear Bursts}
				\\ [1.5ex] \hline
				\midrule
				4U 1724-207 &
$1.81^{+0.25}_{-0.37}$ &
$12.2^{+1.4}_{-1.4}$ &
\cite{2016ApJ...820...28O} &
$3.33$ &$3.1$&
$0.61$& $0.66$
\\ [1.5ex]
4U 1820-30 &
$1.46^{+0.21}_{-0.21}$ &
$11.1^{+1.8}_{-1.8}$ &
\cite{2016ApJ...820...28O} &
$3$ &
$2.6$&$0.45$&$0.36$
\\ [1.5ex]
SAX J1748.9-2021 &
$1.81^{+0.25}_{-0.37}$ &
$11.7^{+1.7}_{-1.7}$ &
\cite{2016ApJ...820...28O} &
$2.02$ &
$1.9$&$0.33$&$0.04$
\\ [1.5ex]
				EXO 1745-268 &
$1.65^{+0.21}_{-0.31}$ &
$10.5^{+1.6}_{-1.6}$ &
\cite{2016ApJ...820...28O} &
$0.77$ &
$0.8$&$0.31$&$0.37$
\\ [1.5ex]
				4U 1608-52 &
$1.57^{+0.30}_{-0.29}$ &
$9.8^{+1.8}_{-1.8}$ &
\cite{2016ApJ...820...28O} &
$1.5$ &
$1.1$&$0.62$&$1.71$
\\  [1.5ex]
				KS 1731-260 &
$1.61^{+0.35}_{-0.37}$ &
$10.0^{+2.2}_{-2.2}$ &
\cite{2016ApJ...820...28O} &
$1.4$ &
$1.13$&$0.52$&$0.74$
\\ \hline
				\bottomrule
			\end{tabular}
		\end{center}
		\caption{The boundary density $\rho_l$ and slope of the radial pressure reported for various observed objects. The density and the derivative of radial pressure are derived for the values $c_1=0$ and $c_1=0.5$ at the stellar boundary $l$ assuming a nuclear density of the order $10^{15}gr/cm^3$.}
		\label{tab:density}
	\end{table}
\end{small}
\section{Discussion  and conclusions } \label{S7}
In this study, we derived an anisotropic spherically symmetric  solution in the framework of $f\mathcal{(R)}$  gravity. To this aim,  we used a line element  containing two unknown functions and applied it to the field equations of $f\mathcal{(R)}$ deriving four non-linear differential equations: three of them are  the field equations   and the fourth one is  the trace of  the field equations.  This system involves six unknown functions, two of them are the metric potentials, one is the form of $f\mathcal{(R)}$  and the other are the density, and radial and tangential pressures.  We solved these differential equations  assuming the forms of the radial component  of the metric  and of the derivative of $f\mathcal{(R)}$. Substituting these quantities in the anisotropy parameter, $\Delta=\hat{p}_t-\hat{p}_r$, we get  the time component of the  metric. From these results,  we obtained the form of $f\mathcal{(R)}$, the density, radial, and tangential pressures. It is interesting to mention that our solution is characterized by a constant $c_1$ which parameterize the deviation with respect to  GR. Specifically, for $c_1=0$,  we return to GR.

Due to the fact that the obtained interior solution has a non-trivial form for the Ricci scalar,  we match it with a spherically-symmetric exterior solution  having also a non-trivial form of Ricci scalar. From this matching,  it is possible to  relate the mass and the maximal radius   of the compact object with the compactness parameter, that is   $C=\frac{2GM}{c^2l}$ which must be $C<<0.5$ for the physical consistency of the model.  The further step is to see if this solution represents a realistic star. To this aim, we discussed the necessary conditions  required by any physically consistent compact stellar object and show  that our model satisfies  all these conditions.  In particular, we considered two cases: one for $c_1=0$ and another  for $c_1=0.5$.

Finally, we checked if the  model satisfies the stability conditions. To this aim, we  derived  the TOV equation  and showed that it is in static equilibrium either when $c_1=0$ or $c_1=0.5$.  We also studied the adiabatic index to complete the picture of the stability and showed that the model satisfies the adiabatic index  $\gamma >4/3$.

These results can be applied to the  observational data of different millisecond pulsars with white dwarf companions and pulsars presenting thermonuclear bursts as reported in Table I.

For these objects, it is possible to  calculate the density at the center and at the surface and then derive  the EoS at the center and at the surface. It is interesting to see that, in the case of millisecond pulsars with a white dwarf companion, and, in the case of pulsars presenting thermonuclear bursts,  our solution is compatible with observations. From the observational data in literature, as reported in the Table I, it is possible to derive the total mass and the radius of the given pulsar.

To conclude, we derived an interior solution in the framework  of $f\mathcal{(R)}$  assuming physically motivated conditions to obtain the form of $f\mathcal{(R)}$. This solution has a non-trivial form of the Ricci scalar. The internal  solution can be  matched with the external one considering  compacteness and anisotropy parameters. In a forthcoming paper, we will consider a similar situation assuming   a charged interior solution.

\section*{Acknowledgements}

SC is  supported  by the INFN {\it sez. di Napoli, iniziative specifiche}  MOONLIGHT2 and QGSKY.

\newpage
{\centerline{\bf Appendix A:  Energy density, radial and tangential pressures.}}\vspace{0.3cm}
Let us report now the quantities related to the interior solution discussed in this paper.
\renewcommand{\eqref}{}
\begin{eqnarray*}
 && \hat{\epsilon}(y)=\frac{1}{2y^2[c_3{}^4e^{2c_0y^2}+4c_3{}^3c_2c_0e^{3c_0y^2/2}+6c_3{}^2c_2{}^2c_0{}^2e^{c_0y^2}+4c_3c_2{}^3c_0{}^3e^{c_0y^2/2}+c_2{}^4c_0{}^4]}
 \Big[c_0c_2c_3e^{c_0y^2/2}(4c_2{}^2c_0{}^2[c_1y^2+1]\nonumber\\
 &&+c_3{}^2c_0y^2[30c_1y^2+17]-4c_3{}^2[4c_1y^2+1])+c_0{}^3c_2{}^3c_3e^{-c_0y^2/2}[c_0y^2(11+18c_1y^2)-4-16c_1y^2]+
 4c_3{}^3c_2c_0e^{3c_0y^2/2}(c_1y^2+1)\nonumber\\
 &&+c_0{}^4c_2{}^4e^{-c_0y^2}[y^2c_0(2+3c_1y^2)-4c_1y^2-1]+c_3{}^4e^{2c_0y^2}(c_1y^2+1)+c_3{}^2e^{c_0y^2}[6c_0{}^2c_2{}^2
 (c_1y^2+1)+c_0{}^2c_3{}^2y^2(9c_1y^2+5) \nonumber\\
 &&-c_3{}^2(4c_1y^2+1)]+c_0{}^2c_2{}^2[c_0{}^2c_2{}^2(c_1y^2+1)+3c_0c_3{}^2y^2(12c_1y^2+7)-6c_3{}^2(4c_1y^2+1)]\Big]\,, \nonumber\\
 &&\hat{p}(y)=\frac{e^{-c_0y^2}\Big\{c_3e^{c_0y^2/2}[1+c_0y^2(3+7c_1y^2)]-c_3e^{3c_0y^2/2}[1+c_1y^2]+c_0c_2[1-e^{c_0y^2}(1+c_1y^2)+c_0y^2(2+c_1y^2)]\Big\}}{2y^2[c_0c_2+c_3e^{c_0y^2/2]}}\,, \nonumber\\
 && \hat{p}_{_{_t}}(y)=\frac{1}{2y^2[c_3{}^4e^{2c_0y^2}+4c_3{}^3c_2c_0e^{3c_0y^2/2}+6c_3{}^2c_2{}^2c_0{}^2e^{c_0y^2}+4c_3c_2{}^3c_0{}^3e^{c_0y^2/2}+c_2{}^4c_0{}^4]}\Big[c_0c_2c_3e^{c_0y^2/2}[4c_0{}^2c_2{}^2(1+c_1y^2)\nonumber\\
 && +c_0c_3{}^2y^2(3+2c_1y^2)-4c_3{}^2(1+2c_1y^2)]-c_0{}^3c_2{}^3c_3e^{-c_0y^2/2}[4(1+2c_1y^2)-c_0y^2(1-2c_1y^2)]+4c_0c_2c_3{}^3e^{3c_0y^2/2}[1+c_1y^2]\nonumber\\
 && -c_0{}^4c_2{}^4e^{-c_0y^2}[1+2c_1y^2+c_1c_0y^4]+c_3{}^4e^{2c_0y^2}(1+c_1y^2)+c_3{}^2e^{c_0y^2}[6c_0{}^2c_2{}^2(1+c_1y^2)+c_3{}^2c_0y^2(1+c_1y^2)-  c_3{}^2(1+2c_1y^2)]\nonumber\\
 && +c_0{}^2c_2{}^2[c_2{}^2c_0{}^2(1+c_1y^2)+3c_0c_3{}^2y^2-6c_3{}^2(1+2c_1y^2]\Big]\,.  \qquad \qquad \qquad \qquad \qquad \qquad \qquad \qquad \qquad \qquad   (\eqref{\mathbf A}) \nonumber
  \end{eqnarray*}
  {\centerline{\bf Appendix B:  The gradients of   energy--density, radial and transverse  pressures.}}\vspace{0.3cm}
  The same as above for gradients.
\renewcommand{\eqref}{}
  \begin{eqnarray*}
 && \hat{\epsilon}'=\frac{1}{2y^3[c_0c_2+c_3e^{c_0y^2/2}][c_0{}^2c_2{}^2+c_3{}^2e^{c_0y^2}+2c_0c_2c_3e^{c_0y^2/2}]^2}
 \Bigg[c_0{}^2c_2{}^2c_3e^{c_0r^2/2}\Big(20c_3{}^2+4c_0c_3{}^2y^2(53c_1y^2+5)-c_0{}^2\Big[10c_2{}^2\nonumber\\
 &&+c_3{}^2y^4(114c_1y^2+67)\Big]\Big)+2e^{3c_0y^2/2}[c_3{}^5+
 c_0c_3{}^5y^2(1+13c_1y^2)-c_0{}^2c_3{}^3(c_2{}^2+c_3{}^2y^4\{5+9c_1y^2\})]+c_0{}^4c_2{}^4c_3e^{-c_0y^2/2}\Big[10\nonumber\\
 &&-c_0{}^2y^4(23+36c_1y^2)+2c_0y^2(41c_1y^2+5)\Big]
 -2c_3{}^5e^{5c_0y^2/2}+c_0c_2\Big\{2c_0{}^4c_2{}^4e^{-c_0y^2}(1-y^4c_0{}^2[2+3c_1y^2]+c_0y^2[1+7c_1y^2])\nonumber\\
 &&+10c_3{}^4e^{2c_0y^2}+c_3{}^2e^{c_0y^2}[\{c_3{}^2y^4
 (72c_1y^2+41)+10c_2{}^2\}c_0{}^2-2c_3{}^2y^2(5+59c_1y^2)-10c_3{}^2]+c_0{}^2c_2{}^2\Big[\Big\{5c_3{}^2y^4\Big(18c_1y^2+11\Big)\nonumber\\
 &&+2c_2{}^2\Big\}c_0{}^2-4c_0c_3{}^2y^2(5+47c_1y^2)-20c_3{}^2\Big]\Big\}\Bigg]\,,\nonumber\\
 &&\hat{p}'=-\frac{1}
 {2y^3[c_0{}^2c_2{}^2+c_3{}^2e^{c_0y^2}+2c_0c_2c_3e^{c_0y^2/2}]}\Bigg[e^{-c_0y^2}\Big[c_0c_2c_3e^{c_0y^2/2}\{4+c_0y^4[9+22c_1y^2]+4c_0y^2[1-6c_1y^2]\}-2c_3{}^2e^{2c_0y^2}
\nonumber\\
 && -4c_0c_2c_3e^{3c_0y^2/2}
 +2e^{c_0y^2}(c_3{}^2+c_0c_3{}^2y^2[1-c_1y^2]+c_0{}^2[c_3{}^2y^4(3+7c_1y^2)-c_2{}^2])+2c_0{}^2c_2{}^2\Big\{1+c_0{}^2y^4[2+5c_1y^2]\nonumber\\
 &&+c_0y^2[1-5c_1y^2]\Big\}\Big]\Bigg]\,,\nonumber\\
 && \hat{p}'_t=\frac{1}{2y^3[c_0c_2+c_3e^{c_0y^2/2}][c_0{}^2c_2{}^2+c_3{}^2e^{c_0y^2}+2c_0c_2c_3e^{c_0y^2/2}]^2}\Bigg[c_0{}^2c_2{}^2c_3e^{c_0r^2/2}
 \Big(20c_3{}^2+4c_0c_3{}^2y^2(11c_1y^2+5)-c_0{}^2\Big[10c_2{}^2\nonumber\\
 &&-c_3{}^2y^4(2c_1y^2-9)\Big]\Big)+2e^{3c_0y^2/2}c_3{}^3[c_3{}^2+c_0c_3{}^2y^2(1+3c_1y^2)-c_0{}^2\{10c_2{}^2+c_3{}^2y^4(1+c_1y^2)\}]
 +c_0{}^4c_2{}^4c_3e^{-c_0y^2/2}[10\nonumber\\
 &&+c_0{}^2y^4(8c_1y^2-1)+2c_0y^2(5+7c_1y^2)]-2c_3{}^5e^{5c_0y^2/2}+c_0c_2\Big\{2c_0{}^4c_2{}^4e^{-c_0y^2}(1+c_0y^2)
 (1+c_0c_1y^4)-10c_3{}^2e^{2c_0y^2}\nonumber\\
 &&+e^{c_0y^2}[10c_3{}^4+2c_0c_3{}^4y^2(5+13c_1y^2)-c_0{}^2(20c_2{}^2c_3{}^2+c_3{}^4y^4\{7+4c_1y^2\})]+c_0{}^2c_2{}^2\Big[5c_3{}^2y^4
 (2c_1y^2-1)-2c_2{}^2\nonumber\\
 &&+4c_0c_3{}^2y^2(5+9c_1y^2)+20c_3{}^2\Big]\Big\}\Bigg]\,, \hspace*{10cm}(\eqref{\mathbf B})
  \end{eqnarray*}
  where $\hat{\epsilon}'=\frac{d\hat{\epsilon}}{dy}$,\,  $\hat{p}'=\frac{d\hat{p}}{dy}$ \, and $\hat{p}'_t=\frac{d\hat{p}_t}{dy}$.
 

  {\centerline{\bf Appendix C:   Derivation of Eq. (\ref{dfas(r)})}}\vspace{0.3cm}
  \renewcommand{\eqref}{}
 Eqs. (\ref{const}) can be  inserted in  the last equation of system (\ref{fes}), the trace  (\ref{f3ss}). We get a  lengthy differential equation in $y$ which we report here  for the sake of completeness.   It is  
\begin{eqnarray*}
 && f'''(y)\; \Bigg\{ {\frac {75}{2}}\,{{ c_2}}^{5}{ c_3}\, \Bigg(  \left( {y}^{8}{{
c_3}}^{4}+{\frac {58}{25}}\,{{ c_3}}^{2}{{ c_2}}^{2}{y}^{4}+1/15\,{
{ c_2}}^{4} \right) {{ c_0}}^{4}+ \left( -8/5\,{{ c_3}}^{2}{{
 c_2}}^{2}{y}^{2}-{\frac {77}{25}}\,{y}^{6}{{ c_3}}^{4} \right) {{
 c_0}}^{3}+ \left( -8/5\,{{ c_3}}^{2}{{ c_2}}^{2}-7/5\,{{ c_3}
}^{4}{y}^{4} \right) {{ c_0}}^{2}\nonumber\\
&&+{\frac {84}{25}}\,{{ c_3}}^{4}{
 c_0}\,{y}^{2}+{\frac {42}{25}}\,{{ c_3}}^{4} \Bigg) {{ c_0}}^{
5}{{ e}^{-1/2\,{ c_0}\,{y}^{2}}}-{\frac {63}{8}}\,{{ c_2}}^{3}
 \Bigg(  \left( {y}^{8}{{ c_3}}^{4}-{\frac {44}{3}}\,{{ c_3}}^{2}{
{ c_2}}^{2}{y}^{4}-{\frac {80}{21}}\,{{ c_2}}^{4} \right) {{ c_0
}}^{4}+ \left( {\frac {40}{21}}\,{y}^{6}{{ c_3}}^{4}+16\,{{ c_3}}^
{2}{{ c_2}}^{2}{y}^{2} \right) {{ c_0}}^{3}\nonumber\\
&&+ \left( 16\,{{ c_3}}
^{2}{{ c_2}}^{2}-{\frac {40}{21}}\,{{ c_3}}^{4}{y}^{4} \right) {{
 c_0}}^{2}-{\frac {160}{21}}\,{{ c_3}}^{4}{ c_0}\,{y}^{2}-{
\frac {80}{21}}\,{{ c_3}}^{4} \Bigg) {{ c_3}}^{3}{{ c_0}}^{3}{
{ e}^{1/2\,{ c_0}\,{y}^{2}}}+1/4\,{ c_2}\,{{ c_3}}^{5}{ c_0
}\, \Bigg(  \left( 252\,{{ c_2}}^{4}+60\,{{ c_3}}^{2}{{ c_2}}^{2
}{y}^{4}+{y}^{8}{{ c_3}}^{4} \right) {{ c_0}}^{4}\nonumber\\
&&+ \left( -240\,{{
 c_3}}^{2}{{ c_2}}^{2}{y}^{2}+11\,{y}^{6}{{ c_3}}^{4} \right) {{
 c_0}}^{3}+ \left( 21\,{{ c_3}}^{4}{y}^{4}-240\,{{ c_3}}^{2}{{
 c_2}}^{2} \right) {{ c_0}}^{2}+20\,{{ c_3}}^{4}{ c_0}\,{y}^{2
}+10\,{{ c_3}}^{4} \Bigg) {{ e}^{3/2\,{ c_0}\,{y}^{2}}}+{
\frac {477}{8}}\,{{ c_2}}^{7}{ c_3}\, \Bigg(  \Bigg( {\frac {74}{
477}}\,{y}^{4}{{ c_2}}^{2}\nonumber\\
&&+{y}^{8}{{ c_3}}^{2} \Bigg) {{ c_0}}^
{4}+ \left( -{\frac {40}{477}}\,{y}^{2}{{ c_2}}^{2}-{\frac {232}{159
}}\,{y}^{6}{{ c_3}}^{2} \right) {{ c_0}}^{3}+ \left( -{\frac {152}
{159}}\,{{ c_3}}^{2}{y}^{4}-{\frac {40}{477}}\,{{ c_2}}^{2}
 \right) {{ c_0}}^{2}+{\frac {160}{159}}\,{{ c_3}}^{2}{y}^{2}{
c_0}+{\frac {80}{159}}\,{{ c_3}}^{2} \Bigg) {{ c_0}}^{7}{{ e}^{
-3/2\,{ c_0}\,{y}^{2}}}\nonumber\\
&&+17/2\,{{ c_2}}^{9} \left( { c_0}\,{y}^{2
}-1 \right) { c_3}\, \left( { c_0}\,{y}^{2}+1/2 \right)  \left( {y
}^{4}{{ c_0}}^{2}-{\frac {10}{17}}\,{ c_0}\,{y}^{2}-{\frac {10}{17
}} \right) {{ c_0}}^{9}{{ e}^{-5/2\,{ c_0}\,{y}^{2}}}+ \Bigg(
 \left( 30\,{{ c_3}}^{7}{{ c_2}}^{3}-11/4\,{y}^{4}{{ c_3}}^{9}{
 c_2} \right) {{ c_0}}^{3}\nonumber\\
&&-5\,{y}^{2}{{ c_3}}^{9}{ c_2}\,{{
 c_0}}^{2}-5\,{{ c_3}}^{9}{ c_2}\,{ c_0} \Bigg) {{ e}^{5/2
\,{ c_0}\,{y}^{2}}}+5/2\,{{ e}^{7/2\,{ c_0}\,{y}^{2}}}{{ c_3}
}^{9}{ c_2}\,{ c_0}+{\frac {1059}{16}}\,{{ c_2}}^{6} \Bigg(
 \left( {y}^{8}{{ c_3}}^{4}+{\frac {4}{1059}}\,{{ c_2}}^{4}+{
\frac {200}{353}}\,{{ c_3}}^{2}{{ c_2}}^{2}{y}^{4} \right) {{
c_0}}^{4}\nonumber\\
&&-{\frac {672}{353}}\, \left( {\frac {5}{28}}\,{{ c_2}}^{2}+{
{ c_3}}^{2}{y}^{4} \right) {{ c_3}}^{2}{y}^{2}{{ c_0}}^{3}+
 \left( -{\frac {120}{353}}\,{{ c_3}}^{2}{{ c_2}}^{2}-{\frac {392}
{353}}\,{{ c_3}}^{4}{y}^{4} \right) {{ c_0}}^{2}+{\frac {560}{353}
}\,{{ c_3}}^{4}{ c_0}\,{y}^{2}+{\frac {280}{353}}\,{{ c_3}}^{4}
 \Bigg) {{ c_0}}^{6}{{ e}^{-{ c_0}\,{y}^{2}}}\nonumber\\
&&+1/4 \left(
 \left( {y}^{8}{{ c_3}}^{4}+210{{ c_2}}^{4}-12{{ c_3}}^{2}{{
 c_2}}^{2}{y}^{4} \right) {{ c_0}}^{4}+ \left( 2{y}^{6}{{ c_3}
}^{4}-90{{ c_3}}^{2}{{ c_2}}^{2}{y}^{2} \right) {{ c_0}}^{3}+
 \left( -90{{ c_3}}^{2}{{ c_2}}^{2}+3{{ c_3}}^{4}{y}^{4}
 \right) {{ c_0}}^{2}+2{{ c_3}}^{4}{ c_0}\,{y}^{2}+{{ c_3}}^
{4} \right) {{ c_3}}^{6}{{ e}^{2{ c_0}{y}^{2}}}\nonumber\\
&&+{\frac {489
}{16}}\,{{ c_2}}^{8} \Bigg(  \left( {y}^{8}{{ c_3}}^{2}+{\frac {16
}{489}}\,{y}^{4}{{ c_2}}^{2} \right) {{ c_0}}^{4}+ \left( -{\frac
{8}{489}}\,{y}^{2}{{ c_2}}^{2}-{\frac {200}{163}}\,{y}^{6}{{ c_3}}
^{2} \right) {{ c_0}}^{3}+ \left( -{\frac {8}{489}}\,{{ c_2}}^{2}-
{\frac {140}{163}}\,{{ c_3}}^{2}{y}^{4} \right) {{ c_0}}^{2}+{
\frac {120}{163}}\,{{ c_3}}^{2}{y}^{2}{ c_0}\nonumber\\
&&+{\frac {60}{163}}\,{{
 c_3}}^{2} \Bigg) {{ c_0}}^{8}{{ e}^{-2\,{ c_0}\,{y}^{2}}}+{
{ c_2}}^{10} \left( { c_0}\,{y}^{2}-1 \right) ^{2} \left( { c_0}
\,{y}^{2}+1/2 \right) ^{2}{{ c_0}}^{10}{{ e}^{-3\,{ c_0}\,{y}^{
2}}}+{\frac {51}{16}}\,{{ c_3}}^{2} \Bigg\{ -{\frac {8}{51}}\,{{
c_3}}^{6} \Bigg[  \left( {{ c_3}}^{2}{y}^{4}-{\frac {45}{2}}\,{{
c_2}}^{2} \right) {{ c_0}}^{2}+{{ c_3}}^{2}{y}^{2}{ c_0}\nonumber\\
&&+{{ c_3
}}^{2} \Bigg] {{ e}^{3\,{ c_0}\,{y}^{2}}}+{\frac {4}{51}}\,{{
 c_3}}^{8}{{ e}^{4\,{ c_0}\,{y}^{2}}}+{{ c_2}}^{2} \Bigg( -
 \Bigg\{  \left( -{\frac {280}{17}}\,{{ c_2}}^{4}-{\frac {336}{17}}\,
{{ c_3}}^{2}{{ c_2}}^{2}{y}^{4}+{y}^{8}{{ c_3}}^{4} \right) {{
 c_0}}^{4}-{\frac {16}{17}}\,{{ c_3}}^{2}{y}^{2} \left( -35\,{{
 c_2}}^{2}+{{ c_3}}^{2}{y}^{4} \right) {{ c_0}}^{3}\nonumber\\
&&+ \left( {
\frac {560}{17}}\,{{ c_3}}^{2}{{ c_2}}^{2}-{\frac {76}{17}}\,{{
 c_3}}^{4}{y}^{4} \right) {{ c_0}}^{2}-{\frac {120}{17}}\,{{ c_3
}}^{4}{ c_0}\,{y}^{2}-{\frac {60}{17}}\,{{ c_3}}^{4} \Bigg\} {{
 c_3}}^{2}{{ e}^{{ c_0}\,{y}^{2}}}+{{ c_2}}^{2} \Bigg[
 \left( {y}^{8}{{ c_3}}^{4}+{\frac {60}{17}}\,{{ c_2}}^{4}+{\frac
{672}{17}}\,{{ c_3}}^{2}{{ c_2}}^{2}{y}^{4} \right) {{ c_0}}^{4}\nonumber\\
&&
-{\frac {336}{17}}\,{{ c_3}}^{2}{y}^{2} \left( {{ c_3}}^{2}{y}^{4}
+5/3\,{{ c_2}}^{2} \right) {{ c_0}}^{3}-{\frac {56}{17}}\,{{ c_3
}}^{2} \left( {{ c_3}}^{2}{y}^{4}+10\,{{ c_2}}^{2} \right) {{
c_0}}^{2}+{\frac {560}{17}}\,{{ c_3}}^{4}{ c_0}\,{y}^{2}+{\frac {
280}{17}}\,{{ c_3}}^{4} \Bigg] {{ c_0}}^{2} \Bigg) {{ c_0}}^{2
} \Bigg\}  \Bigg\} {y}^{3} \Bigg\{ 5/4\,{ c_2}\, \Bigg\{ -4/5\,{ c_0
}\,{y}^{2}\nonumber\\
&&+{y}^{4}{{ c_0}}^{2}-4/5 \Bigg\} { c_3}{ c_0}
{ e}^{-\frac{{ c_0}{y}^{2}}{2}}+{ c_3}{{ e}^{1/2{ c_0}
{y}^{2}}}{ c_2}{ c_0}+{{ c_2}}^{2} \left( { c_0}{y}^{2}-1
 \right)  \left( { c_0}{y}^{2}+1/2 \right) {{ c_0}}^{2}{{ e}^
{-{ c_0}{y}^{2}}}+1/2{{ c_3}}^{2}{{ e}^{{ c_0}{y}^{2}}}
+ \left( 1/2{{ c_2}}^{2}-1/2{{ c_3}}^{2}{y}^{4} \right) {{
c_0}}^{2}\nonumber\\
&&-1/2\,{{ c_3}}^{2}{y}^{2}{ c_0}-1/2\,{{ c_3}}^{2}
 \Bigg\} ^{-3} \left( 2\,{ c_3}\,{{ e}^{1/2\,{ c_0}\,{y}^{2}}}{
 c_2}\,{ c_0}+{{ c_3}}^{2}{{ e}^{{ c_0}\,{y}^{2}}}+{{ c_2
}}^{2}{{ c_0}}^{2} \right) ^{-2}\nonumber\\
&&
-{\frac {9f''(y){y}^{2} }{8}}\;\Bigg\{ {\frac {669}{8}}\, \Bigg( {y}^{2}
 \left({\frac {394}{669}}\,{{ c_3
}}^{2}{{ c_2}}^{2}{y}^{4} -{\frac {2}{223}}\,{{ c_2}}^{4}+{{ c_3}}^{4}{y}^{8} \right) {{ c_0}}^{
5}+ \left({\frac {160}{
2007}}\,{{ c_2}}^{4}-{\frac {462}{223}}\,{{ c_3}}^{4}{y}^{8}+{\frac {240}{223}}\,{{ c_3}}^{2}{{ c_2}}^{2
}{y}^{4} \right) {{ c_0}}^{4}- \Bigg( {\frac {1168}{669}}\,{{ c_3
}}^{2}{{ c_2}}^{2}{y}^{2}\nonumber\\
&&+{\frac {1414}{669}}\,{{ c_3}}^{4}{y}^{6}
 \Bigg) {{ c_0}}^{3}+ \left( -{\frac {1280}{669}}\,{{ c_3}}^{2}{{
 c_2}}^{2}+{\frac {616}{669}}\,{{ c_3}}^{4}{y}^{4} \right) {{
c_0}}^{2}+{\frac {868}{223}}\,{{ c_3}}^{4}{ c_0}\,{y}^{2}+{\frac {
448}{223}}\,{{ c_3}}^{4} \Bigg) {{ c_0}}^{5}{ c_3}\,{{ c_2}}^
{5}{{ e}^{-1/2\,{ c_0}\,{y}^{2}}}+ \Bigg\{\Bigg({\frac {385}{4}}\,{y}^{6}{{
 c_3}}^{5}{{ c_2}}^{5}\nonumber\\
&& -{\frac {107}{
16}}\,{{ c_3}}^{7}{y}^{10}{{ c_2}}^{3}-7\,{y}^{2}{{ c_3}}^{3}{{ c_2}}^{7}
 \Bigg) {{ c_0}}^{8}+ \left( -{\frac {151}{4}}\,{{ c_3}}^{7}{y}^{
8}{{ c_2}}^{3}+80\,{{ c_3}}^{3}{{ c_2}}^{7}+91\,{y}^{4}{{ c_3}
}^{5}{{ c_2}}^{5} \right) {{ c_0}}^{7}+ \left( -315\,{y}^{2}{{
c_3}}^{5}{{ c_2}}^{5}-{\frac {43}{4}}\,{{ c_3}}^{7}{y}^{6}{{ c_2}
}^{3} \right) {{ c_0}}^{6}\nonumber\\
\end{eqnarray*}
\begin{eqnarray*}
&&+ \left(94{{ c_3}}^{7}{y}^{4}{{ c_2}}^{3} -336\,{{ c_3}}^{5}{{ c_2}}^{5
} \right) {{ c_0}}^{5}+157
{{ c_0}}^{4}{{ c_3}}^{7}{y}^{2}{{ c_2}}^{3}+80{{ c_0}}^{3}{{
 c_3}}^{7}{{ c_2}}^{3} \Bigg\} { e}^{\frac{{ c_0}{y}^{2}}{2}}-
1/2{ c_0}{{ c_3}}^{5} \Bigg\{ {y}^{2} \left( {{ c_3}}^{4}{y}^
{8}-{\frac {111}{2}}{{ c_3}}^{2}{{ c_2}}^{2}{y}^{4}+21{{ c_2
}}^{4} \right) {{ c_0}}^{5}\nonumber\\
&&+ \left( 28\,{{ c_3}}^{2}{{ c_2}}^{2}
{y}^{4}-336\,{{ c_2}}^{4}-11/2\,{{ c_3}}^{4}{y}^{8} \right) {{
c_0}}^{4}+ \left( 308\,{{ c_3}}^{2}{{ c_2}}^{2}{y}^{2}-{\frac {47}{
3}}\,{{ c_3}}^{4}{y}^{6} \right) {{ c_0}}^{3}+ \left( -27\,{{
c_3}}^{4}{y}^{4}+320\,{{ c_3}}^{2}{{ c_2}}^{2} \right) {{ c_0}}^{
2}-{\frac {53}{2}}\,{{ c_3}}^{4}{ c_0}\,{y}^{2}\nonumber\\
&&-{\frac {40}{3}}{
{ c_3}}^{4} \Bigg\} { c_2}{{ e}^{3/2\,{ c_0}{y}^{2}}}+{
\frac {1353}{16}}{{ c_0}}^{7}{ c_3} \Bigg\{ {y}^{6} \left( {{
 c_3}}^{2}{y}^{4}+{\frac {16}{369}}{{ c_2}}^{2} \right) {{ c_0
}}^{5}+ \left( {\frac {56}{451}}{{ c_2}}^{2}{y}^{4}-{\frac {52}{41
}}{{ c_3}}^{2}{y}^{8} \right) {{ c_0}}^{4}-\left( {\frac {568}
{4059}}{{ c_2}}^{2}{y}^{2}+{\frac {2116}{1353}}{y}^{6}{{ c_3}}
^{2} \right) {{ c_0}}^{3}\nonumber\\
&&+ \Bigg\{ -{\frac {640}{4059}}\,{{ c_2}}^{
2}-{\frac {160}{1353}}\,{{ c_3}}^{2}{y}^{4} \Bigg\} {{ c_0}}^{2}+{
\frac {816}{451}}\,{{ c_3}}^{2}{y}^{2}{ c_0}+{\frac {1280}{1353}}
\,{{ c_3}}^{2} \Bigg\} {{ c_2}}^{7}{{ e}^{-3/2\,{ c_0}\,{y}^{
2}}}+{\frac {28}{3}}\, \Bigg\{{\frac {151}{112}}\,{ c_0}\,{y}^{2}-{
\frac {59}{56}}\,{{ c_0}}^{4}{y}^{8}+5/7-{\frac {23}{56}}\,{{ c_0}
}^{2}{y}^{4}\nonumber\\
&&-{\frac {23}{16}}\,{{ c_0}}^{3}{y}^{6}+{{ c_0}}^{5}{y}
^{10} \Bigg\} {{ c_0}}^{9}{ c_3}\,{{ c_2}}^{9}{{ e}^{-5/2\,{
 c_0}\,{y}^{2}}}-{\frac {11}{12}}\,{ c_0}\,{{ c_3}}^{7}{ c_2}
\, \Bigg\{  \left( {\frac {36}{11}}\,{{ c_2}}^{2}{y}^{2}+{y}^{6}{{
 c_3}}^{2} \right) {{ c_0}}^{3}+ \left( -{\frac {960}{11}}\,{{
c_2}}^{2}+{\frac {82}{11}}\,{{ c_3}}^{2}{y}^{4} \right) {{ c_0}}^{2
}+{\frac {158}{11}}\,{{ c_3}}^{2}{y}^{2}{ c_0}\nonumber\\
&&+{\frac {160}{11}}\,
{{ c_3}}^{2} \Bigg\} {{ e}^{5/2\,{ c_0}\,{y}^{2}}}-1/12\,{
c_0}\,{{ c_3}}^{9}{ c_2}\, \left( -80+{ c_0}\,{y}^{2} \right) {
{ e}^{7/2\,{ c_0}\,{y}^{2}}}+112\,{{ c_0}}^{6} \Bigg\{ {y}^{2}
 \left( -{\frac {1}{1344}}\,{{ c_2}}^{4}+{\frac {71}{448}}\,{{ c_3
}}^{2}{{ c_2}}^{2}{y}^{4}+{{ c_3}}^{4}{y}^{8} \right) {{ c_0}}^{
5}\nonumber\\
&&+ \left( {\frac {1}{168}}
{{ c_2}}^{4}-{\frac {97}{64}}{{ c_3}}^{4}{y}^{8}+{\frac {41}{112}}{{ c_3}}^{2}{{ c_2}}^{2}{y}^{4}
 \right) {{ c_0}}^{4}+ \left( -{\frac {27}{56}}{{ c_3}}^{2}{{
 c_2}}^{2}{y}^{2}-{\frac {111}{64}}{{ c_3}}^{4}{y}^{6} \right) {
{ c_0}}^{3}+ \left( -{\frac {15}{28}}{{ c_3}}^{2}{{ c_2}}^{2}+
3/16{{ c_3}}^{4}{y}^{4} \right) {{ c_0}}^{2}+{\frac {77}{32}}{
{ c_3}}^{4}{ c_0}{y}^{2}\nonumber\\
&&+5/4\,{{ c_3}}^{4} \Bigg\} {{ c_2}}^
{6}{{ e}^{-{ c_0}\,{y}^{2}}}+1/3\, \Bigg\{{y}^{2} \left( {{ c_3
}}^{4}{y}^{8}+{\frac {51}{4}}\,{{ c_3}}^{2}{{ c_2}}^{2}{y}^{4}-21
\,{{ c_2}}^{4} \right) {{ c_0}}^{5}+ \left( -{\frac {105}{2}}\,{{
 c_3}}^{2}{{ c_2}}^{2}{y}^{4}+420\,{{ c_2}}^{4}+2\,{{ c_3}}^{4
}{y}^{8} \right) {{ c_0}}^{4}\nonumber\\
&&+ \left(4{{ c_3}}^{2}{y}^{4} -{\frac {351}{2}}{{ c_2}}^{2} \right){y}^{2}{ c_3}^
{2} { c_0}^{
3}+ \left( 5{{ c_3}}^{2}{y}^{4}-180{{ c_2}}^{2}
 \right) {{ c_3}}^{2}{{ c_0}}^{2}+4{{ c_3}}^{4}{ c_0}{y}^{2}+2{{ c_3
}}^{4} \Bigg\} {{ c_3}}^{6}{{ e}^{2{ c_0}{y}^{2}}}+{\frac {
603{c_0}^{8}}{16}} \Bigg\{  \left( {\frac {16}{1809}}{{ c_2}
}^{2}+{{ c_3}}^{2}{y}^{4} \right) {y}^{6}{{ c_0}}^{5}\nonumber\\
&&+ \left( {
\frac {56}{1809}}\,{{ c_2}}^{2}-{\frac {76}{67}}\,{{ c_3}}^
{2}{y}^{4} \right){y}^{4} {{ c_0}}^{4}+ \left( -{\frac {56}{1809}}\,{{
c_2}}^{2}-{\frac {892}{603}}\,{y}^{4}{{ c_3}}^{2} \right){y}^{2} {{
 c_0}}^{3}- \left({\frac {64}{1809}}\,{{ c_2}}^{2}+{\frac {176}{
603}}\,{{ c_3}}^{2}{y}^{4} \right) {{ c_0}}^{2}+{\frac {304}{201}}
\,{{ c_3}}^{2}{y}^{2}{ c_0}+{\frac {160}{201}}\,{{ c_3}}^{2}
 \Bigg\}\nonumber\\
&&\times {{ c_2}}^{8}{{ e}^{-2\,{ c_0}\,{y}^{2}}}+ \left( {{
 c_0}}^{3}{y}^{6}-1/2\,{{ c_0}}^{2}{y}^{4}-7/6\,{ c_0}\,{y}^{2}-
4/3 \right) {{ c_0}}^{10} \left( { c_0}\,{y}^{2}-1 \right)
 \left( { c_0}\,{y}^{2}+1/2 \right) {{ c_2}}^{10}{{ e}^{-3\,{
 c_0}\,{y}^{2}}}+{\frac {411}{16}}\,{{ c_3}}^{2} \Bigg\{ -{\frac {
16}{1233}}\,{{ c_3}}^{6}\nonumber\\
&& \left(  \left( 9/4\,{{ c_2}}^{2}{y}^{2}+{
y}^{6}{{ c_3}}^{2} \right) {{ c_0}}^{3}+ \left( -90\,{{ c_2}}^{2
}+3\,{{ c_3}}^{2}{y}^{4} \right) {{ c_0}}^{2}+4\,{{ c_3}}^{2}{y}
^{2}{ c_0}+4\,{{ c_3}}^{2} \right) {{ e}^{3\,{ c_0}\,{y}^{2}}
}+{\frac {32}{1233}}\,{{ c_3}}^{8}{{ e}^{4\,{ c_0}\,{y}^{2}}}+{
{ c_0}}^{2} \Bigg[ -{\frac {34}{137}}\, \Bigg\{ {y}^{2} \Bigg[ {{
c_3}}^{4}{y}^{8}\nonumber\\
&&-{\frac {182}{17}}\,{{ c_3}}^{2}{{ c_2}}^{2}{y}^{4}
+{\frac {28}{17}}\,{{ c_2}}^{4} \Bigg] {{ c_0}}^{5}- \left( {
\frac {224}{51}}\,{{ c_3}}^{2}{{ c_2}}^{2}{y}^{4}+{\frac {1120}{51
}}\,{{ c_2}}^{4}-{\frac {6}{17}}\,{{ c_3}}^{4}{y}^{8} \right) {{
 c_0}}^{4}+{\frac {112}{51}}\,{y}^{2}{{ c_3}}^{2} \left( 19\,{{
 c_2}}^{2}-{{ c_3}}^{2}{y}^{4} \right) {{ c_0}}^{3}- \Bigg\{ {
\frac {380}{51}}\,{{ c_3}}^{4}{y}^{4}\nonumber\\
&&-{\frac {2240}{51}}\,{{ c_3}}
^{2}{{ c_2}}^{2} \Bigg\} {{ c_0}}^{2}-{\frac {158}{17}}\,{{ c_3}
}^{4}{ c_0}\,{y}^{2}-{\frac {80}{17}}\,{{ c_3}}^{4} \Bigg\} {{
c_3}}^{2}{{ e}^{{ c_0}\,{y}^{2}}}+{{ c_0}}^{2} \Bigg[ {y}^{2}
 \left( {{ c_3}}^{4}{y}^{8}+{\frac {1372}{411}}\,{{ c_3}}^{2}{{
 c_2}}^{2}{y}^{4}-{\frac {16}{137}}\,{{ c_2}}^{4} \right) {{ c_0
}}^{5}+ \Bigg\{ {\frac {1904}{411}}\,{{ c_3}}^{2}{{ c_2}}^{2}{y}^{4
}\nonumber\\
&&+{\frac {160}{137}}{{ c_2}}^{4}-{\frac {588}{137}}{{ c_3}}^{4
}{y}^{8} \Bigg\}{{ c_0}}^{4}+ \left( -{\frac {4144}{411}}\,{{ c_3
}}^{2}{{ c_2}}^{2}{y}^{2}-{\frac {476}{137}}\,{{ c_3}}^{4}{y}^{6}
 \right) {{ c_0}}^{3}+{\frac {1792}{411}}\,{{ c_3}}^{2} \left( {{
 c_3}}^{2}{y}^{4}-5/2\,{{ c_2}}^{2} \right) {{ c_0}}^{2}+{\frac
{1456}{137}}{{ c_3}}^{4}{ c_0}{y}^{2}+{\frac {2240}{411}}{{
 c_3}}^{4} \Bigg]\nonumber\\
&& {{ c_2}}^{2} \Bigg] {{ c_2}}^{2} \Bigg\}
 \Bigg\}\Bigg[ { c_3}
\,{{ e}^{1/2\,{ c_0}\,{y}^{2}}}{ c_2}\,{ c_0}+5/4\,{ c_0}\,
 \left( -4/5\,{ c_0}\,{y}^{2}+{{ c_0}}^{2}{y}^{4}-4/5 \right) {
 c_3}\,{ c_2}\,{{ e}^{-1/2\,{ c_0}\,{y}^{2}}}+ \left( { c_0
}\,{y}^{2}-1 \right) {{ c_0}}^{2} \left( { c_0}\,{y}^{2}+1/2
 \right) {{ c_2}}^{2}{{ e}^{-{ c_0}\,{y}^{2}}}\nonumber\\
&&+1/2\,{{ c_3}}^
{2}{{ e}^{{ c_0}\,{y}^{2}}}+ \left( 1/2\,{{ c_2}}^{2}-1/2\,{{
 c_3}}^{2}{y}^{4} \right) {{ c_0}}^{2}-1/2\,{{ c_3}}^{2}{y}^{2}{
 c_0}-1/2\,{{ c_3}}^{2} \Bigg]^{-3} \left( 2\,{ c_3}\,{{ e}
^{1/2\,{ c_0}\,{y}^{2}}}{ c_2}\,{ c_0}+{{ c_3}}^{2}{{ e}^{{
 c_0}\,{y}^{2}}}+{{ c_2}}^{2}{{ c_0}}^{2} \right)^{-2}\nonumber\\
&& 
f'(y)y\; \Bigg\{ {\frac {1683}{8}}\,{{ c_0}}^{5}{ c_3}\, \Bigg\{  \left( -{\frac {677}{1122}}\,{{ c_3}}^{2}{{ c_2}}^{2}{y}^{4}+{\frac {74}{
5049}}\,{{ c_2}}^{4}+{{ c_3}}^{4}{y}^{8} \right) {y}^{4}{{ c_0}}
^{6}-{\frac {3521}{3366}}\,{y}^{2} \left( {\frac {58}{3521}}\,{{ c_2
}}^{4}+{{ c_3}}^{4}{y}^{8}-{\frac {14804}{3521}}\,{{ c_3}}^{2}{{
 c_2}}^{2}{y}^{4} \right) {{ c_0}}^{5}\nonumber\\
&& + \left( -{\frac {12697}{
3366}}\,{{ c_3}}^{4}{y}^{8}+{\frac {100}{1683}}\,{{ c_2}}^{4}-{
\frac {6176}{1683}}\,{{ c_3}}^{2}{{ c_2}}^{2}{y}^{4} \right) {{
 c_0}}^{4}+ \left( -{\frac {217}{153}}\,{{ c_3}}^{4}{y}^{6}-{
\frac {248}{187}}\,{{ c_3}}^{2}{{ c_2}}^{2}{y}^{2} \right) {{
c_0}}^{3}+ \left( {\frac {12740}{1683}}\,{{ c_3}}^{4}{y}^{4}-{\frac {
80}{51}}\,{{ c_3}}^{2}{{ c_2}}^{2} \right) {{ c_0}}^{2}\nonumber\\
&& +{\frac {
602}{187}}\,{{ c_3}}^{4}{ c_0}\,{y}^{2}+{\frac {56}{33}}\,{{ c_3
}}^{4} \Bigg\} {{ c_2}}^{5}{{ e}^{-1/2\,{ c_0}\,{y}^{2}}}+{
\frac {501}{32}}\,{{ c_0}}^{3}{ c_3}\, \Bigg\{  \left( {{ c_3}}^{
6}{y}^{12}+{\frac {928}{501}}\,{{ c_3}}^{2}{{ c_2}}^{4}{y}^{4}-{
\frac {7562}{501}}\,{{ c_3}}^{4}{{ c_2}}^{2}{y}^{8}+{\frac {80}{
1503}}\,{{ c_2}}^{6} \right) {{ c_0}}^{6}\nonumber\\
&& -{\frac {895}{167}}\,
 \left( -{\frac {49028}{2685}}\,{{ c_3}}^{2}{{ c_2}}^{2}{y}^{4}+{{
 c_3}}^{4}{y}^{8}+{\frac {432}{895}}\,{{ c_2}}^{4} \right) {y}^{2}
{{ c_3}}^{2}{{ c_0}}^{5}-{\frac {5750}{501}}\, \left( {{ c_3}}^{
4}{y}^{8}-{\frac {96}{115}}\,{{ c_2}}^{4}+{\frac {20048}{2875}}\,{{
 c_3}}^{2}{{ c_2}}^{2}{y}^{4} \right) {{ c_3}}^{2}{{ c_0}}^{4}\nonumber\\
 &&
+ \left( -{\frac {6384}{167}}\,{{ c_3}}^{4}{{ c_2}}^{2}{y}^{2}-{
\frac {200}{167}}\,{{ c_3}}^{6}{y}^{6} \right) {{ c_0}}^{3}+
 \left( {\frac {17632}{501}}\,{{ c_3}}^{6}{y}^{4}-{\frac {7392}{167}
}\,{{ c_3}}^{4}{{ c_2}}^{2} \right) {{ c_0}}^{2}+{\frac {3472}{
167}}\,{{ c_3}}^{6}{ c_0}\,{y}^{2}+{\frac {5440}{501}}\,{{ c_3}}
^{6} \Bigg\} {{ c_2}}^{3}{{ e}^{1/2\,{ c_0}\,{y}^{2}}}\nonumber\\
\end{eqnarray*}
\begin{eqnarray*}
&&-{\frac {
23}{16}}\,{ c_0}\, \Bigg\{  \left( {{ c_3}}^{6}{y}^{12}-{\frac {160
}{23}}\,{{ c_2}}^{6}-{\frac {616}{23}}\,{{ c_3}}^{2}{{ c_2}}^{4}
{y}^{4}+{\frac {1233}{23}}\,{{ c_3}}^{4}{{ c_2}}^{2}{y}^{8}
 \right) {{ c_0}}^{6}-{\frac {158}{69}}\,{y}^{2}{{ c_3}}^{2}
 \left( {\frac {9822}{79}}\,{{ c_3}}^{2}{{ c_2}}^{2}{y}^{4}-{
\frac {1890}{79}}\,{{ c_2}}^{4}+{{ c_3}}^{4}{y}^{8} \right) {{
c_0}}^{5}\nonumber\\
 &&+ \left( {\frac {6016}{23}}\,{{ c_3}}^{4}{{ c_2}}^{2}{y}^{
4}-{\frac {172}{23}}\,{{ c_3}}^{6}{y}^{8}-{\frac {5040}{23}}\,{{
c_3}}^{2}{{ c_2}}^{4} \right) {{ c_0}}^{4}+ \left( {\frac {4656}{23
}}\,{{ c_3}}^{4}{{ c_2}}^{2}{y}^{2}-{\frac {949}{69}}\,{{ c_3}}^
{6}{y}^{6} \right) {{ c_0}}^{3}+ \Bigg\{ {\frac {5280}{23}}\,{{ c_3
}}^{4}{{ c_2}}^{2}\nonumber\\
 &&-{\frac {1204}{69}}\,{{ c_3}}^{6}{y}^{4}
 \Bigg\} {{ c_0}}^{2}-{\frac {438}{23}}\,{{ c_3}}^{6}{ c_0}\,{y}
^{2}-{\frac {680}{69}}\,{{ c_3}}^{6} \Bigg\} {{ c_3}}^{3}{ c_2}
\,{{ e}^{3/2\,{ c_0}\,{y}^{2}}}+{\frac {4509}{32}}\,{{ c_0}}^{7
}{ c_3}\, \Bigg[ {y}^{8} \left( -{\frac {1084}{13527}}\,{{ c_2}}^{
2}+{{ c_3}}^{2}{y}^{4} \right) {{ c_0}}^{6}+ \Bigg\{ {\frac {8354}{
13527}}\,{{ c_2}}^{2}{y}^{6}\nonumber\\
 &&-{\frac {145}{501}}{{ c_3}}^{2}{y}^{
10} \Bigg\} {{ c_0}}^{5}+ \left( -{\frac {15674}{4509}}{{ c_3}}^
{2}{y}^{8}-{\frac {7216}{13527}}{{ c_2}}^{2}{y}^{4} \right) {{
c_0}}^{4}+ \left( -{\frac {728}{4509}}{{ c_2}}^{2}{y}^{2}-{\frac {
1928}{1503}}{y}^{6}{{ c_3}}^{2} \right) {{ c_0}}^{3}+ \left( -{
\frac {880}{4509}}{{ c_2}}^{2}+{\frac {29536}{4509}}{{ c_3}}^{
2}{y}^{4} \right) {{ c_0}}^{2}\nonumber\\
 &&+{\frac {1136}{501}}\,{{ c_3}}^{2}{y
}^{2}{ c_0}+{\frac {5440}{4509}}\,{{ c_3}}^{2} \Bigg] {{ c_2}}^
{7}{ e}^{-\frac{3{ c_0}{y}^{2}}{2}}+{\frac {13{ c_0}{{
 c_3}}^{5} }{48}}\Bigg\{  \left( 12\,{{ c_2}}^{2}+{{ c_3}}^{2}{y}^{4}
 \right)  \left( {\frac {84}{13}}\,{{ c_2}}^{2}+{{ c_3}}^{2}{y}^{4
} \right) {{ c_0}}^{4}- \left( {\frac {1656}{13}}{{ c_3}}^{2}{{
 c_2}}^{2}{y}^{2}+{\frac {335}{13}}{{ c_3}}^{4}{y}^{6} \right) {
{ c_0}}^{3}\nonumber\\
 &&+ \left( -{\frac {440}{13}}\,{{ c_3}}^{4}{y}^{4}+{
\frac {7200}{13}}{{ c_3}}^{2}{{ c_2}}^{2} \right) {{ c_0}}^{2}
-{\frac {1188}{13}}\,{{ c_3}}^{4}{ c_0}{y}^{2}-{\frac {1320}{13}
}\,{{ c_3}}^{4} \Bigg\} { c_2}{{ e}^{5/2{ c_0}\,{y}^{2}}}
-11\,{{ c_0}}^{9}{ c_3}\Bigg\{{\frac {731}{176}}{{ c_0}}^{
4}{y}^{8}+{\frac {85}{66}}+{\frac {731}{528}}{{ c_0}}^{3}{y}^{6}-{
\frac {211}{88}}{ c_0}{y}^{2}\nonumber\\
 &&-{\frac {151}{264}}\,{{ c_0}}^{5}
{y}^{10}-{{ c_0}}^{6}{y}^{12}-{\frac {95}{12}}\,{{ c_0}}^{2}{y}^{4
} \Bigg\} {{ c_2}}^{9}{{ e}^{-5/2\,{ c_0}\,{y}^{2}}}-{\frac {11
}{12}}\,{ c_0}\,{{ c_3}}^{7} \left(  \left( {{ c_3}}^{2}{y}^{4}-
{\frac {120}{11}}\,{{ c_2}}^{2} \right) {{ c_0}}^{2}+{\frac {63}{
22}}\,{{ c_3}}^{2}{y}^{2}{ c_0}-{\frac {150}{11}}\,{{ c_3}}^{2}
 \right) { c_2}\,{{ e}^{7/2\,{ c_0}\,{y}^{2}}}\nonumber\\
 &&+\frac{5{ c_0}{
{ c_3}}^{9}{ c_2}{ e}^{\frac{9{ c_0}{y}^{2}}{2}}}{6}+{\frac {7155
}{32}}{{ c_0}}^{6} \Bigg\{  \left({\frac {32}{21465}}{{ c_2}}^{4}+{{
c_3}}^{4}{y}^{8}-{\frac {1576}{7155}}{{ c_3}
}^{2}{{ c_2}}^{2}{y}^{4} \right) {y}^{4}{{ c_0}}^{6}-{\frac {4573}{7155}}{y
}^{2} \left( {\frac {12}{4573}}{{ c_2}}^{4}+{{ c_3}}^{4}{y}^{8}-
{\frac {11896}{4573}}{{ c_3}}^{2}{{ c_2}}^{2}{y}^{4} \right) {{
 c_0}}^{5}\nonumber\\
 &&+ \left( -{\frac {24796}{7155}}\,{{ c_3}}^{4}{y}^{8}+{
\frac {8}{1431}}\,{{ c_2}}^{4}-{\frac {10112}{7155}}\,{{ c_3}}^{2}
{{ c_2}}^{2}{y}^{4} \right) {{ c_0}}^{4}+ \left( -{\frac {9464}{
7155}}\,{{ c_3}}^{4}{y}^{6}-{\frac {368}{795}}\,{{ c_3}}^{2}{{
c_2}}^{2}{y}^{2} \right) {{ c_0}}^{3}+ \left( {\frac {5264}{795}}\,{{
 c_3}}^{4}{y}^{4}-{\frac {88}{159}}\,{{ c_3}}^{2}{{ c_2}}^{2}
 \right) {{ c_0}}^{2}\nonumber\\
 &&+{\frac {5992}{2385}}\,{{ c_3}}^{4}{ c_0}\,
{y}^{2}+{\frac {1904}{1431}}\,{{ c_3}}^{4} \Bigg\} {{ c_2}}^{6}{
{ e}^{-{ c_0}\,{y}^{2}}}+1/2\, \Bigg[  \left( {{ c_3}}^{6}{y}^{
12}+42\,{{ c_3}}^{2}{{ c_2}}^{4}{y}^{4}-{\frac {133}{4}}\,{{ c_3
}}^{4}{{ c_2}}^{2}{y}^{8}+35\,{{ c_2}}^{6} \right) {{ c_0}}^{6}+
7/3\,{y}^{2} \Bigg\{ {\frac {843}{14}}\,{{ c_3}}^{2}{{ c_2}}^{2}{y}
^{4}\nonumber\\
 &&+{{ c_3}}^{4}{y}^{8}-54\,{{ c_2}}^{4} \Bigg\} {{ c_3}}^{2}{{
 c_0}}^{5}+ \left( -{\frac {373}{2}}\,{{ c_3}}^{4}{{ c_2}}^{2}{y
}^{4}+525\,{{ c_3}}^{2}{{ c_2}}^{4}+11/2\,{{ c_3}}^{6}{y}^{8}
 \right) {{ c_0}}^{4}+ \left( -{\frac {441}{2}}\,{{ c_3}}^{4}{{
 c_2}}^{2}{y}^{2}+{\frac {43}{6}}\,{{ c_3}}^{6}{y}^{6} \right) {{
 c_0}}^{3}+ \Bigg\{ 8/3\,{{ c_3}}^{6}{y}^{4}\nonumber\\
 &&-{\frac {495}{2}}\,{{
 c_3}}^{4}{{ c_2}}^{2} \Bigg\} {{ c_0}}^{2}+11/2\,{{ c_3}}^{6}
{ c_0}\,{y}^{2}+{\frac {17}{6}}\,{{ c_3}}^{6} \Bigg] {{ c_3}}^{
4}{{ e}^{2\,{ c_0}\,{y}^{2}}}+{\frac {423}{8}}\, \Bigg\{ {y}^{8}
 \left( -{\frac {28}{1269}}\,{{ c_2}}^{2}+{{ c_3}}^{2}{y}^{4}
 \right) {{ c_0}}^{6}+ \left( {\frac {218}{1269}}\,{{ c_2}}^{2}{y}
^{6}+{\frac {40}{423}}\,{{ c_3}}^{2}{y}^{10} \right) {{ c_0}}^{5}\nonumber\\
 &&-
 \left( {\frac {190}{1269}}\,{{ c_2}}^{2}{y}^{4}+{\frac {1570}{423}
}{{ c_3}}^{2}{y}^{8} \right) {{ c_0}}^{4}+ \left( -{\frac {2}{47
}}{{ c_2}}^{2}{y}^{2}-{\frac {550}{423}}{y}^{6}{{ c_3}}^{2}
 \right) {{ c_0}}^{3}+ \left( -{\frac {22}{423}}{{ c_2}}^{2}+{
\frac {2968}{423}}\,{{ c_3}}^{2}{y}^{4} \right) {{ c_0}}^{2}+{
\frac {106}{47}}\,{{ c_3}}^{2}{y}^{2}{ c_0}+{\frac {170}{141}}\,{{
 c_3}}^{2} \Bigg\}\nonumber\\
 &&\times {{ c_0}}^{8}{{ c_2}}^{8}{{ e}^{-2\,{ c_0
}\,{y}^{2}}}+{\frac {7}{12}}\, \Bigg\{  \left({{ c_3}}^{4}{y}^{8} -{\frac {12}{7}}\,{{
c_3}}^{2}{{ c_2}}^{2}{y}^{4}+30\,{{ c_2}}^{4}
 \right) {{ c_0}}^{4}- \left( {\frac {297}{14}}\,{{ c_3}}^{2}{{
 c_2}}^{2}{y}^{2}+{\frac {41}{7}}\,{{ c_3}}^{4}{y}^{6} \right) {{
 c_0}}^{3}+ \left(\frac{4}{7}\,{{ c_3}}^{4}{y}^{4}+{\frac {675}{7}}\,{{
 c_3}}^{2}{{ c_2}}^{2} \right) {{ c_0}}^{2}\nonumber\\
 &&-{\frac {30}{7}}\,{{
 c_3}}^{4}{ c_0}\,{y}^{2}-{\frac {33}{7}}\,{{ c_3}}^{4} \Bigg\}
{{ c_3}}^{6}{{ e}^{3\,{ c_0}\,{y}^{2}}}+{{ c_0}}^{10} \left(
{\frac {17}{12}}-{\frac {19}{4}}\,{{ c_0}}^{4}{y}^{8}-{\frac {37}{24
}}\,{{ c_0}}^{3}{y}^{6}+{\frac {21}{8}}\,{ c_0}\,{y}^{2}+\frac{7{{
 c_0}}^{5}{y}^{10}}{6}+{{ c_0}}^{6}{y}^{12}+{\frac {109}{12}}\,{{
c_0}}^{2}{y}^{4} \right) {{ c_2}}^{10}{{ e}^{-3\,{ c_0}\,{y}^{2}
}}\nonumber\\
 &&-\frac{\,{{ c_3}}^{8}}{6} \left(  \left( -{\frac {45}{2}}\,{{ c_2}}^{2}
+{{ c_3}}^{2}{y}^{4} \right) {{ c_0}}^{2}+3/2{{ c_3}}^{2}{y}^{
2}{ c_0}-\frac{15{{ c_3}}^{2}}{2} \right) {{ e}^{4\,{ c_0}{y}^{2}
}}+\frac{{{ c_3}}^{10}{{ e}^{5{ c_0}{y}^{2}}}}{12}+{\frac {1689}
{16}}\,{{ c_0}}^{2} \Bigg[ -{\frac {75}{1126}}{{ c_3}}^{2}
 \Bigg\{  \Bigg[ -{\frac {448}{75}}{{ c_3}}^{2}{{ c_2}}^{4}{y}^{4
}{{ c_0}}^{5}{y}^{10}\nonumber\\
 &&+{\frac {5528}{225}}\,{{ c_3}}^{4}{{ c_2}}^{2}{y}^{8}-{\frac {8}{
15}}\,{{ c_2}}^{6}+{{ c_3}}^{6}{y}^{12} \Bigg] {{ c_0}}^{6}+
 \left( {\frac {17}{9}}\,{{ c_3}}^{6}{y}^{10}+{\frac {728}{75}}\,{{
 c_3}}^{2}{{ c_2}}^{4}{y}^{2}-{\frac {10976}{75}}\,{{ c_3}}^{4}{
{ c_2}}^{2}{y}^{6} \right) {{ c_0}}^{5}+ \Bigg\{ {\frac {9184}{75}}
\,{{ c_3}}^{4}{{ c_2}}^{2}{y}^{4}+{\frac {484}{225}}\,{{ c_3}}^{
6}{y}^{8}\nonumber\\
 &&-{\frac {112}{3}}\,{{ c_3}}^{2}{{ c_2}}^{4} \Bigg\} {{
 c_0}}^{4}+ \left( {\frac {1792}{25}}\,{{ c_3}}^{4}{{ c_2}}^{2}{
y}^{2}-{\frac {1076}{225}}\,{{ c_3}}^{6}{y}^{6} \right) {{ c_0}}^{
3}+ \left( -{\frac {5176}{225}}\,{{ c_3}}^{6}{y}^{4}+{\frac {1232}{
15}}\,{{ c_3}}^{4}{{ c_2}}^{2} \right) {{ c_0}}^{2}-{\frac {436}
{25}}\,{{ c_3}}^{6}{ c_0}\,{y}^{2}-{\frac {136}{15}}\,{{ c_3}}^{
6} \Bigg\} {{ e}^{{ c_0}\,{y}^{2}}}\nonumber\\
 &&+{{ c_0}}^{2}{{ c_2}}^{2}
 \Bigg\{  \left( {{ c_3}}^{6}{y}^{12}+{\frac {4}{5067}}\,{{ c_2}}^{
6}+{\frac {200}{1689}}\,{{ c_3}}^{2}{{ c_2}}^{4}{y}^{4}-{\frac {
3394}{1689}}\,{{ c_3}}^{4}{{ c_2}}^{2}{y}^{8} \right) {{ c_0}}^{
6}+ \left( -{\frac {84}{563}}\,{{ c_3}}^{2}{{ c_2}}^{4}{y}^{2}-{
\frac {2981}{1689}}\,{{ c_3}}^{6}{y}^{10}+{\frac {7840}{563}}\,{{
 c_3}}^{4}{{ c_2}}^{2}{y}^{6} \right) {{ c_0}}^{5}\nonumber\\
 &&+ \left( -{
\frac {6440}{563}}\,{{ c_3}}^{4}{{ c_2}}^{2}{y}^{4}-{\frac {8156}{
1689}}\,{{ c_3}}^{6}{y}^{8}+{\frac {300}{563}}\,{{ c_3}}^{2}{{
c_2}}^{4} \right) {{ c_0}}^{4}-{\frac {2632}{1689}}\,{{ c_3}}^{4}{y
}^{2} \left( 3\,{{ c_2}}^{2}+{{ c_3}}^{2}{y}^{4} \right) {{ c_0}
}^{3}+{\frac {6160}{563}}\,{{ c_3}}^{4} \left( {{ c_3}}^{2}{y}^{4}
-1/2\,{{ c_2}}^{2} \right) {{ c_0}}^{2}\nonumber\\
 &&+{\frac {3024}{563}}\,{{
 c_3}}^{6}{ c_0}\,{y}^{2}+{\frac {4760}{1689}}\,{{ c_3}}^{6}
 \Bigg\}  \Bigg] {{ c_2}}^{2} \Bigg\}  \Bigg[ { c_3}\,{{ e}^{1/2\,{ c_0}
\,{y}^{2}}}{ c_2}\,{ c_0}+5/4\,{ c_0}\, \left( -4/5\,{ c_0}\,{
y}^{2}+{{ c_0}}^{2}{y}^{4}-4/5 \right) { c_3}\,{ c_2}\,{{ e}^
{-1/2\,{ c_0}\,{y}^{2}}}+ \left( { c_0}\,{y}^{2}-1 \right) {{
c_0}}^{2}\times\nonumber\\
 && \left( { c_0}{y}^{2}+\frac{1}{2} \right) {{ c_2}}^{2}{{ e}^{
-{ c_0}{y}^{2}}}+\frac{{{ c_3}}^{2}{{ e}^{{ c_0}{y}^{2}}}}{2}+
 \frac{\left({{ c_2}}^{2}-{{ c_3}}^{2}{y}^{4} \right) {{
c_0}}^{2}}{2}-1/2{{ c_3}}^{2}{y}^{2}{ c_0}-1/2{{ c_3}}^{2}
 \Bigg] ^{-3} \left( 2{ c_3}{ e}^{\frac{{ c_0}{y}^{2}}{2}}{
 c_2}{ c_0}+{{ c_3}}^{2}{{ e}^{{ c_0}{y}^{2}}}+{{ c_2
}}^{2}{{ c_0}}^{2} \right) ^{-2}
\end{eqnarray*}
\begin{eqnarray*}
&&-2 f(y)\; \Bigg\{ {\frac {9}{32}}\,{{ c_0}}^{5} \Bigg[
{y}^{4} \left( {{ c_3}}^{4}{y}^{8}+{\frac {74}{3}}\,{{ c_2}}^{4}+
318\,{{ c_3}}^{2}{{ c_2}}^{2}{y}^{4} \right) {{ c_0}}^{6}+
 \Bigg( -200\,{{ c_3}}^{4}{y}^{10}-464\,{{ c_3}}^{2}{{ c_2}}^{2}
{y}^{6}-{\frac {40}{3}}\,{{ c_2}}^{4}{y}^{2} \Bigg) {{ c_0}}^{5}+
 \Bigg[ 108\,{{ c_3}}^{4}{y}^{8}-{\frac {40}{3}}\,{{ c_2}}^{4}\nonumber\\
 &&-304
\,{{ c_3}}^{2}{{ c_2}}^{2}{y}^{4} \Bigg] {{ c_0}}^{4}+ \left(
504\,{{ c_3}}^{4}{y}^{6}+320\,{{ c_3}}^{2}{{ c_2}}^{2}{y}^{2}
 \right) {{ c_0}}^{3}+ \left( -28\,{{ c_3}}^{4}{y}^{4}+160\,{{
c_3}}^{2}{{ c_2}}^{2} \right) {{ c_0}}^{2}-336\,{{ c_3}}^{4}{
c_0}\,{y}^{2}-112\,{{ c_3}}^{4} \Bigg] { c_3}\,{{ c_2}}^{5}{
{ e}^{-1/2\,{ c_0}\,{y}^{2}}}\nonumber\\
 &&-{\frac {195}{64}}\,{{ c_0}}^{3}{
 c_3}\, \Bigg\{  \left( -{\frac {16}{39}}\,{{ c_2}}^{6}+{{ c_3}}^
{6}{y}^{12}-{\frac {240}{13}}\,{{ c_3}}^{4}{{ c_2}}^{2}{y}^{8}-{
\frac {1392}{65}}\,{{ c_3}}^{2}{{ c_2}}^{4}{y}^{4} \right) {{
c_0}}^{6}-{\frac {252}{65}}\, \left( {{ c_3}}^{4}{y}^{8}-{\frac {80}{
21}}\,{{ c_2}}^{4}-{\frac {44}{3}}\,{{ c_3}}^{2}{{ c_2}}^{2}{y}^
{4} \right) {y}^{2}{{ c_3}}^{2}{{ c_0}}^{5}\nonumber\\
 &&-{\frac {492}{65}}\,
 \left( {{ c_3}}^{4}{y}^{8}-{\frac {80}{41}}\,{{ c_2}}^{4}-{\frac
{140}{41}}\,{{ c_3}}^{2}{{ c_2}}^{2}{y}^{4} \right) {{ c_3}}^{2}
{{ c_0}}^{4}+ \left( -{\frac {4032}{65}}\,{{ c_3}}^{4}{{ c_2}}^{
2}{y}^{2}-{\frac {32}{13}}\,{{ c_3}}^{6}{y}^{6} \right) {{ c_0}}^{
3}+ \left( -{\frac {2016}{65}}\,{{ c_3}}^{4}{{ c_2}}^{2}+{\frac {
144}{13}}\,{{ c_3}}^{6}{y}^{4} \right) {{ c_0}}^{2}\nonumber\\
 &&+{\frac {192}{
13}}\,{{ c_3}}^{6}{ c_0}\,{y}^{2}+{\frac {64}{13}}\,{{ c_3}}^{6}
 \Bigg\} {{ c_2}}^{3}{{ e}^{1/2\,{ c_0}\,{y}^{2}}}+{\frac {7}{
16}}\,{ c_0}\, \Bigg\{  \left( {\frac {240}{7}}\,{{ c_2}}^{6}+{{
 c_3}}^{6}{y}^{12}-27\,{{ c_3}}^{4}{{ c_2}}^{2}{y}^{8}+198\,{{
 c_3}}^{2}{{ c_2}}^{4}{y}^{4} \right) {{ c_0}}^{6}-6/7\,{y}^{2}{
{ c_3}}^{2} \Bigg\{ 252\,{{ c_2}}^{4}\nonumber\\
 &&+{{ c_3}}^{4}{y}^{8}+60{{
 c_3}}^{2}{{ c_2}}^{2}{y}^{4} \Bigg\} {{ c_0}}^{5}+ \left( -{
\frac {39{{ c_3}}^{6}{y}^{8}}{7}}-216\,{{ c_3}}^{2}{{ c_2}}^{4
}+{\frac {360}{7}}{{ c_3}}^{4}{{ c_2}}^{2}{y}^{4} \right) {{
c_0}}^{4}+ \left( {\frac {1440{{ c_3}}^{4}{{ c_2}}^{2}{y}^{2
}}{7}}-{\frac {86{{ c_3}}^{6}{y}^{6}}{7}} \right) {{ c_0}}^{3}+
 \Bigg\{ {\frac {720{{ c_3}}^{4}{{ c_2}}^{2}}{7}}\nonumber\\
 &&-{\frac {93}{7}}
{{ c_3}}^{6}{y}^{4} \Bigg\} {{ c_0}}^{2}-{\frac {60}{7}}{{
c_3}}^{6}{ c_0}{y}^{2}-{\frac {20{{ c_3}}^{6} }{7}}\Bigg\} {{
 c_3}}^{3}{ c_2}{ e}^{\frac{3{ c_0}{y}^{2}}{2}}+{\frac {2397{{ c_0}}^{7}{ c_3}{{ c_2}}^{7}}
{64}} \Bigg\{ {y}^{8} \left( {
\frac {16{{ c_2}}^{2}}{47}}+{{ c_3}}^{2}{y}^{4} \right) {{ c_0
}}^{6}- \left( {\frac {296{{ c_2}}^{2}{y}^{6}}{799}}+{\frac {1908
{{ c_3}}^{2}{y}^{10}}{799}} \right) {{ c_0}}^{5}\nonumber\\
 &&- \left( \frac {216{{ c_2}}^{2}{y}^{4}}{799}+{\frac {516{{ c_3}
}^{2}{y}^{8} }{799}}\right) {{ c_0}}^{4}+ \left( {\frac {2464{y}^{
6}{{ c_3}}^{2}}{799}}+{\frac {160{{ c_2}}^{2}{y}^{2} }{799}}\right) {{
 c_0}}^{3}+ \left( {\frac {432{{ c_3}}^{2}{y}^{4}}{799}}+{\frac
{80{{ c_2}}^{2} }{799}}\right) {{ c_0}}^{2}-{\frac {960{{ c_3}}^{2}{y}^{2}{ c_0}}{799}}
-{\frac {320{{ c_3}}^{2}}{799}}
 \Bigg\}{ e}^{-\frac{3{ c_0}{y}^{2}}{2}}\nonumber\\
 &&+3/8\,{ c_0}\, \Bigg\{
 \left( 84\,{{ c_2}}^{4}+{{ c_3}}^{4}{y}^{8}+30\,{{ c_3}}^{2}{{
 c_2}}^{2}{y}^{4} \right) {{ c_0}}^{4}+ \left( -120\,{{ c_3}}^{2
}{{ c_2}}^{2}{y}^{2}+11\,{{ c_3}}^{4}{y}^{6} \right) {{ c_0}}^{3
}+ \left( 21\,{{ c_3}}^{4}{y}^{4}-120\,{{ c_3}}^{2}{{ c_2}}^{2}
 \right) {{ c_0}}^{2}+20\,{{ c_3}}^{4}{ c_0}\,{y}^{2}\nonumber\\
 &&+10\,{{
c_3}}^{4} \Bigg\} {{ c_3}}^{5}{ c_2}\,{{ e}^{5/2\,{ c_0}\,{y}^
{2}}}-{\frac {31}{4}}\,{{ c_0}}^{9} \left( { c_0}\,{y}^{2}-1
 \right) ^{2} \left( { c_0}\,{y}^{2}+1/2 \right) ^{2} \left( {
\frac {20}{31}}+{\frac {20}{31}}\,{ c_0}\,{y}^{2}-{{ c_0}}^{2}{y}^
{4} \right) { c_3}\,{{ c_2}}^{9}{{ e}^{-5/2\,{ c_0}\,{y}^{2}}
}+{\frac {33}{16}}\,{ c_0}\,{{ c_3}}^{7} \Bigg\{  \Bigg[ {\frac {
80}{11}}\,{{ c_2}}^{2}\nonumber\\
 &&-{{ c_3}}^{2}{y}^{4} \Bigg] {{ c_0}}^{2}-
{\frac {20}{11}}\,{{ c_3}}^{2}{y}^{2}{ c_0}-{\frac {20}{11}}\,{{
 c_3}}^{2} \Bigg\} { c_2}\,{{ e}^{7/2\,{ c_0}\,{y}^{2}}}+5/4
\,{ c_0}\,{{ c_3}}^{9}{ c_2}\,{{ e}^{9/2\,{ c_0}\,{y}^{2}}}
+{\frac {843}{32}}\,{{ c_0}}^{6} \Bigg\{  \left( {{ c_3}}^{4}{y}^{8
}+{\frac {8}{281}}\,{{ c_2}}^{4}+{\frac {489}{281}}\,{{ c_3}}^{2}{
{ c_2}}^{2}{y}^{4} \right) {y}^{4}{{ c_0}}^{6}\nonumber\\
 &&- \left( {\frac {
1059{{ c_3}}^{4}{y}^{10}}{281}}+{\frac {600{{ c_3}}^{2}
{{ c_2}}^{2}{y}^{6}}{281}}+{\frac {4{{ c_2}}^{4}{y}^{2} }{281}}\right)
{c_0}^{5}- \left( {\frac {51{{ c_3}}^{4}{y}^{8}}{281}}+{
\frac {4{{ c_2}}^{4}}{281}}+{\frac {420{{ c_3}}^{2}{{
 c_2}}^{2}{y}^{4}}{281}} \right) {{ c_0}}^{4}+ \left( {\frac {1736{{ c_3}}^{4}{y}^{6}}{281}}
+{\frac {360{{ c_3}}^{2}{{ c_2}}^{
2}{y}^{2}}{281}} \right) {{ c_0}}^{3}\nonumber\\
 &&+ \left( {\frac {168}{281}}\,{{ c_3}
}^{4}{y}^{4}+{\frac {180}{281}}\,{{ c_3}}^{2}{{ c_2}}^{2} \right)
{{ c_0}}^{2}-{\frac {840}{281}}\,{{ c_3}}^{4}{ c_0}\,{y}^{2}-{
\frac {280}{281}}\,{{ c_3}}^{4} \Bigg\} {{ c_2}}^{6}{{ e}^{-{
 c_0}\,{y}^{2}}}-1/8\,{{ c_3}}^{4} \Bigg\{  \Bigg[ -210\,{{ c_2}}
^{6}+{{ c_3}}^{6}{y}^{12}+{\frac {153}{4}}\,{{ c_3}}^{4}{{ c_2}}
^{2}{y}^{8}\nonumber\\
 &&-378\,{{ c_3}}^{2}{{ c_2}}^{4}{y}^{4} \Bigg] {{ c_0}
}^{6}+ \left( 3\,{{ c_3}}^{6}{y}^{10}-36\,{{ c_3}}^{4}{{ c_2}}^{
2}{y}^{6}+630\,{{ c_3}}^{2}{{ c_2}}^{4}{y}^{2} \right) {{ c_0}}^
{5}+ \left( 6\,{{ c_3}}^{6}{y}^{8}+630\,{{ c_3}}^{2}{{ c_2}}^{4}
-171\,{{ c_3}}^{4}{{ c_2}}^{2}{y}^{4} \right) {{ c_0}}^{4}+
 \Bigg\{ 7\,{{ c_3}}^{6}{y}^{6}\nonumber\\
 &&-270\,{{ c_3}}^{4}{{ c_2}}^{2}{y}^
{2} \Bigg\} {{ c_0}}^{3}+ \left( -135\,{{ c_3}}^{4}{{ c_2}}^{2}+
6\,{{ c_3}}^{6}{y}^{4} \right) {{ c_0}}^{2}+3\,{{ c_3}}^{6}{
c_0}\,{y}^{2}+{{ c_3}}^{6} \Bigg\} {{ e}^{2\,{ c_0}\,{y}^{2}}}+{
\frac {387}{16}}\, \left( { c_0}\,{y}^{2}-1 \right)  \Bigg\{  \left(
{{ c_3}}^{2}{y}^{8}+{\frac {8}{129}}\,{{ c_2}}^{2}{y}^{4} \right)
{{ c_0}}^{4}\nonumber\\
 &&+ \left( -{\frac {4}{129}}\,{{ c_2}}^{2}{y}^{2}-{
\frac {60}{43}}\,{y}^{6}{{ c_3}}^{2} \right) {{ c_0}}^{3}+ \left(
-{\frac {40}{43}}\,{{ c_3}}^{2}{y}^{4}-{\frac {4}{129}}\,{{ c_2}}^
{2} \right) {{ c_0}}^{2}+{\frac {40}{43}}\,{{ c_3}}^{2}{y}^{2}{
 c_0}+{\frac {20}{43}}\,{{ c_3}}^{2} \Bigg\}  \left( { c_0}\,{y}
^{2}+1/2 \right) {{ c_0}}^{8}{{ c_2}}^{8}{{ e}^{-2\,{ c_0}\,{
y}^{2}}}\nonumber\\
 &&+3/8\, \left(  \left( 70{{ c_2}}^{4}+{{ c_3}}^{4}{y}^{8}
-6{{ c_3}}^{2}{{ c_2}}^{2}{y}^{4} \right) {{ c_0}}^{4}+
 \left( -45{{ c_3}}^{2}{{ c_2}}^{2}{y}^{2}+2{{ c_3}}^{4}{y}^
{6} \right) {{ c_0}}^{3}+ \left( 3{{ c_3}}^{4}{y}^{4}-45{{
c_3}}^{2}{{ c_2}}^{2} \right) {{ c_0}}^{2}+2{{ c_3}}^{4}{ c_0
}{y}^{2}+{{ c_3}}^{4} \right) {{ c_3}}^{6}{{ e}^{3{ c_0}
{y}^{2}}}\nonumber\\
 &&+{{ c_0}}^{10} \left( { c_0}\,{y}^{2}-1 \right) ^{3}
 \left( { c_0}\,{y}^{2}+1/2 \right) ^{3}{{ c_2}}^{10}{{ e}^{-3
\,{ c_0}\,{y}^{2}}}-3/8\,{{ c_3}}^{8} \left(  \left( {{ c_3}}^{2
}{y}^{4}-15\,{{ c_2}}^{2} \right) {{ c_0}}^{2}+{{ c_3}}^{2}{y}^{
2}{ c_0}+{{ c_3}}^{2} \right) {{ e}^{4\,{ c_0}\,{y}^{2}}}+1/8
\,{{ c_3}}^{10}{{ e}^{5\,{ c_0}\,{y}^{2}}}\nonumber\\
 &&-{\frac {159}{16}}\,{
{ c_0}}^{2} \Bigg[ -{\frac {15}{106}}\,{{ c_3}}^{2} \Bigg\{
 \left( 4\,{{ c_2}}^{6}+{{ c_3}}^{6}{y}^{12}+{\frac {17}{5}}\,{{
 c_3}}^{4}{{ c_2}}^{2}{y}^{8}+{\frac {336}{5}}\,{{ c_3}}^{2}{{
 c_2}}^{4}{y}^{4} \right) {{ c_0}}^{6}+ \left( {\frac {17}{5}}\,{{
 c_3}}^{6}{y}^{10}-{\frac {336}{5}}\,{{ c_3}}^{4}{{ c_2}}^{2}{y}
^{6}-56\,{{ c_3}}^{2}{{ c_2}}^{4}{y}^{2} \right) {{ c_0}}^{5}\nonumber\\
 &&+
 \left( 9/5\,{{ c_3}}^{6}{y}^{8}-56\,{{ c_3}}^{2}{{ c_2}}^{4}-{
\frac {56}{5}}\,{{ c_3}}^{4}{{ c_2}}^{2}{y}^{4} \right) {{ c_0}}
^{4}+ \left( 112\,{{ c_3}}^{4}{{ c_2}}^{2}{y}^{2}-{\frac {36}{5}}
\,{{ c_3}}^{6}{y}^{6} \right) {{ c_0}}^{3}+ \left( 56\,{{ c_3}}^
{4}{{ c_2}}^{2}-{\frac {68}{5}}\,{{ c_3}}^{6}{y}^{4} \right) {{
 c_0}}^{2}-12\,{{ c_3}}^{6}{ c_0}\,{y}^{2}\nonumber\\
 &&-4{{ c_3}}^{6}
 \Bigg\} {{ e}^{{ c_0}{y}^{2}}}-{{ c_0}}^{2} \Bigg\{  \left( 
{\frac {2}{159}}{{ c_2}}^{6}-{{ c_3}}^{6}{y}^{12}+{\frac {1059}{
106}}{{ c_3}}^{4}{{ c_2}}^{2}{y}^{8}+{\frac {150}{53}}{{ c_3
}}^{2}{{ c_2}}^{4}{y}^{4} \right) {{ c_0}}^{6}+{\frac {51}{106}}
 \left( {{ c_3}}^{4}{y}^{8}+{\frac {60}{17}}{{ c_2}}^{4}+{\frac
{672}{17}}{{ c_3}}^{2}{{ c_2}}^{2}{y}^{4} \right) {y}^{2}{{
c_3}}^{2}{{ c_0}}^{5}\nonumber\\
 &&+ \left( -{\frac {453}{106}}\,{{ c_3}}^{6}{y}^
{8}+{\frac {90}{53}}\,{{ c_3}}^{2}{{ c_2}}^{4}+{\frac {588}{53}}\,
{{ c_3}}^{4}{{ c_2}}^{2}{y}^{4} \right) {{ c_0}}^{4}+ \left( -{
\frac {840}{53}}\,{{ c_3}}^{4}{{ c_2}}^{2}{y}^{2}-{\frac {364}{53}
}\,{{ c_3}}^{6}{y}^{6} \right) {{ c_0}}^{3}+ \left( -{\frac {420}{
53}}\,{{ c_3}}^{4}{{ c_2}}^{2}+{\frac {168}{53}}\,{{ c_3}}^{6}{y
}^{4} \right) {{ c_0}}^{2}\nonumber\\
 &&+{\frac {420}{53}}\,{{ c_3}}^{6}{ c_0}
\,{y}^{2}+{\frac {140}{53}}\,{{ c_3}}^{6} \Bigg\} {{ c_2}}^{2}
 \Bigg] {{ c_2}}^{2} \Bigg\}  \Bigg[ { c_3}\,{{ e}^{1/2\,{
c_0}\,{y}^{2}}}{ c_2}\,{ c_0}+5/4\,{ c_0}\, \left( -4/5\,{ c_0}
\,{y}^{2}+{{ c_0}}^{2}{y}^{4}-4/5 \right) { c_3}\,{ c_2}\,{
{ e}^{-1/2\,{ c_0}\,{y}^{2}}}+ \left( { c_0}\,{y}^{2}-1
 \right) {{ c_0}}^{2}\nonumber\\
\end{eqnarray*}
\begin{eqnarray*}
 && \left( { c_0}\,{y}^{2}+1/2 \right) {{ c_2}
}^{2}{{ e}^{-{ c_0}\,{y}^{2}}}+1/2\,{{ c_3}}^{2}{{ e}^{{
c_0}\,{y}^{2}}}+ \left( 1/2\,{{ c_2}}^{2}-1/2\,{{ c_3}}^{2}{y}^{4}
 \right) {{ c_0}}^{2}-1/2\,{{ c_3}}^{2}{y}^{2}{ c_0}-1/2\,{{
c_3}}^{2} \Bigg] ^{-3} \Bigg[ 2\,{ c_3}\,{{ e}^{1/2\,{ c_0}\,{y
}^{2}}}{ c_2}\,{ c_0}+{{ c_3}}^{2}{{ e}^{{ c_0}\,{y}^{2}}}\nonumber\\
 &&+
{{ c_2}}^{2}{{ c_0}}^{2} \Bigg] ^{-2}=0\,.
\end{eqnarray*}
The above lengthy differential equation can be rewritten in a compact form as
\begin{eqnarray*}
f'''(y)N(y)+f''(y)N_1(y)+f'(y)N_2(y)+f(y)N_3(y)=0\,,
\end{eqnarray*}
where $N(y),\, N_1(y),\,N_2(y),\, N_3(y)$ are complicated functions of $y$ obtained by summing up the coefficients of the same derivatives. The asymptotic form of this differential equation, up to $O\left(\frac{1}{y}\right)$, gives  Eq. (\ref{dfas(r)}).
%
\end{document}